\documentclass[aps,twocolumn,superscriptaddress,tightenlines,nofootinbib,floatfix,longbibliography,preprintnumbers]{revtex4-1}
\usepackage[left=18mm,right=19mm,top=23mm,bottom=16mm]{geometry}

\usepackage{helvet}
\usepackage{sansmath}
\usepackage[fontsize=12pt]{scrextend}
\usepackage[12pt]{moresize}
\usepackage{ragged2e}

\setcitestyle{super}


\usepackage{setspace}

\usepackage{amsmath,amssymb}
\usepackage{bm}
\usepackage{comment}
\usepackage{graphicx}
\usepackage{color}
\usepackage{cancel}
\usepackage{tikz}
\usepackage{MnSymbol}
\usepackage{slashed}
\usetikzlibrary{shapes.misc}

\usepackage{mathtools}
\usepackage{CJKutf8}
\usepackage{soul}

\makeatletter
\renewcommand\@caption@fignum@sep{ $|$}
\renewcommand\@make@capt@title[2]{%
\@ifx@empty\float@link{\@firstofone}{\expandafter\href\expandafter{\float@link}}%
{\raggedright\ssmall\fontfamily{\sfdefault}\selectfont{\textbf{#1}}}\@caption@fignum@sep#2}%
\makeatother

\usepackage[titletoc]{appendix}
\usepackage{etoolbox}

\usepackage[hypertexnames=false]{hyperref}
\hypersetup{
    colorlinks=true,
    linkcolor=blue,
    citecolor=blue,
    filecolor=blue, 
    urlcolor=blue
}

\usepackage{multibib}
\newcites{M}{Methods References}

\def\a{{\alpha}}

\def\d{{\delta}}
\def\D{{\Delta}}
\def\e{{\epsilon}}
\def\g{{\gamma}}

\def\l{{\lambda}}

\def\S{{\Sigma}}
\def\s{{\sigma}}
\def\t{{\tau}}

\def\Dslash{D\hskip-0.65em /}

\def\eqref#1{{(\ref{#1})}}

\newcount\hour \newcount\hourminute \newcount\minute
\hour=\time \divide \hour by 60
\hourminute=\hour \multiply \hourminute by 60
\minute=\time \advance \minute by -\hourminute

\def\ga{1.271 \pm 0.013}
\def\gaM{1.271(11)(06)}
\def\gaindividual{1.2711(103)^{\textrm{s}}(39)^{\chi}(15)^{\textrm{a}}(19)^{\textrm{v}}(04)^{\textrm{I}}(55)^{\textrm{M}}}

\begin{document}
\newcommand{\docfont}{\scriptsize}
\docfont
\renewcommand{\figurename}{{\bf Figure}}

\makeatletter
\def\@affil@script#1#2#3#4{
	\@ifnum{#1=\z@}{}{
		\raggedright
		\begingroup
		\@ifnum{\c@affil<\affil@cutoff}{}{%
			\def\@thefnmark{#1}\@makefnmark
		}~\hspace{-4pt}#3
		\@if@empty{#4}{}{\frontmatter@footnote{#4}}
		\endgroup
	}
}
\makeatother

\preprint{\scriptsize
		BNL-203631-2018-JAAM, 
		INT-PUB-18-021, 
		LLNL-JRNL-747003, 
		RBRC-1283, 
		RIKEN-iTHEMS-Report-18
}
\maketitle

\onecolumngrid{
{\Large \noindent \textbf{A percent-level determination of the nucleon axial \\ coupling from Quantum Chromodynamics}}}

\vspace{5pt}
\noindent \raggedright {C.C.~Chang \begin{CJK*}{UTF8}{bsmi}(張家丞)\end{CJK*},${}^{1,2}$ A.N.~Nicholson,${}^{3,1,4}$ E.~Rinaldi,${}^{5,6,1}$ E.~Berkowitz,${}^{7,6}$ N.~Garron,${}^{8}$ D.A.~Brantley,${}^{9,1}$ H.~Monge-Camacho,${}^{9,1}$ C.~Monahan,${}^{10,11}$ C.~Bouchard,${}^{12,9}$ M.A.~Clark,${}^{13}$ B.~Jo\'o,${}^{14}$ T.~Kurth,${}^{15,1}$ K.~Orginos,${}^{9,16}$ P.~Vranas,${}^{6,1}$ and A.~Walker-Loud${}^{1,6}$}

\vspace{5pt}
\noindent  \scriptsize{\raggedright
	${}^1$Nuclear Science Division, Lawrence Berkeley National Laboratory, Berkeley, CA 94720, USA\\
	${}^2$Interdisciplinary Theoretical and Mathematical Sciences Program (iTHEMS), RIKEN, 2-1 Hirosawa, Wako, Saitama 351-0198, Japan\\
	${}^3$Department of Physics, University of California, Berkeley, CA 94720, USA\\
	${}^4$Department of Physics and Astronomy, University of North Carolina, Chapel Hill, NC 27516-3255, USA\\
	${}^5$RIKEN-BNL Research Center, Brookhaven National Laboratory, Upton, NY
	11973, USA\\
	${}^6$Physics Division, Lawrence Livermore National Laboratory, Livermore, CA 94550, USA\\
	${}^7$Institut f\"ur Kernphysik and Institute for Advanced Simulation, Forschungszentrum J\"ulich, 54245 J\"ulich Germany\\
	${}^8$Theoretical Physics Division, Department of Mathematical Sciences, University of Liverpool, Liverpool L69 3BX, UK\\
	${}^9$Department of Physics, The College of William \& Mary, Williamsburg, VA 23187, USA\\
	${}^{10}$Department of Physics and Astronomy, Rutgers, The State University of
	New Jersey, Piscataway, NJ 08854, USA\\
	${}^{11}$Institute for Nuclear Theory, University of Washington, Seattle, WA 98195, USA 12School of Physics and Astronomy,\\
	${}^{12}$University of Glasgow, Glasgow G12 8QQ, UK\\
	${}^{13}$NVIDIA Corporation, 2701 San Tomas Expressway, Santa Clara, CA 95050, USA\\
	${}^{14}$Scientific Computing Group, Thomas Jefferson National Accelerator Facility, Newport News, VA 23606, USA\\
	${}^{15}$NERSC, Lawrence Berkeley National Laboratory, Berkeley, CA 94720, USA\\
	${}^{16}$Theory Center, Thomas Jefferson National Accelerator Facility, Newport News, VA 23606, USA}

\vspace{5pt}
\twocolumngrid

\justify
\setul{}{0.75pt}

{\textbf{
The \textit{axial coupling of the nucleon}, $g_A$, is the strength of its coupling to the \textit{weak} axial current of the Standard Model of particle physics, in much the same way as the electric charge is the strength of the coupling to the electromagnetic current.
This axial coupling dictates the rate at which neutrons decay to protons, the strength of the attractive long-range force between nucleons and other features of nuclear physics. 
Precision tests of the Standard Model in nuclear environments require a quantitative understanding of nuclear physics rooted in Quantum Chromodynamics, a pillar of the Standard Model. The prominence of $g_A$ makes it a benchmark quantity to determine theoretically -- a difficult task because quantum chromodynamics is non-perturbative, precluding known analytical methods. Lattice Quantum Chromodynamics provides a rigorous, non-perturbative definition of quantum chromodynamics that can be implemented numerically.
It has been estimated that a precision of two percent would be possible by 2020 if two challenges are overcome\cite{usqcd_doe_2016,Bhattacharya:2016zcn}: contamination of $g_A$ from excited states must be controlled in the calculations and statistical precision must be improved markedly\cite{Edwards:2005ym, Capitani:2012gj, Horsley:2013ayv, Bali:2014nma, Abdel-Rehim:2015owa, Bhattacharya:2016zcn, Bhattacharya:2016zcn, Alexandrou:2017hac, Capitani:2017qpc, Lin:2017snn}. Here we report a calculation of $\mathbf{g_A^{QCD} = \ga}$, using an unconventional method  inspired by the Feynman--Hellmann theorem\cite{Bouchard:2016heu} that overcomes these challenges.
}}

\begin{figure*}
\begin{tabular}{c}
\makebox[\textwidth]{\includegraphics[width=0.8\paperwidth]{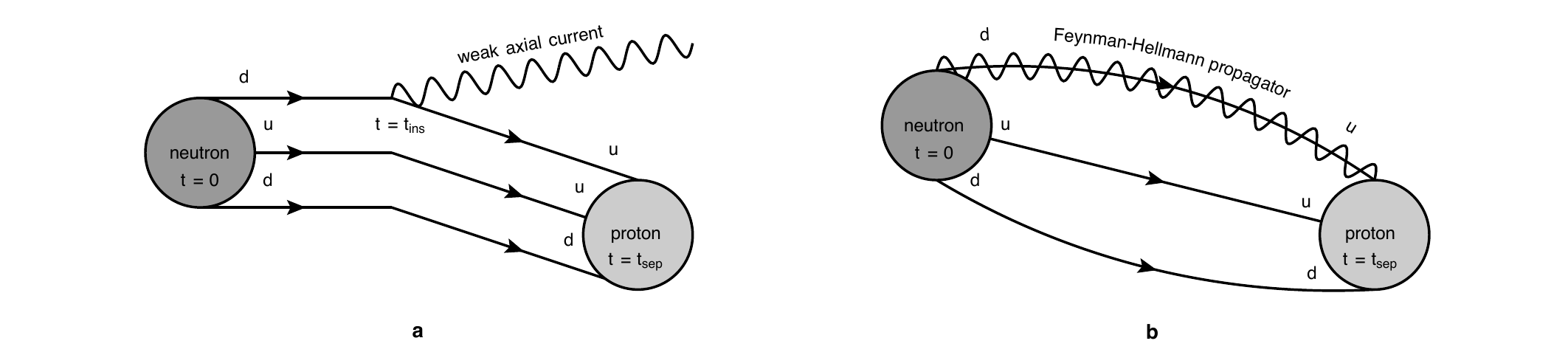}}
\end{tabular}
\caption{\label{fig:feyn_diagram}
\textbf{Feynman diagrams of $g_A$.} 
The decay of a neutron to a proton occurs when one of the \textit{down} quarks ($d$) in the neutron is converted to an \textit{up} ($u$) quark via the vector and axial components of the weak current. Not depicted in these figures are the infinite set of diagrams describing the coupling of the gluons to the quarks, gluons to gluons and the dynamical quark/anti-quark pairs popping in and out of the vacuum -- it is this infinite set of graphs that requires the use of a computational approach to QCD. The time, $t$, refers to calculational details discussed in the text. \textbf{a,} The standard method of computing $g_A$ relies upon three different times, the creation time $t=0$, the current insertion time $t_{\textrm{ins}}$ and the separation time $t_{\textrm{sep}}$.  Controlling the excited state systematics requires varying both $t_{\textrm{ins}}$ and $t_{\textrm{sep}}$. \textbf{b}, Our Feynman-Hellmann method\cite{Bouchard:2016heu} sums over all possible interaction times ($t_\textrm{ins}$) of the external weak axial current, leading to an exponential enhancement of the signal.
}
\end{figure*}

To demonstrate the efficacy of lattice Quantum Chromodynamics (LQCD) for the broad nuclear physics research program, one must begin by demonstrating control over the simplest quantities, such as $g_A$. In addition to those mentioned above, there are a number of challenges in using LQCD to compute properties of nucleons and nuclei. The first challenge arises from the non-perturbative features of Quantum Chromodynamics (QCD) itself. QCD describes the interactions between quarks and gluons, the basic constituents of nucleons, through the Lagrangian, $\mathcal{L}_{QCD} = -G^2 / (4g) + \sum_q \bar{\psi}_q [\Dslash + m_q  ]\psi_q$, where the quark fields, $\psi_q$, come in flavours $q=\{u,d,s,...\}$ with masses $m_q = \{m_u, m_d, m_s, \dots\}$. $G^2$ describes the non-linear gluon self-interactions while $\Dslash\,$ includes the quark-gluon interactions, both with a strength determined by the coupling, $g$. Most of nuclear physics depends on only three or four input parameters from QCD: $g$, the light quark masses, $m_u$ and $m_d$ and in some cases, the strange quark mass, $m_s$. Once these parameters are fixed, and electroweak corrections are added, all of nuclear physics -- from the kiloelectronvolt energy levels in nuclei to the energy densities of the neutron star equation of state (a few hundred megaelectronvolts per cubic fermi (fm), where 1 fm = $10^{-15}$~m) -- can in principle be predicted from QCD.

At short distances (high energies), such as those explored by the Large Hadron Collider at CERN, QCD has been rigorously tested, because in this energy regime, $g \ll 1$ and \textit{perturbative} methods are applicable. At long distances of approximately 1~fm (low energies) characteristic of nuclear physics, $g$ is large and perturbation theory fails to converge. Consequently, quarks and gluons are confined in protons, neutrons and other \textit{hadrons} observed experimentally. Fortunately, \textit{non-perturbative} calculations can be carried out in the strong-coupling regime using LQCD, the only first-principles approach known to control all sources of systematic uncertainty.

LQCD is the formulation of QCD on a finite four-dimensional spacetime lattice following the Feynman path integral description. Monte Carlo methods are used to sample the resulting high-dimensional integrals stochastically. The values of the \textit{lattice spacing}, $a$, and \textit{finite size}, $L$, are chosen to encompass the characteristic length scales emergent from QCD, such as the proton radius $r_p \sim 0.8$~fm. In state-of-the-art calculations, typical lattice spacings are $0.04 \lesssim a/r_p \lesssim 0.2$ and typical spatial extents are $3 \lesssim L/r_p \lesssim 8$. The continuum and infinite volume limits are recovered through the extrapolation in several values of $a$ and $L$. Additionally, the input values of the quark masses in the LQCD calculations do not typically reproduce their physical values. State-of-the-art calculations now regularly access $m^{\text{calc.}}_\pi \sim 130$~MeV, where the pion $(\pi)$, the lightest hadron, is used to calibrate the input light-quark mass under the guidance of $m^{\text{nature}}_\pi = 139.6$~MeV\cite{Olive:2016xmw}. In this work, $m_\pi$ ranges between $130$ and $400$~MeV, allowing for a fully controlled interpolation (input parameters of the calculation are provided in Extended Data Tab.~\ref{tab:hisq}). The continuum and infinite volume extrapolations and pion mass interpolations are necessary for all LQCD calculations in order to compare with nature and make predictions.

Over the past decade, the LQCD community has determined hadronic properties for \textit{mesons} (which are QCD eigenstates composed of one quark and one antiquark) with fully controlled systematic uncertainties at the subpercent level, yielding some of the most stringent tests on the structure of the Standard Model (SM). The Flavour Lattice Averaging Group produces a world average of meson properties determined from LQCD\cite{Aoki:2016frl}, similar to the Particle Data Group (PDG) averages of experimental results\cite{Olive:2016xmw}. In contrast to mesons, stochastic sampling of the nucleon path integral results in an exponentially smaller signal-to-noise ratio, hence requiring exponentially more computational resources to replicate the precision achieved for meson properties. In fact, only a handful of LQCD calculations involving nucleons have demonstrated control over all sources of uncertainty\cite{Durr:2008zz,Borsanyi:2014jba}. Insight provided by previous LQCD calculations of $g_A$ also identifies the contamination of \textit{excited states} as another major source of uncertainty \cite{Edwards:2005ym, Capitani:2012gj, Horsley:2013ayv, Bali:2014nma, Abdel-Rehim:2015owa, Bhattacharya:2016zcn, Bhattacharya:2016zcn, Alexandrou:2017hac, Capitani:2017qpc, Lin:2017snn}. Excited state contamination, which results from the imperfect coupling of the chosen creation operators to the state of physical interest, is present in all lattice calculations, and has proven to be particularly problematic for calculations of $g_A$.

The standard method of calculating $g_A$, as shown in Fig.~\ref{fig:feyn_diagram}\textbf{a}, relies upon two independent separation times: the separation of the neutron and proton ($t_{\textrm{sep}}$), and the separation of the neutron and the insertion of the weak axial current ($t_\textrm{ins}$). Although $g_A$ is independent of both times, the contamination from excited states is time dependent. This contamination shifts the calculated values of $g_A$ at small time separations, but vanishes exponentially with respect to $t_\textrm{ins}$ and $t_\textrm{sep}-t_\textrm{ins}$. Computational limitations restrict the calculations to have fixed (and relatively small) neutron-proton separation times, requiring multiple calculations with varying values of $t_\textrm{sep}$ to fully control the excited state contributions. However, the relative stochastic noise grows exponentially with $t_\textrm{sep}$ while only vanishing with the square root of the stochastic sample size. Therefore, overcoming the noise requires exponentially more computing resources, rendering the standard method an expensive strategy.

\begin{figure}
\includegraphics[width=\columnwidth]{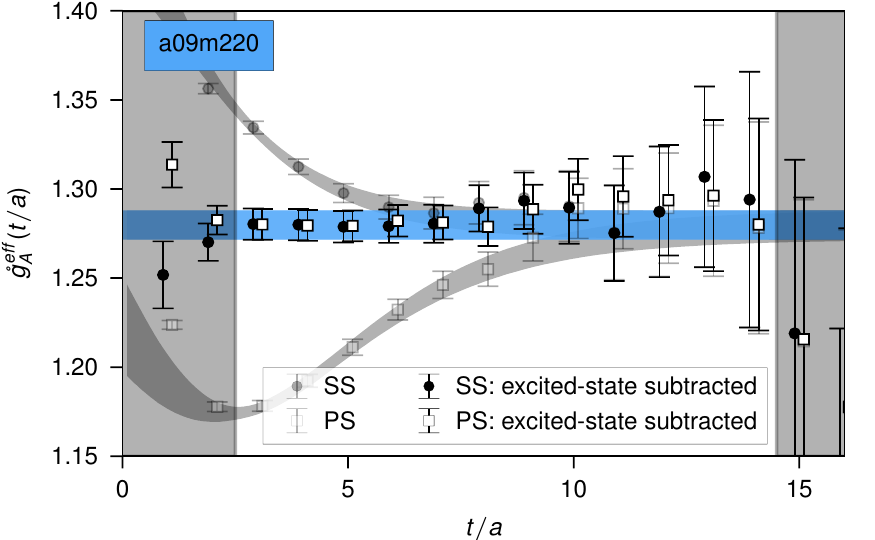}
\caption{\label{fig:fh_method}
\textbf{Demonstration of the improved method\cite{Bouchard:2016heu} on the ensemble with lattice spacing $a\approx0.09$~fm and $m_\pi\approx220$~MeV.} The two sets of results for $\mathring{g}_A^{\rm eff}(t/a)$ correspond to different choices of annihilation operators for the nucleon, denoted as SS and PS. At long times, both values must approach the ground state value of $\mathring{g}_A$ asymptotically, whereas at short times, they couple differently to the excited state contributions. The raw numerical results are shown in grey and the grey bands represent the full fit to the data (points inside the vertical grey bands are not included in the fits). Error bars correspond to one standard error of the mean (s.e.m.). The solid black and white data points are reconstructed from the two datasets, with the excited states (determined by the fit) subtracted from the raw results. The solid blue band is the ground state value of $\mathring{g}_A$ determined by the full fit. We make efficient use of points at small Euclidean times, before the stochastic noise overwhelms the signal. The agreement between the subtracted data and the asymptotic large-time value of $\mathring{g}_A$, even at short times, demonstrates our control over excited state contributions. The time axis is given in dimensionless lattice units, with $a \approx 0.09$~fm corresponding to $3\times10^{-25}$~s, so that $t/a=2$ corresponds to $6\times10^{-25}$~s.
}
\end{figure}

In contrast, the method we use in this work\cite{Bouchard:2016heu}, inspired by the Feynman-Hellmann Theorem, uses an explicit sum over all current insertion times, $t_{\textrm{ins}}$ (Fig.~\ref{fig:feyn_diagram}\textbf{b}), with the ability to vary $t=t_\textrm{sep}$, at the numerical cost of a \emph{single} separation time of the standard method: all excited state contributions depend only upon $t$ and the computation must asymptote to $g_A$ in the large $t$ limit (Fig.~\ref{fig:fh_method}). By analysing the spectrum and $g_A$ matrix element calculations simultaneously with nonlinear regression, we demonstrate the ability to fully control excited state contributions and determine precise values of $g_A$, as suggested by the agreement between the data (gray points with error bars) and the fit \textit{Ans\"{a}tze} (gray bands). In Supplemental Material Sec.~\ref{sec:correlators}, Extended Data Fig.~\ref{fig:correlator_fitcurves} and Supplemental Figs.~\ref{fig:a15m310a15m220_curve}--\ref{fig:a09m310a09m220_curve}, we show this is true for all ensembles (different choices of $a$, $L$, and $m_\pi$) in our calculation. In summary, this Feynman-Hellman inspired method\cite{Bouchard:2016heu} provides access to more data ($t=t_\textrm{sep}$ in Fig.~\ref{fig:fh_method}) with a reduced computational cost, allowing us to remove the unwanted excited state contamination and utilize data at early times, where the signal-to-noise ratio is exponentially more precise, thus resolving both of the aforementioned major challenges to determining $g_A$.

What remains is to extrapolate the values of $g_A$ obtained from our lattice calculations to the physical parameters. Effective Field Theory (EFT)\cite{Weinberg:1978kz} is employed to provide a rigorous prescription for performing the continuum and infinite volume extrapolations along with the interpolation to the physical pion mass. First, one identifies the relevant degrees of freedom for low-energy nuclear physics, which are the nucleons and pions. Second, one identifies a small expansion parameter, $\e$, which often emerges through a ratio of length scales; for pions, this is $\e_\pi = m_\pi / (4\pi F_\pi)$, where $F_\pi$ is the quantity known as the pion decay constant. $F_\pi^{\text{nature}}$ is measured to be $92.1(1.2)$~MeV\cite{Olive:2016xmw}, and $\e^{\text{nature}}_\pi \sim 0.12$. The resulting EFT may be systematically improved: when working to $\mathrm{O}(\e_\pi^n)$ (where $n = 0$ denotes leading order, $n = 1$ next-to-leading order, and so on), the truncation errors enter at $\mathrm{O}(\e_\pi^{n+1})$.

Chiral Perturbation Theory ($\chi$PT) is the EFT of pions\cite{Gasser:1983yg} and their interactions with nucleons\cite{Jenkins:1990jv}, and describes all possible interactions between them that are consistent with the symmetries of QCD, ordered by increasing powers of $\e_\pi$. Although the forms of the interactions are known, the strengths of the interactions are emergent low-energy couplings, and can be determined only from experiment or LQCD calculations.  However, once the couplings are known, $\chi$PT can be used to make predictions of new quantities, and in particular, can be used to describe the simulated universes where the quark masses differ from those in nature.  This allows for a model-independent interpolation of LQCD results to $m_\pi^{\text{nature}}$.

\begin{figure*}{\docfont
\begin{tabular}{cc}
\includegraphics[width=\columnwidth]{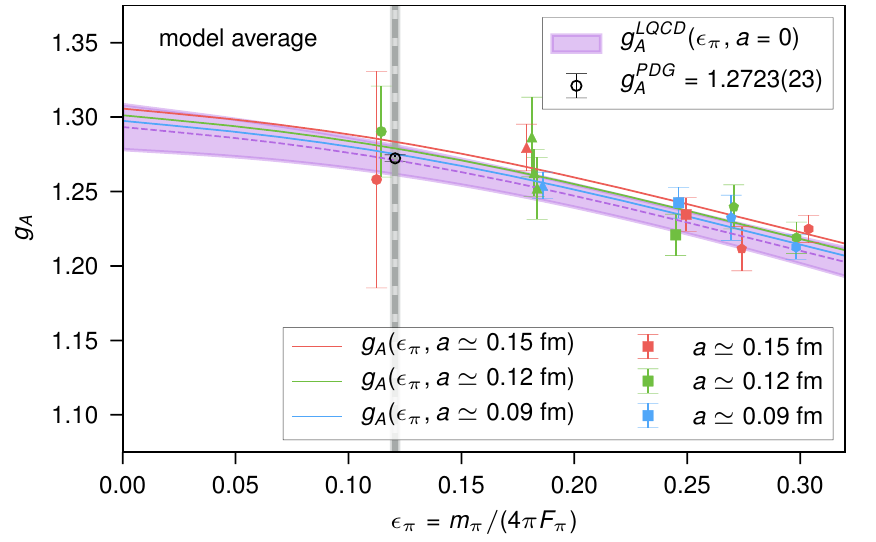}
&
\includegraphics[width=\columnwidth]{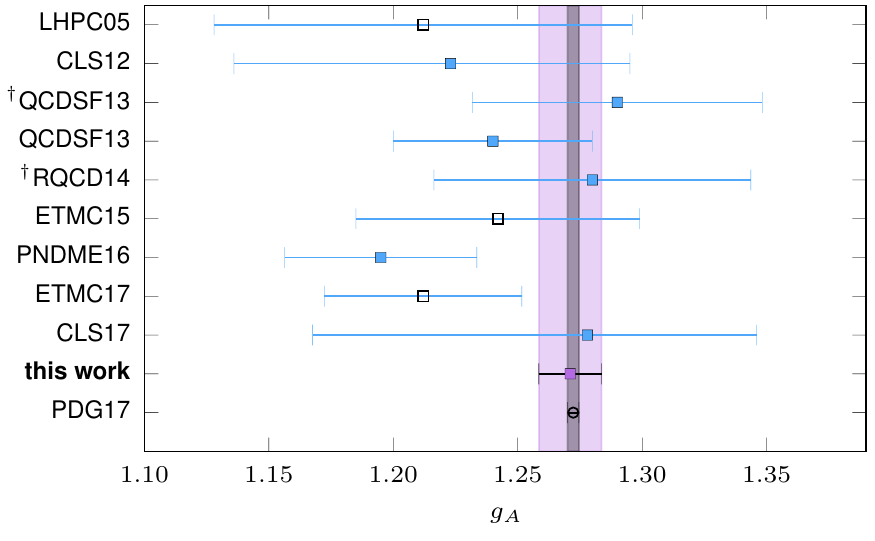}
\\
\docfont\textbf{a}&\docfont\textbf{b}
\end{tabular}
\caption{\label{fig:results}
\textbf{Physical-point extrapolation for this work and summary of $g_A$ from LQCD.} \textbf{a,} The solid red, green and blue curves are the central values of $g_A$ as a function of $\e_\pi$ at fixed lattice spacings and infinite volume, and the black circle represents the experimental value.  The magenta band represents the central 68\% confidence band of the continuum and infinite volume extrapolated value of $g_A$ as a function $\e_\pi$, and its range at the physical pion mass, given by its intersection with the gray band, is our main result.  The numerical results have been adjusted to their infinite volume values. Some of the results are slightly shifted horizontally for visual clarity. \textbf{b,} Summary of select LQCD calculations of $g_A$ along with the result of this work and the experimental determination (PDG17). The vertical magenta band is our full uncertainty to guide the eye, while the vertical gray band is the experimental uncertainty. Results with closed symbols have also included an extrapolation to the continuum limit, while results with open symbols have only included an extrapolation/interpolation to the physical pion mass. When provided separately, the statistical and systematic uncertainties are added in quadrature. Labels marked with $\dagger$ are indirectly obtained from an extrapolation of $g_A/F_\pi$. Previous peer reviewed works are available for LHPC05\cite{Edwards:2005ym}, CLS12\cite{Capitani:2012gj}, QCDSF13\cite{Horsley:2013ayv}, RQCD14\cite{Bali:2014nma}, ETMC15\cite{Abdel-Rehim:2015owa}, PNDME16\cite{Bhattacharya:2016zcn}, ETMC17\cite{Alexandrou:2017hac} and CLS17\cite{Capitani:2017qpc}.  The averaged experimental determination is obtained from the Particle Data Group\cite{Olive:2016xmw}. Uncertainties are one s.e.m.}
}\end{figure*}

$\chi$PT is also extended to account for artefacts arising from the finite periodic volume\cite{Gasser:1987zq}. For the large volumes used in our calculation, the small parameter controlling the finite-volume corrections scales approximately as $\e_{L} = e^{-m_\pi L}$. Extended Data Fig.~\ref{fig:IV_extrapolation} shows consistency between the predicted finite-volume corrections and our results at fixed pion mass.

Artefacts introduced by our calculation at non-zero lattice spacing are also accounted for with EFT. Unlike dependence on $\e_\pi$ and $\e_L$, which are governed purely by the long-distance dynamics of QCD, the continuum extrapolation depends upon the specific discretization of the QCD Lagrangian, or \textit{lattice action}, employed in the calculation. To parameterize these artefacts, one uses Symanzik's EFT\cite{Symanzik:1983dc} and expands the non-local discretized action around small lattice spacings, giving a series of purely local interactions. The resulting effects in low-energy dynamics can be systematically understood. The dependence on the choice of discretization must vanish in the continuum limit since the only interactions remaining are those of QCD. The lattice action we have chosen\cite{Berkowitz:2017opd} was designed to minimise the leading discretization errors, such that the leading corrections scale as $\mathrm{O}(a^2)$, and to preserve more of the underlying symmetries of QCD. This choice of lattice action yields a mild continuum extrapolation (Extended Data Fig.~\ref{fig:continuum_extrapolation}).

The final extrapolation of our results (Extended Data Table~\ref{tab:ga_extrap}) is presented in Fig.~\ref{fig:results}\textbf{a}. For quantities with mild pion mass dependence, such as $g_A$, a simple Taylor expansion in $\e_\pi$ or $\e_\pi^2$ in addition to the $\chi$PT extrapolation, provides a robust extrapolation/interpolation of the results. We perform the extrapolation with several models and our final result is determined as a \textit{model average}, depicted in Extended Data Figure~\ref{fig:model_selection}, and described in detail in Supplemental Material Secs.~\ref{sec:extrap} and \ref{sec:uncertainty}. Our final result, $g_A = \ga$, with the uncertainties broken down to the different contributions of statistical (s), chiral ($\chi$), continuum (a), infinite volume (v), isospin breaking (I) and model-selection (M) is
\begin{equation}\label{eq:ga_breakdown}
g_A = \gaindividual\, .
\end{equation}
This value that is commensurate with the experimentally determined value, $g_A^{\rm PDG} = 1.2723(23)$\cite{PhysRevLett.56.919, YEROZOLIMSKY1997240, LIAUD199753, Mostovoi:2001ye, PhysRevLett.100.151801, Mund:2012fq, Mendenhall:2012tz,Olive:2016xmw}.

Figure~\ref{fig:results}\textbf{b} summarises the improvement of the LQCD determination of $g_A$ achieved by this work. These results are derived from three lattice spacings, five values of the pion (quark) masses and multiple volumes, which control the three standard extrapolations (the input values of parameters used in our calculation are provided in Extended Data Table~\ref{tab:hisq}). Additionally, we demonstrate that our result is robust under different truncations/variations in the extrapolation function (Extended Data Fig.~\ref{fig:final_stability}) and that the perturbative expansion converges over the range of parameters used, as discussed in the Supplemental Material Sec.~\ref{sec:uncertainty} and shown in Extended Data Fig.~\ref{fig:chipt_convergence}. Details on the individual contributions to our total uncertainty may be found in Supplemental Material Sec.~\ref{sec:error_budget}.

Our result, Eq.~\eqref{eq:ga_breakdown}, is predominantly limited by statistics. This signifies a straightforward path for improvement: more precise results at the physical pion mass will reduce the statistical, extrapolation and model-selection uncertainties, which are the three largest. An uncertainty comparable to that of measurements may offer insight into the upward trending value of $g_A$ observed in the most recent set of experiments\cite{Olive:2016xmw}. At the present, our result has a noticeable phenomenological impact, as depicted in Extended Data Fig.~\ref{fig:right_handed_couplings}. Using EFT, experimental results from collider and low-energy experiments can be used to place bounds on right-handed BSM currents\cite{Alioli:2017ces} with our result placing one of the most stringent bounds.

\def\bibfont{\scriptsize}
\bibliographystyle{revtex_nature}
\bibliography{c51_bib}

\bigskip\noindent
\textbf{Supplementary Information} is available in the online version of this paper.

\smallskip\noindent
\textbf{Code and data availability:}
The majority of the software used for this work is publicly available on github. The \texttt{Chroma} software suite (developed by USQCD) is available at \href{https://github.com/JeffersonLab/chroma}{https://github.com/JeffersonLab/chroma}. This was linked against \texttt{Quda} (the optimised library for performing calculations on NVIDIA GPU-enable machines) which is available at \href{https://github.com/lattice/quda}{https://github.com/lattice/quda}. A private C++ layer is compiled on top of \texttt{Chroma}, which will become available at \href{https://github.com/callat-qcd/lalibe}{https://github.com/callat-qcd/lalibe} within one year. The original lattice QCD correlation functions, as well as our analysis results are available at \href{https://github.com/callat-qcd/project_gA}{https://github.com/callat-qcd/project\_gA} and \href{https://zenodo.org/record/1241374}{https://zenodo.org/record/1241374}.

\smallskip\noindent
\textbf{Acknowledgements}
We thank C.~Bernard, A.~Bernstein, P.J.~Bickel, C.~Detar, A.X.~El-Khadra, W.~Haxton, Y.~Hsia, V.~Koch, A.S.~Kronfeld, W.T.~Lee, G.P.~Lepage, E.~Mereghetti, G.~Miller, A.E.~Raftery, D.~Toussaint and F.~Yuan for discussions.
We thank E.~Mereghetti for an updated Extended Data Fig.~\ref{fig:right_handed_couplings}\cite{Alioli:2017ces}.
We thank the MILC Collaboration for providing their HISQ configurations\cite{Bazavov:2012xda} without restriction.
An award of computer time was provided by the Innovative and Novel Computational Impact on Theory and Experiment (INCITE) program to CalLat (2016).
We thank the LLNL Multiprogrammatic and Institutional Computing program for Grand Challenge allocations on the LLNL supercomputers.
This research used the NVIDIA GPU-accelerated Titan supercomputer at the Oak Ridge Leadership Computing Facility at the Oak Ridge National Laboratory, which is supported by the Office of Science of the U.S. Department of Energy under Contract No. DE-AC05-00OR22725, and the GPU-enabled Surface and RZHasGPU and BG/Q Vulcan clusters at LLNL.
This work was supported by the NVIDIA Corporation (MAC), the DFG and the NSFC Sino-German CRC110 (EB), an LBNL LDRD (AWL), the RIKEN Special Postdoctoral Researcher Program (ER), the Leverhulme Trust (NG), the U.S. Department of Energy, Office of Science: Office of Nuclear Physics (EB, CMB, DAB, CCC, TK, CM, HMC, AN, ER, BJ, KO, PV, AWL); Office of Advanced Scientific Computing (EB, BJ, TK, AWL); Nuclear Physics Double Beta Decay Topical Collaboration (DAB, HMC, AWL); and the DOE Early Career Award Program (DAB, CCC, HMC, AWL).
This work was performed under the auspices of the U.S. Department of Energy by LLNL under Contract No. DE-AC52-07NA27344 (EB, ER, PV).
Part of this work was performed at the Kavli Institute for Theoretical Physics supported by NSF Grant No. PHY-1748958.

\smallskip\noindent
\textbf{Author Contributions}
The project management is run by AWL and PV.
The computing allocations were written by EB, PV, AWL, TK, AN and ER.
The FH method was implemented by KO, AWL, CCC, CB and TK.
The lattice action was designed by CM, AWL and KO.
The QUDA MDWF solver was optimised by MAC.
The integration of the QUDA MDWF solver to Chroma was implemented by TK and BJ.
The HMC generation of new ensembles was done with MILC software (v7.8.0) by ER and AWL.
The implementation of code for the mixed mesons using MILC and Chroma was done by ER, BJ and AWL.
The calculations were performed by ER, AWL, EB, CCC, AN, DAB, HMC, using the job management software written by EB.
The non-perturbative renormalisation was performed by DAB, HMC, NG and AWL.
The correlator analysis was performed by CCC, AN, AWL, CB and CM.
The extrapolation analysis was performed by CCC and AWL.

\smallskip\noindent
\textbf{Author Information}
Reprints and permissions information is available at \url{www.nature.com/reprints}.
The authors declare no competing financial interests.
Correspondence and requests for materials should be addressed to A.W-L. \href{mailto:awalker-loud@lbl.gov}{awalker-loud@lbl.gov}.

\renewcommand\appendixname{EXTENDED DATA}
\renewcommand\thesection{}
\docfont

\clearpage
\renewcommand{\figurename}{{\bf Extended Data Figure}}
\renewcommand{\tablename}{{\bf Extended Data Table}}
\setcounter{figure}{0}
\setcounter{table}{0}
\onecolumngrid

\begin{table*}{\docfont
\begin{tabular}{lccccc}
		\hline\hline
		ensemble& $\e_\pi$& $m_\pi L$& $a/w_0$& $\alpha_S$& $g_A$\\
		\hline
		a15m400   & 0.30374(53) & 4.8451(49) & 0.8804(3) & 0.58801 & 1.216(06) \\
		a15m350   & 0.27411(50) & 4.2359(47) & 0.8804(3) & 0.58801 & 1.198(13) \\
		a15m310   & 0.24957(36) & 3.7772(48) & 0.8804(3) & 0.58801 & 1.215(12) \\
		a15m220   & 0.18084(30) & 3.9673(45) & 0.8804(3) & 0.58801 & 1.274(14)\\
		a15m130   & 0.11340(74) & 3.227(19)\phantom{0} & 0.8804(3) & 0.58801 & 1.270(72) \\
		\hline
		a12m400   & 0.29841(52) & 5.8428(39) & 0.7036(5) & 0.53796 & 1.217(10) \\
		a12m350   & 0.27063(69) & 5.1352(49) & 0.7036(5) & 0.53796 & 1.236(14) \\
		a12m310   & 0.24485(50) & 4.5282(41) & 0.7036(5) & 0.53796 & 1.214(13) \\
		a12m220S & 0.18419(57) & 3.2523(76) & 0.7036(5) & 0.53796 & 1.272(28) \\
		a12m220   & 0.18221(42) & 4.2959(56) & 0.7036(5) & 0.53796 & 1.259(15) \\
		a12m220L & 0.18156(44) & 5.3604(61) & 0.7036(5) & 0.53796 & 1.252(21) \\
		a12m130   & 0.11347(50) & 3.899(12)\phantom{0} & 0.7036(5) & 0.53796   & 1.292(30) \\
		\hline
		a09m400   & 0.29818(53) & 5.7965(46) & 0.5105(3) & 0.43356 & 1.210(08) \\
		a09m350   & 0.26949(57) & 5.0502(62) & 0.5105(3) & 0.43356 & 1.228(15) \\
		a09m310   & 0.24619(44) & 4.5035(38) & 0.5105(3) & 0.43356 &  1.236(11) \\
		a09m220   & 0.18197(37) & 4.6990(32) & 0.5105(3) & 0.43356 & 1.253(09) \\
		\hline\hline
	\end{tabular}
\caption{\label{tab:ga_extrap} {\textbf{Data and inputs for the chiral-continuum extrapolation.}} $\epsilon_\pi$, $m_\pi L$, and renormalized values of $g_A$ determined in this work. The lattice spacing $a/w_0$ and strong coupling-constant $\alpha_S$ are obtained as described previously\cite{Berkowitz:2017opd}.  We use $a/w_0$ determined at the physical-mass for each lattice spacing. The quantities $\epsilon_\pi$, $a/w_0$, and $m_\pi L$ are used to guide the chiral, continuum, and infinite volume extrapolations respectively.
}
}\end{table*}

\clearpage

\begin{table*}
	{\docfont
		\begin{ruledtabular}
			\begin{tabular}{llrrccc|cccccccc}
				\multicolumn{7} {c|} {\bfseries HISQ gauge configuration parameters} & \multicolumn{8} {c} {\bfseries valence parameters}  \\
				abbr.         & $N_{\text{cfg}}$ & volume & $\begin{matrix}\sim a \\\text{[fm]}\end{matrix}$ & $m_l/m_s$ & $\begin{matrix}\sim m_{\pi_5}\\ \text{[MeV]}\end{matrix}$ & $\sim m_{\pi_5} L$& $N_{\text{src}}$ & $L_5/a$ & $aM_5$ & $b_5$ & $ c_5$ & $am_l^{\text{val.}}$ &  $\s_{\text{smr}}$& $N_{\text{smr}}$ \\
				\hline
				a15m400 & 1000 & $16^3\times48$ & 0.15 & 0.334& 400& 4.8 &
				8& 12& 1.3& 1.5& 0.5& 0.0278& 3.0& 30\\
				a15m350 & 1000 & $16^3\times48$ & 0.15 & 0.255& 350& 4.2 &
				16& 12& 1.3& 1.5& 0.5& 0.0206& 3.0& 30\\
				a15m310  & 1960 & $16^3\times48$ & 0.15 &     0.2 & 310 & 3.8 &
				24& 12 & 1.3 &   1.5 &   0.5 & 0.01580 & 4.2 &   60 \\
				a15m220  & 1000 & $24^3\times48$ & 0.15 &     0.1 & 220 & 4.0 &
				24& 16 & 1.3 & 1.75 & 0.75 & 0.00712 & 4.5 &   60 \\
				a15m130 &  1000 & $32^3\times48$ & 0.15 & 0.036 & 130 & 3.2 &
				5& 24 & 1.3 & 2.25 & 1.25 & 0.00216 & 4.5 &   60 \\
				\hline
				a12m400  & 1000 & $24^3\times64$ & 0.12 & 0.334 & 400 & 5.8 &
				8&   8 & 1.2 & 1.25 & 0.25 & 0.02190 & 3.0 &   30 \\
				a12m350  & 1000 & $24^3\times64$ & 0.12 & 0.255 & 350 & 5.1 &
				8&   8 & 1.2 & 1.25 & 0.25 & 0.01660 & 3.0 &   30 \\
				a12m310  & 1053 & $24^3\times64$ & 0.12 &     0.2 & 310 & 4.5 &
				8&   8 & 1.2 & 1.25 & 0.25 & 0.01260 & 3.0 &   30 \\
				a12m220S& 1000 & $24^3\times64$ & 0.12 &     0.1 & 220 & 3.2 &
				4& 12 & 1.2 &   1.5 &   0.5 & 0.00600 & 6.0 &   90 \\
				a12m220  & 1000 & $32^3\times64$ & 0.12 &     0.1 & 220 & 4.3 &
				4& 12 & 1.2 &   1.5 &   0.5 & 0.00600 & 6.0 &   90 \\
				a12m220L& 1000 & $40^3\times64$ & 0.12 &     0.1 & 220 & 5.4 &
				4& 12 & 1.2 &   1.5 &   0.5 & 0.00600 & 6.0 &   90 \\
				a12m130 &  1000 & $48^3\times64$ & 0.12 & 0.036 & 130 & 3.9 &
				3& 20 & 1.2 &   2.0 &   1.0 & 0.00195 & 7.0 & 150\\
				\hline
				a09m400  &   1201 & $32^3\times64$ & 0.09 & 0.335 & 400 & 5.8 &
				8&   6 & 1.1 & 1.25 & 0.25 & 0.0160 & 3.5 & 45\\
				a09m350  &   1201 & $32^3\times64$ & 0.09 & 0.255& 350 & 5.1 &
				8&   6 & 1.1 & 1.25 & 0.25 & 0.0121 & 3.5 & 45\\
				a09m310  &   784 & $32^3\times96$ & 0.09 &     0.2 & 310 & 4.5 &
				8&   6 & 1.1 & 1.25 & 0.25 & 0.00951 & 7.5 & 167\\
				a09m220 &  1001 & $48^3\times96$ & 0.09 &     0.1 & 220 & 4.7 &
				6&   8 & 1.1 & 1.25 & 0.25 & 0.00449 & 8.0 & 150
			\end{tabular}
		\end{ruledtabular}
		\caption{\label{tab:hisq} {\bf{HISQ gauge configurations and valence sector parameters.}}The HISQ ensembles used in this work (with the abbreviated naming convention\cite{Bhattacharya:2016zcn} (`abbr.'), a15m310 stands for the ensemble with $a\sim0.15$~fm and $m_\pi\sim310$~MeV.
The table also shows the number of configurations $N_{\text{cfg}}$, lattice volume, approximate lattice spacing $a$, ratio of the input light and strange sea quark masses ($m_l/m_s$), approximate HISQ taste-5 pion mass, and approximate value of $m_{\pi,5} L$. The values were obtained from Table I of ref. \cite{Bazavov:2012xda} with increased number of configurations. With the HISQ gauge configurations, we generate Mobius domain-wall propagators at a number of sources per configuration $N_{\text{src}}$, with the fifth dimensional extent $L_5/a$, such that $m_{\text{res}}$ is minimized at $aM_5$, with the Mobius kernel defined by $b_5$ and  $c_5$, and valence light-quark masses $am_l^{\text{val.}}$. We also list the width $\sigma_{\text{smr}}$ and iteration count $N_{\text{smr}}$ of the \texttt{SHELL\_SOURCE} and the \texttt{GAUGE\_INV\_GAUSSIAN} smearing algorithm in \texttt{Chroma}.}
	}
\end{table*}

\clearpage

\begin{figure*}[h]{\docfont
\begin{tabular}{cc}
\includegraphics[width=0.49\textwidth]{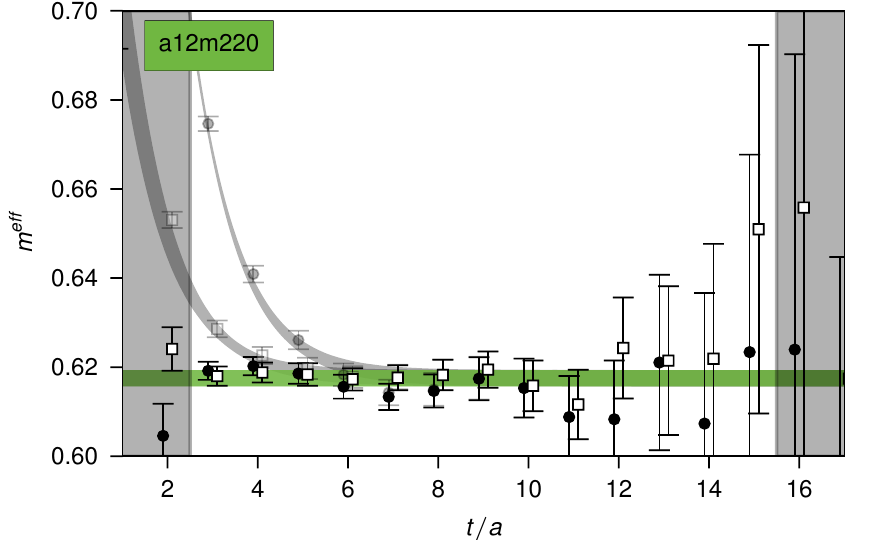}
&
\includegraphics[width=0.49\textwidth]{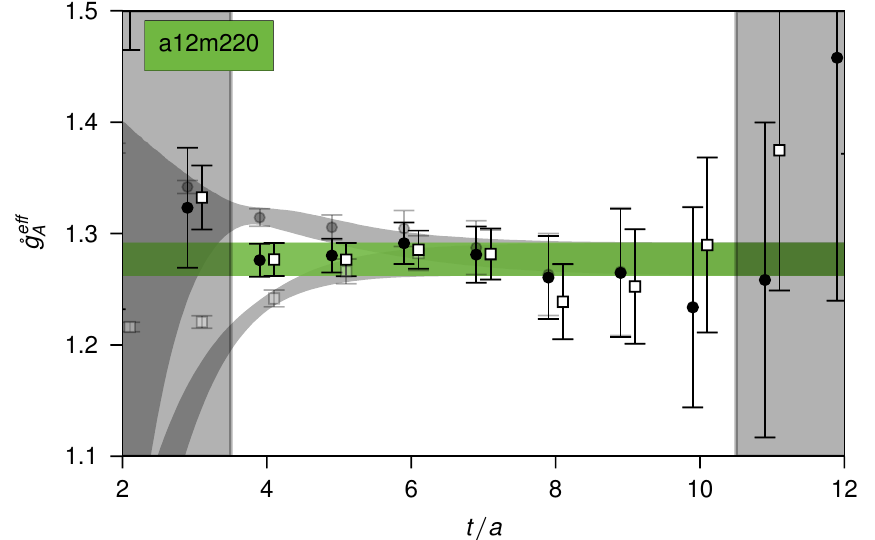}
\\
\textbf{a}& \textbf{b}
\\
\includegraphics[width=0.49\textwidth]{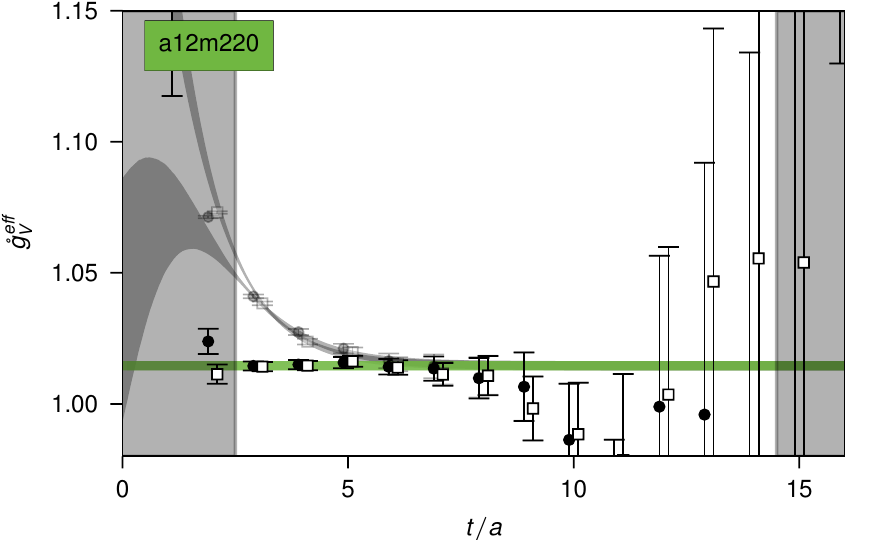}
&
\includegraphics[width=0.49\textwidth]{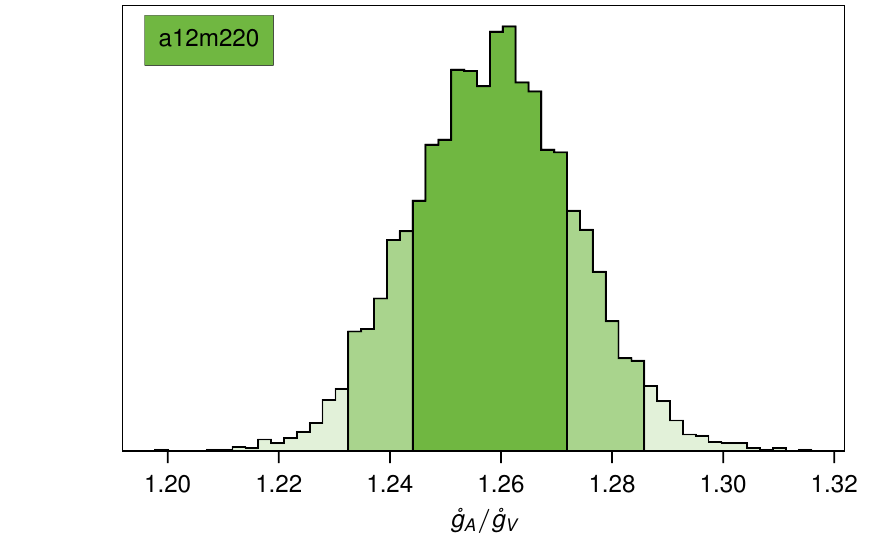}
\\
\textbf{c}& \textbf{d}
\\
\includegraphics[width=0.49\textwidth]{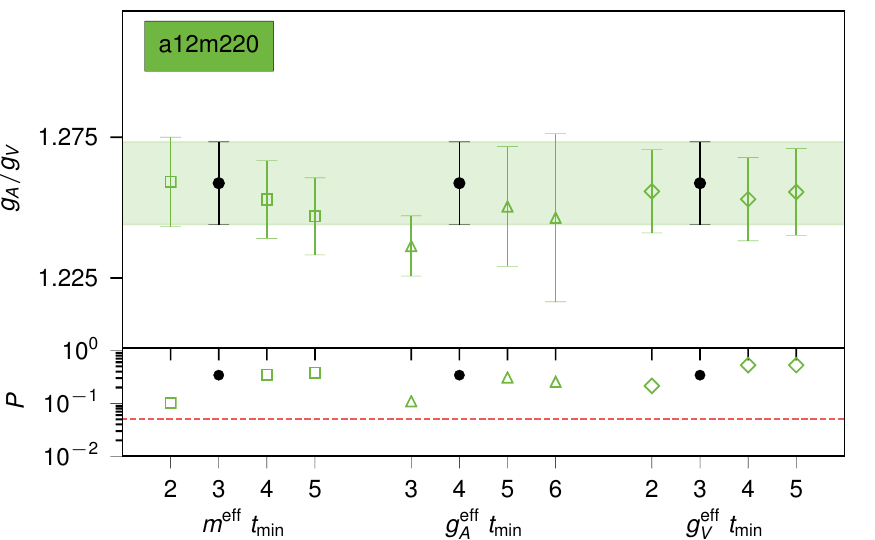}
 &
 \includegraphics[width=0.49\textwidth]{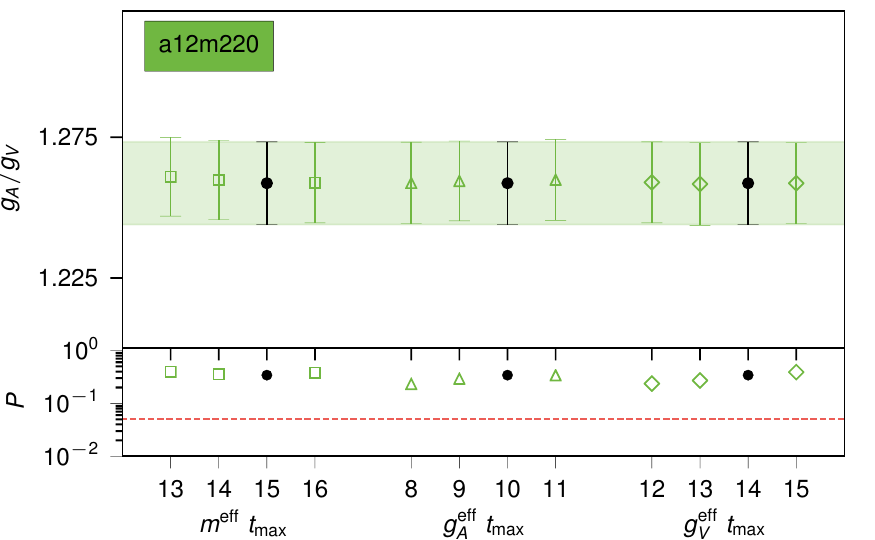}
 \\
 \textbf{e}& \textbf{f}
\end{tabular}
\caption{\label{fig:correlator_fitcurves}
{{\textbf{Correlator fit quality and stability.}} {\textbf{a, b, c}} Fit result for the effective mass ($m^{\text{eff}}$), axial ($\mathring{g}_A^{\text{eff}}$) and vector ($\mathring{g}_V^{\text{eff}}$) FH ratios overlayed on top of correlator data. The black and white filled results are constructed ground state values determined by subtracting the excited state contributions from the raw correlation functions under bootstrap resampling.  {\textbf{d,}} The distribution of $\mathring{g}_A/\mathring{g}_V$ under 5000 bootstrap resamples. The inner shaded green regions correspond to the 68\% (dark green) and 95\% (light green) confidence intervals. All 5000 bootstraps are shown with no evidence of outliers. {\textbf{e, f}} Stability of the correlation function analysis under varying $t_{\textrm{min}}$ and $t_{\textrm{max}}$ for $m^{\mathrm{eff}}$, $\mathring{g}_A^{\mathrm{eff}}$, and $\mathring{g}_V^{\mathrm{eff}}$. The corresponding $P$-values are shown in the bottom panel. The preferred simultaneous fit is highlighted by the solid black symbol.} Uncertainties are one s.e.m.}
}\end{figure*}

\clearpage

\begin{figure*}{\docfont
\begin{tabular}{cc}
\includegraphics[width=0.49\columnwidth]{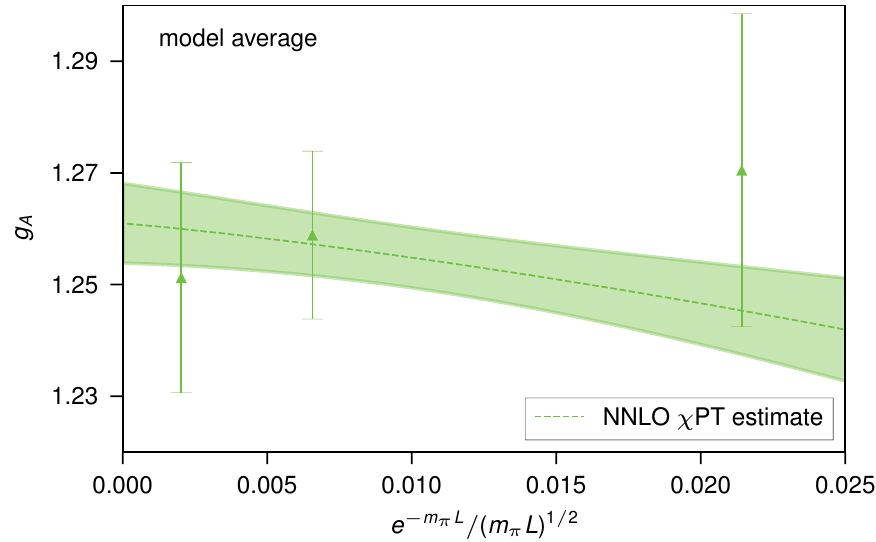}&
\includegraphics[width=0.49\columnwidth]{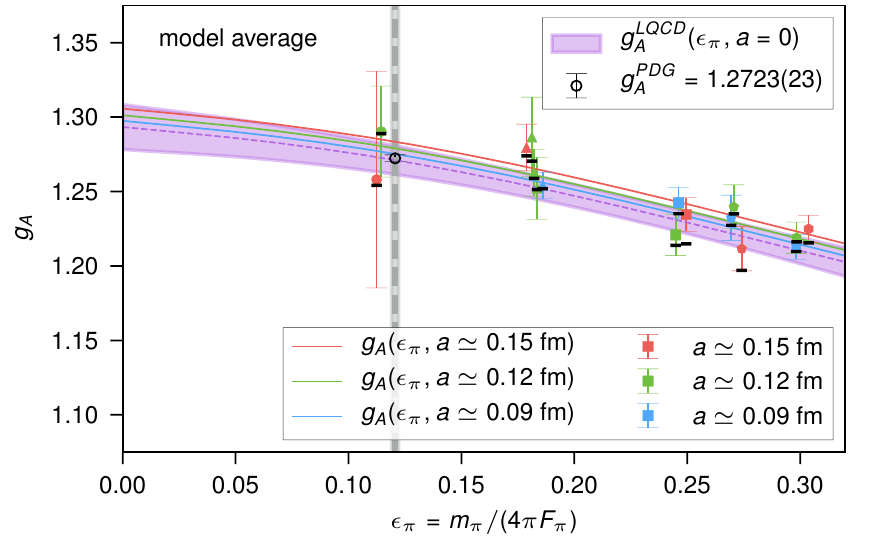}
\\
\textbf{a}& \textbf{b}
\end{tabular}
\caption{\label{fig:IV_extrapolation}
{\textbf{Infinite volume extrapolation of $g_A$.}} \textbf{a}, The three data points correspond to the $a\sim0.12$~fm and $m_\pi\sim220$~MeV ensembles with $m_\pi L=\{5.36, 4.30, 3.25\}$. The NLO finite-volume dependence predicted from the model averaged extrapolation (to all 16 data points) is displayed by the green band with the central value indicated by the dashed green curve.
\textbf{b}, The model average extrapolation with finite volume adjusted (coloured) data.  The central values of the raw data are denoted with a small black dash which in all but one case, lie within one standard deviation of the finite-volume adjusted result.
Uncertainties are one s.e.m.}
}\end{figure*}

\begin{figure*}
	\includegraphics[width=0.5\columnwidth]{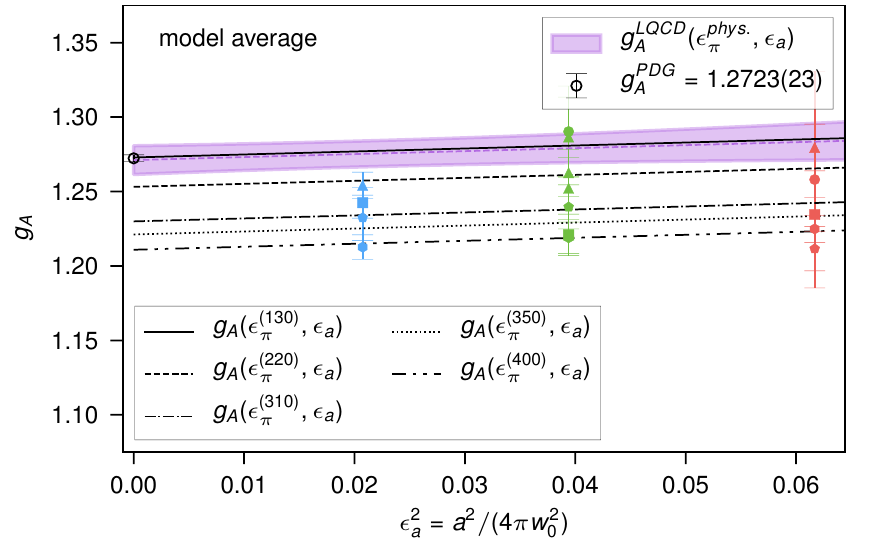}
	\caption{\label{fig:continuum_extrapolation}
		{\textbf{Continuum extrapolation of $g_A$.}} The nucleon axial coupling as a function of $\epsilon_a^2 = a^2/(4\pi w_0)^2$, where $a$ is the lattice spacing and $w_0$ is a hadronic length scale used to normalize LQCD calculations. The physical pion-mass limit is displayed by the magenta band with the central value indicated by the dashed magenta curve. Additional curves with suppressed uncertainty bands are plotted for $m_\pi\approx130$~MeV (solid) $m_\pi\approx220$~MeV (dashed), $m_\pi\approx310$~MeV(dot-dashed), $m_\pi\approx350$~MeV (dotted), and $m_\pi\approx400$~MeV (dot-dot-dashed).
Uncertainties are one s.e.m.}
\end{figure*}

\clearpage

\begin{figure*}[t]{\docfont
\begin{tabular}{cc}
\includegraphics[width=0.49\textwidth]{figs_main/chiral_modelavg}
&
\includegraphics[width=0.49\textwidth]{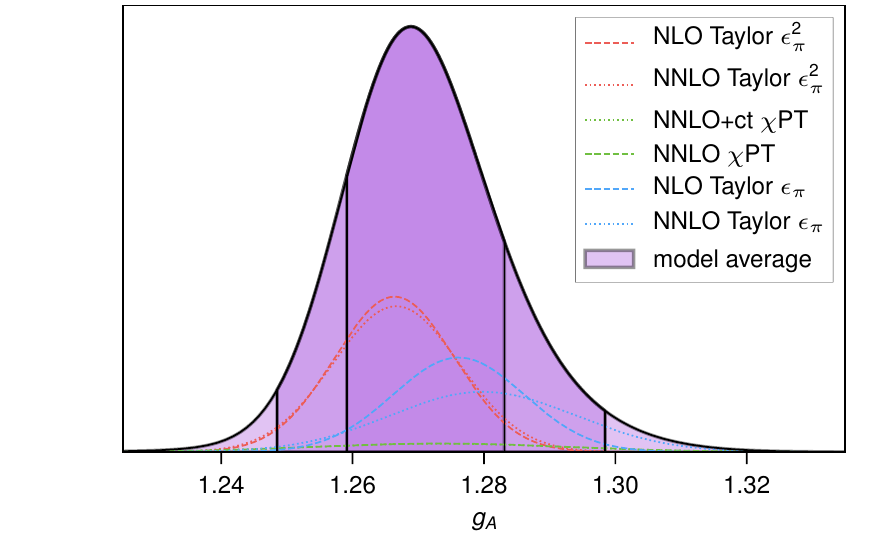}
\\
\textbf{a} & \textbf{b}
\end{tabular}
\begin{tabular}{ccc}
\includegraphics[width=0.327\textwidth]{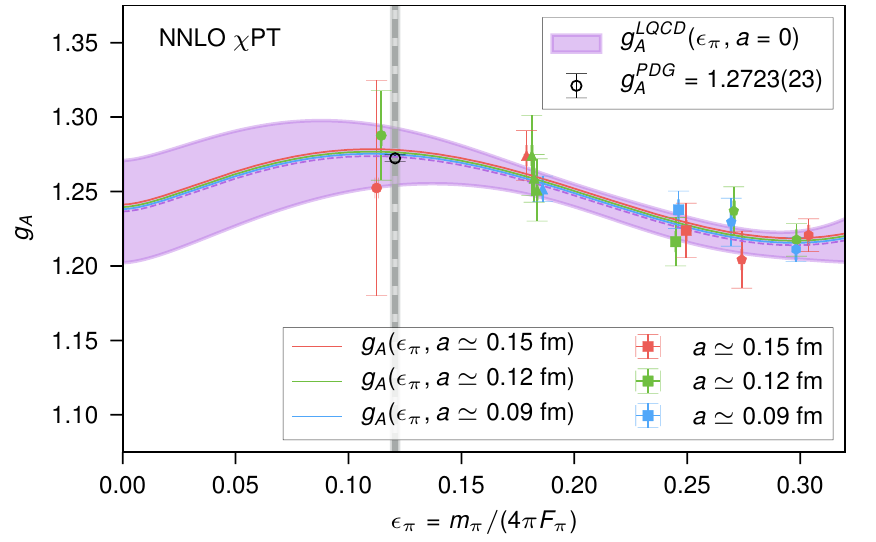}
&
\includegraphics[width=0.327\textwidth]{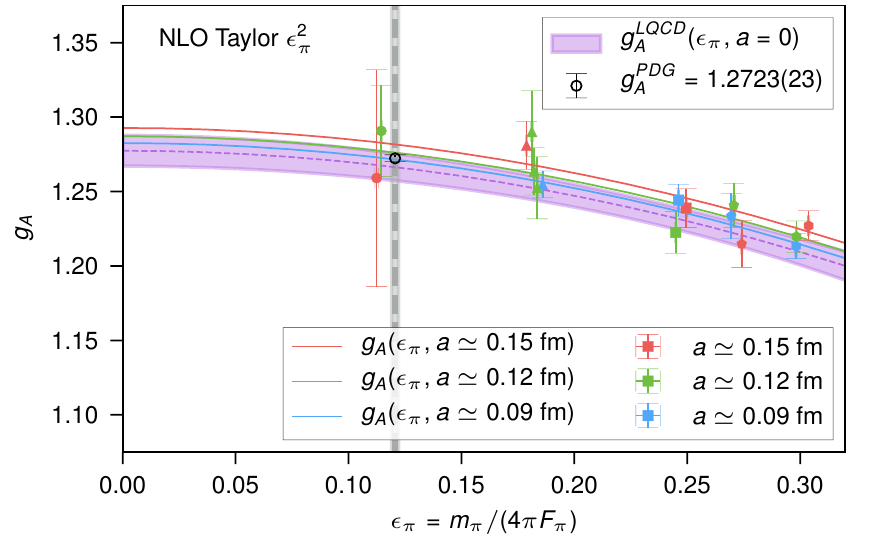}
&
\includegraphics[width=0.327\textwidth]{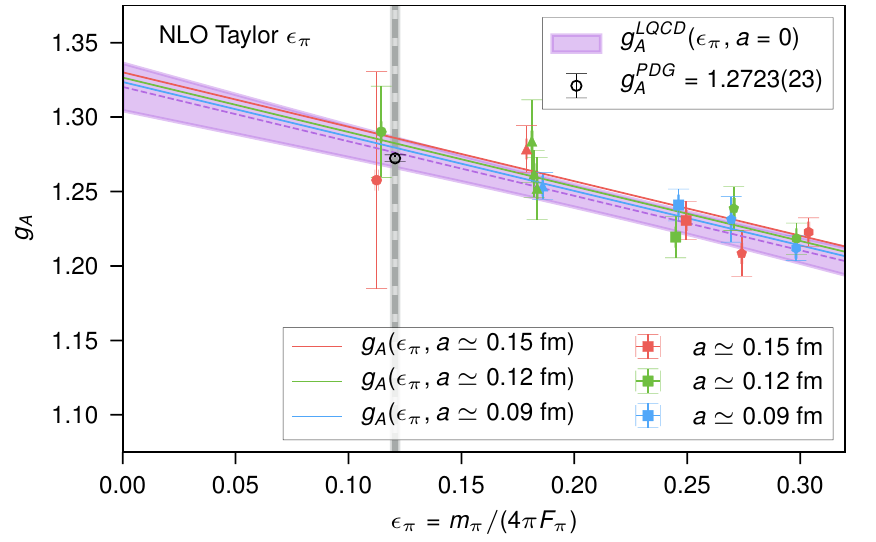}
\\
\textbf{c} & \textbf{d} & \textbf{e}
\\
\includegraphics[width=0.327\textwidth]{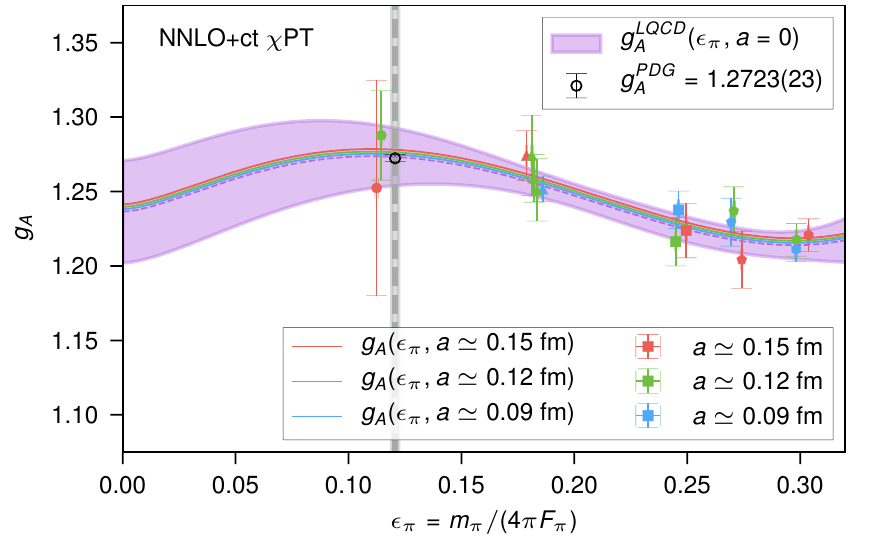}
&
\includegraphics[width=0.327\textwidth]{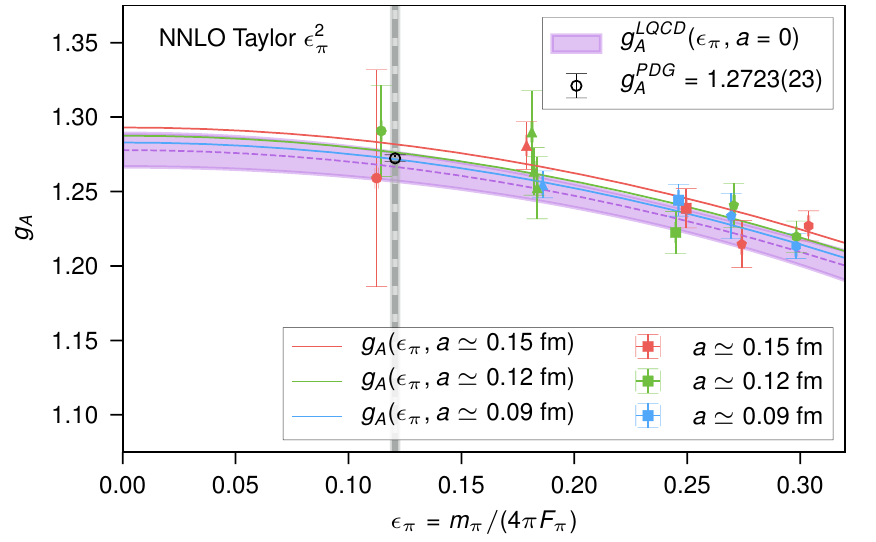}
&
\includegraphics[width=0.327\textwidth]{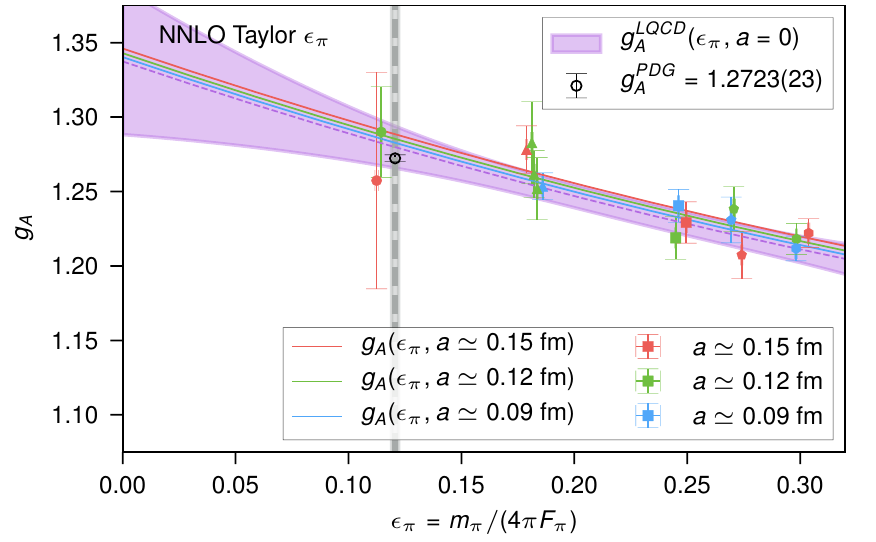}
\\
\textbf{f} & \textbf{g} & \textbf{h}
\end{tabular}
\caption{\label{fig:model_selection}
{\textbf{Model Extrapolation Plots.}} \textbf{a,} The model averaged extrapolation of $g_A$ as a function of $\e_\pi$, determined as described in Supplemental Material Sec.~\ref{sec:uncertainty}. \textbf{b,} The determination of $g_A$ at the physical point from the model averaging procedure. The magenta histogram is the final determination of $g_A$ constructed from a weighted average of the various models used in the extrapolation, appearing as the various distributions lying inside the final histogram. \textbf{c--h,} The resulting extrapolation of $g_A$ as a function of $\e_\pi$ for each of the six models (Supplemental Material Sec.~\ref{sec:extrap}) used in the averaging procedure. In all extrapolation curves, the magenta band is the resulting 68\% confidence interval of the continuum, infinite volume extrapolated value of $g_A$ as a function of $\e_\pi$. The red, green and blue curves are the central values of $g_A$ versus $\e_\pi$ at fixed lattice spacings of $0.15, 0.12, 0.09$~fm respectively.
Uncertainties are one s.e.m.}
}\end{figure*}

\begin{figure*}[h!]
\begin{tabular}{c}
\includegraphics{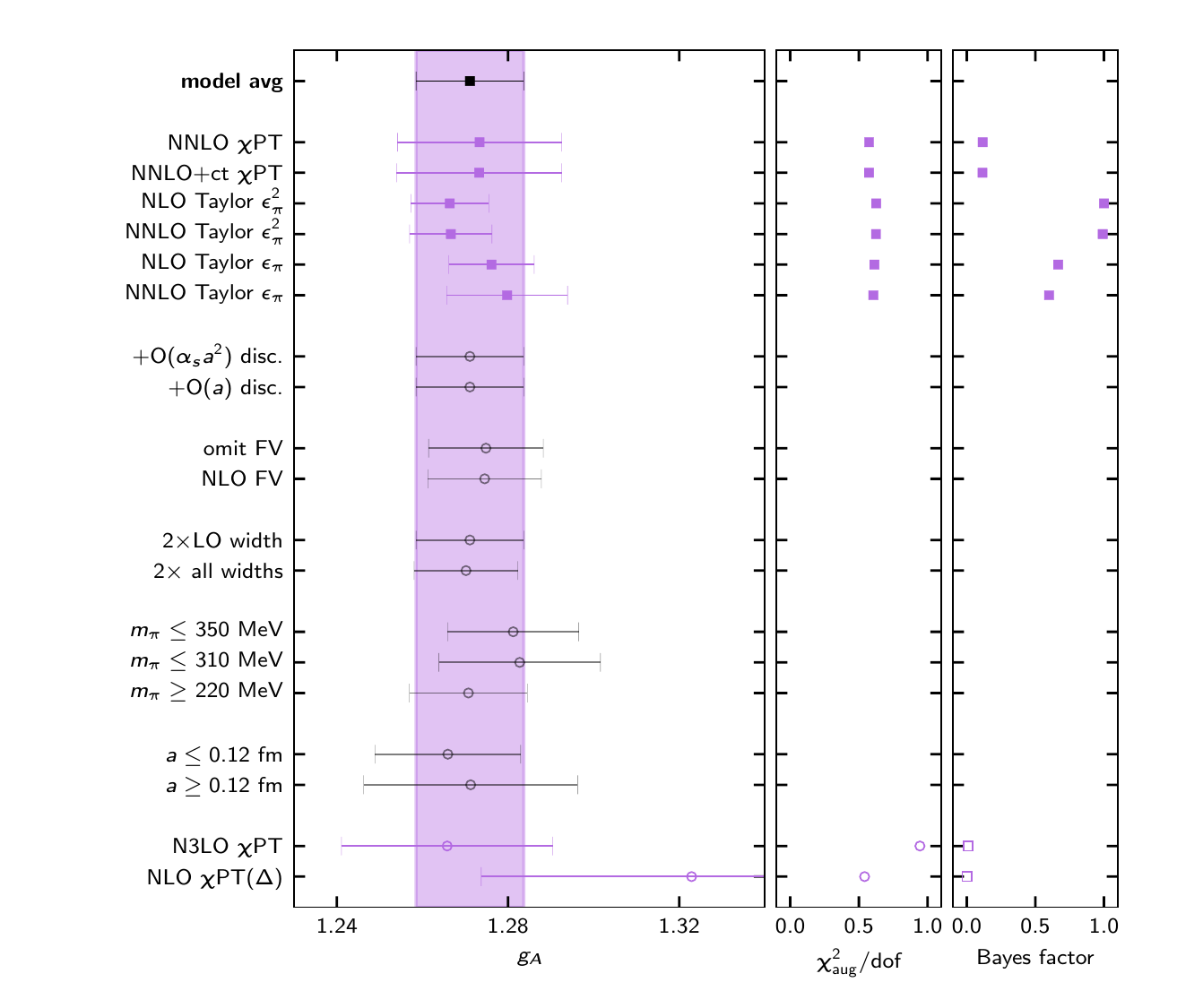}
\end{tabular}
\caption{\label{fig:final_stability}
{
\textbf{Stability and convergence of the chiral-continuum extrapolation.}} The model average result is the black square at the top.  The magenta vertical band is the resulting 68\% confidence band. The next 6 values are the results from the individual extrapolations that go into the model average, described in Supplemental Material Sec.~\ref{sec:uncertainty}. Uncertainties are one s.e.m.  `ct', counter-term; `FV', finite-volume; `disc.', discretization; $\a_S = g^2/(4\pi)$, where $g$ is the quark--gluon coupling of QCD. The middle panel shows the augmented $\chi^2/dof$ where the augmented $\chi^2$ is the sum of the $\chi^2$ from the data and from the priors. All fits have 16 degrees of freedom as each prior is counted as a data point. The right panel shows the resulting Bayes factors normalized by the NLO Taylor $\e_\pi^2$ Bayes factor, which is found to be the largest among them.  These normalized Bayes factors are used as relative weights in the model averaging procedure. The stability of the extrapolation analysis is tested by adding additional discretization terms, omitting the predicted NLO finite volume corrections, increasing the prior widths on the LO and all LECs, performing cuts on the pion masses considered as well as cuts on the discretization scales included. All variations are contained within 1$\s$ of the model average value, with most being significantly less than 1$\s$ from the central value. Finally, we show the resulting extrapolation from the complete N3LO $\chi$PT analysis and the NLO chiral perturbation theory analysis with $\D$ degrees of freedom ($\chi$PT($\D$)). The N3LO fit is not included in the average as it has 5 unknown LECs and we have only 5 different pion mass values. The NLO $\chi$PT($\D$) is not included as it requires input from phenomenology and is thus not a pure lattice QCD prediction, and also the NNLO $\chi$PT($\D$) extrapolation function is not known, so a test of stability and convergence is not possible. Uncertainties are one s.e.m.}
\end{figure*}

\begin{figure*}{\docfont
	\begin{tabular}{cc}
		\includegraphics[width=0.49\textwidth]{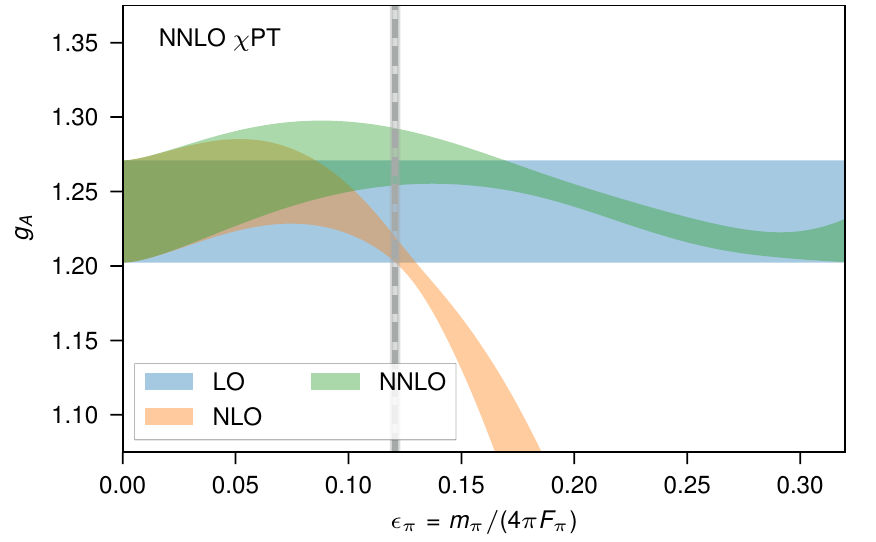}
		&
		\includegraphics[width=0.49\textwidth]{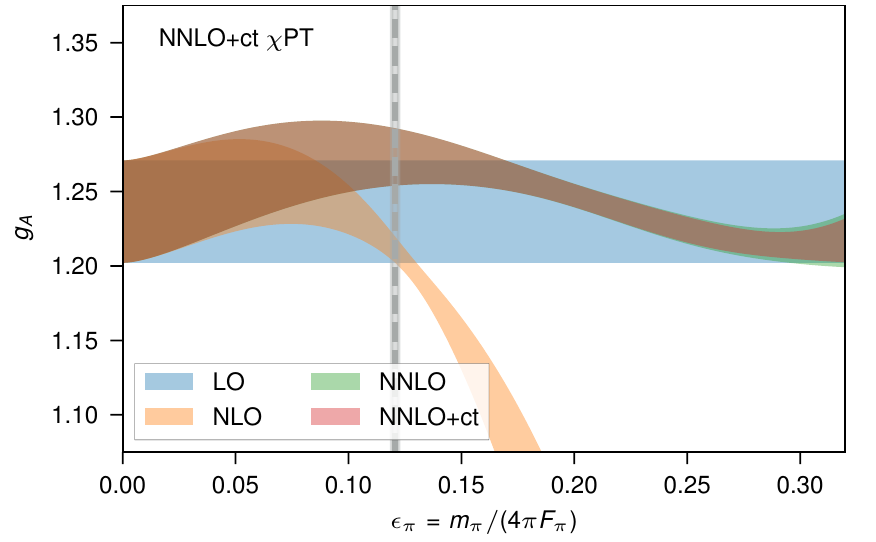}
		\\
		\textbf{a} & \textbf{b}
		\\
		\includegraphics[width=0.49\textwidth]{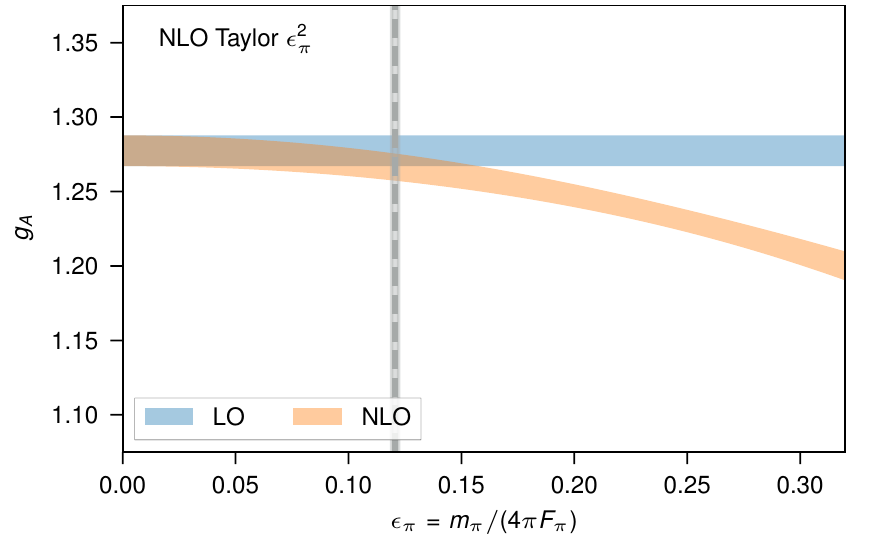}
		&
		\includegraphics[width=0.49\textwidth]{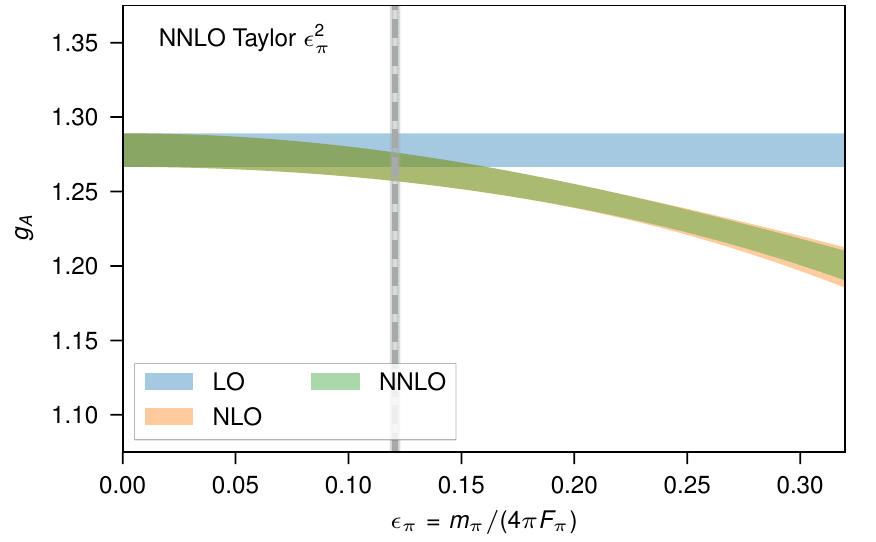}
		\\
		\textbf{c} & \textbf{d}
		\\
		\includegraphics[width=0.49\textwidth]{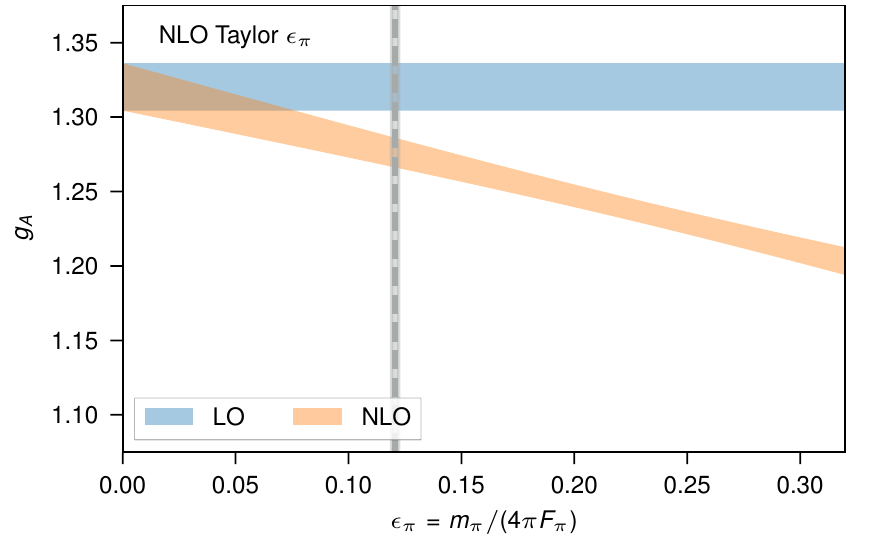}
		&
		\includegraphics[width=0.49\textwidth]{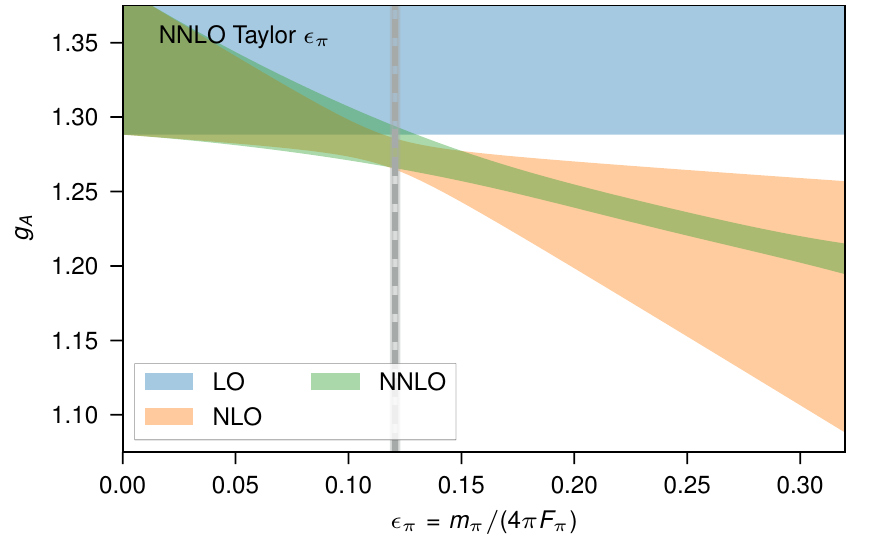}
		\\
		\textbf{e} & \textbf{f}
	\end{tabular}
	\caption{\label{fig:chipt_convergence}
		{\textbf{Convergence of $g_A$.}}
		Order-by-order contribution to the extrapolation of $g_A$ for the six different models that enter in the final model averaged result (see Supplemental Material Sec.~\ref{sec:extrap}). The LECs are determined by the full fit from each model. Higher orders are successively additively included, producing the final reconstruction of the extrapolation when all contributions up to a given order are included.
	}
}\end{figure*}

\begin{figure*}{\docfont
\includegraphics[width=0.49\textwidth]{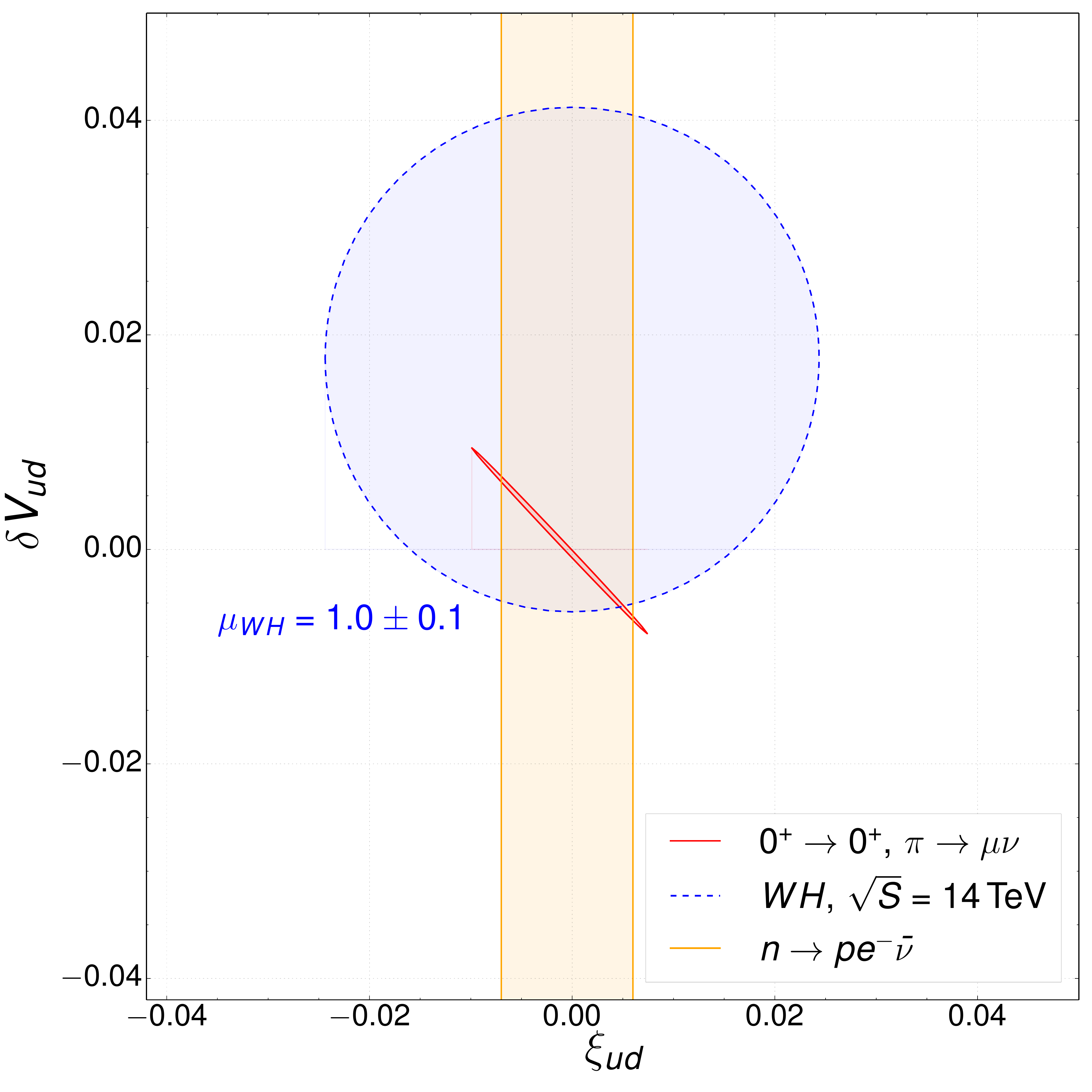}
\caption{\label{fig:right_handed_couplings} {\textbf{Constraint on right-handed beyond the Standard Model currents.} Measurements of cold neutron decays ($n\rightarrow p e^- \bar{\nu}$; $n$, neutron; $p$, proton; $e^-$, electron; $\bar{\nu}$, antineutrino) provide some of the most stringent constraints on new physics. A recent comparison of constraints from low-energy experiments and colliders found comparable constrains on right-handed BSM currents\cite{Alioli:2017ces}. The authors of this work generously provided an update of their Fig.~12 using our determination of $g_A$. The vertical orange band is the constraint on the right handed coupling ($\xi_{ud}$) from our result.  The blue circle arises from collider constraints on $W-$ and Higgs-boson production (WH) at collision energy $\sqrt{S}=14$~TeV, and the diagonal red band is from pion decays (long direction; $\pi\rightarrow\mu\bar{\nu}$) and superallowed $0^+\rightarrow 0^+$ nuclear decays, which constrain corrections to the axial (left - right) and vector (left + right) BSM currents respectively.}
}
}\end{figure*}

\clearpage
\normalsize
\appendix
\twocolumngrid

\renewcommand\appendixname{\normalsize{S}}
\renewcommand\thesection{}
\renewcommand\thesubsection{S.\arabic{subsection}}
\renewcommand\thesubsubsection{\Alph{subsubsection}}
\renewcommand\theequation{S\arabic{equation}}
\setcounter{equation}{0}
\renewcommand{\figurename}{{\bf Supplemental Data Figure}}
\renewcommand{\tablename}{{\bf Supplemental Data Table}}
\setcounter{figure}{0}
\setcounter{table}{0}

\patchcmd{\section}
  {\centering}
  {\raggedright\footnotesize}
  {}
  {}
\patchcmd{\subsection}
  {\centering}
  {\raggedright\footnotesize\vspace{-15pt}}
  {}
  {}
\patchcmd{\subsubsection}
  {\centering}
  {\raggedright\scriptsize\vspace{-15pt}}
  {}
  {}

\setcounter{secnumdepth}{5}
\section*{SUPPLEMENTAL INFORMATION}
\docfont
\subsection{Data and software availability \label{sec:data}}
An HDF5 file containing the lattice QCD correlation functions generated for this work is made available with the publication of this Letter. Additionally, bootstrap distributions from the correlation function analysis are published in a comma-separated-values (CSV) file along with the Jupyter notebook and chiral-continuum extrapolation Python library used to perform the physical-point extrapolation analysis.  These files along with installation instructions for the package may be found at \url{https://github.com/callat-qcd/project_gA}\protect\citeM{zenodo:1241374}.

The software used for this work was built on top of the USQCD Chroma software suite\protect\citeM{Edwards:2004sx} and the highly optimised QCD GPU library QUDA\protect\citeM{Clark:2009wm,Babich:2011np}.
We utilize the highly efficient HDF5 I/O Library\protect\citeM{hdf5} with an interface to HDF5 in the USQCD QDP++ package, added with SciDAC 3 support (CalLat)\protect\citeM{Kurth:2015mqa}, as well as the MILC software for solving for HISQ propagators and generating new ensembles.
The HPC jobs were efficiently managed with a \texttt{bash} job manager, \texttt{METAQ}\protect\citeM{Berkowitz:2017vcp}, capable of intelligently backfilling idle nodes in sets of nodes bundled into larger jobs submitted to HPC systems.
\texttt{METAQ} was also developed with SciDAC 3 support (CalLat) and is available at the \texttt{git} repository \url{https://github.com/evanberkowitz/metaq}.

\subsection{Correlation functions from the Feynman-Hellman theorem\label{sec:FH}}
The significant advancement of our work can be attributed to an unconventional computational method.  This method was developed and implemented on one of the ensembles used in this work (a15m310)\cite{Bouchard:2016heu} and demonstrated on a larger subset in preliminary calculations\protect\citeM{Berkowitz:2017gql,Chang:2017oll}.  We briefly summarise the method here as it is central to our final result.

The Feynman-Hellmann (FH) theorem relates matrix elements to linear variations in the energy eigenvalue with respect to an external source, $\partial E_n/\partial\l=\langle n | H_\l | n \rangle\, ,$ where the Hamiltonian of the system is modified appropriately as $H=H_0 + \l	 H_\l$. The Heisenberg representation illuminates the fact that correlation functions must be exponentially damped by the energy eigenvalue under time evolution in Euclidean spacetime. Therefore, on the lattice the energy can be determined from the \textit{effective mass},
\begin{equation}\label{eq:m_eff}
m^{\text{eff}}(t)=\ln\left( \frac{C(t)}{C(t+1)}\right)\, ,
\end{equation}
where the spectral decomposition of the two-point correlation function is given by
\begin{equation}\label{eq:C_2pt}
C(t) = \sum_{n=0}^{N_\text{states}} z_n z_n^\dagger e^{-E_n t}\, .
\end{equation}
Applying the FH theorem to Eq.~\eqref{eq:m_eff} yields the correlation function we construct to obtain the results in this Letter, which we denote the FH ratio:
\begin{equation}
\left.\frac{\partial m^{\text{eff}}_\l (t)}{\partial \l}\right|_{\l=0}=\left.\left[\frac{\partial_\l C_\l(t)}{C(t)}-\frac{\partial_\l C_\l(t+1)}{C(t+1)}\right]\right|_{\l=0}.
\label{eq:methods_fhcorrelator}
\end{equation}
The path integral representation of $C(t)$ and $\partial_\l C(t)$ for this calculation may be derived by sourcing the nucleon and current operators into the generating functional, while applying the appropriate derivatives with respect to each source.

There are other implementations of methods motivated by the FH theorem\protect\citeM{Chambers:2014qaa, Chambers:2015bka, Savage:2016kon} which are similar, but our method is the most economical implementation, at least for single nucleon properties. Our method directly calculates the $-\partial_\l C(t)$ correlation function\cite{Bouchard:2016heu}, without the need to numerically implement the derivative or to disentangle the different orders of the response of the correlation function to the perturbation, as is required by other implementations.

While we are interested in the axial coupling of the ground state nucleon, the nucleon operators couple to an infinite tower of states, and therefore it is customary to filter out the ground state signal by exponentially damping the excited state signals at large time separations. By going to the Heisenberg representation, we derive the complete spectral decomposition of $-\left.\partial_\l C_\l(t)\right|_{\l=0}$,
\begin{align}\label{eq:dl_C_spectral}
-\partial_\l& C_\l(t) =
	\sum_{n=0}^{N_{\text{states}}} \left[(t-1)z_n g_{nn}z_n^\dagger+d_n\right]e^{-E_nt}
\nonumber\\&
	+\sum_{n\neq m=0}^{N_{\text{states}}}z_n g_{nm}z^\dagger_m\frac{e^{-E_nt}e^{\frac{\D_{nm}}{2}}-e^{-E_mt}e^{\frac{\D_{mn}}{2}}}{e^{\frac{\D_{mn}}{2}}-e^{\frac{\D_{nm}}{2}}},
\end{align}
allowing us to analyse correlation functions even at small time separations. Here, $g_{nn}$ for $n=0$ is the ground state coupling of the nucleon, $z_n z^\dagger_n$ is the non-relativistically normalized relative probability of finding the nucleon in the $n^\text{th}$ state and $\D_{nm} = E_n-E_m$. The linear enhancement of $g_{nn}$ is a direct manifestation of the FH theorem, in which the first derivative of the spectrum (described by the two-point correlation function) is taken, thereby generating a linear moment. Additionally, excited state contributions in the linearly enhanced $n>0$ terms are analogous to contamination present in standard two-point correlation functions, which are generically well under control. The remaining contamination from lattice artefacts, $d_n$ and the sum over $n\neq m$, are not linearly enhanced and therefore are functionally distinct from the signal of interest and can be cleanly removed. The artefacts, $d_n$ arise from contact terms where the current insertion is at the same time as the nucleon creation or annihilation operators, and also from the time region where the current is earlier or later than the nucleon creation or annihilation operators respectively\cite{Bouchard:2016heu}. At $t=1$ the contribution from all terms aside from $d_n$ exactly vanish, allowing for a robust estimate of the contributions to $-\partial_\l C_\l(t)$ from these undesired artefacts.

Inserting Eqs.~\eqref{eq:C_2pt} and \eqref{eq:dl_C_spectral} in to Eq.~\eqref{eq:methods_fhcorrelator}, it is straightforward to show that in the long-time limit, we recover the ground-state matrix element of interest
\begin{equation}
\left.
\lim_{t\rightarrow \infty} \frac{\partial m_\l^{\rm eff}(t)}{\partial \l}\right|_{\l=0} = g_{00}\, .
\end{equation}
The difference in Eq.~\eqref{eq:methods_fhcorrelator} leads to an additional suppression of the excited states (and contact terms) beyond the standard exponential suppression by the mass gap. This allows us to make use of numerical data very early in Euclidean time, before the stochastic noise overwhelms the signal\protect\citeM{Parisi:1983ae,Lepage:1989hd}, providing an effective exponential enhancement of the signal for a fixed number of stochastic samples as compared to the standard methods. In Extended Data Fig.~\ref{fig:correlator_fitcurves} and Supplemental Figs.~\ref{fig:a15m310a15m220_curve}--\ref{fig:a09m310a09m220_curve}, we demonstrate our ability to fit early in Euclidean time on all ensembles used in this work.

One potential drawback of our Feynman-Hellmann strategy is that it requires new calculations of Feynman-Hellmann quark propagators for each matrix element or momentum injection of interest. By contrast, the standard methods used in the literature provide the flexibility to study arbitrary quark bi-linear matrix elements between the proton and neutron, with arbitrary momentum injection by the current, without need for additional computational cost. On the other hand, our Feynman-Hellmann method comes with an additional flexibility not present in the standard methods: we can compute the matrix elements of the same quark bi-linear currents in various hadronic states, such as hyperons or multi-nucleon systems, without needing to recompute the quark propagators coupling to the current. Using the standard methods, the sink interpolating operator is fixed, so computing a new sequential propagator is necessary for each final state.

\subsection{Lattice action \label{sec:action}}
For this work, we have chosen a \textit{mixed action} (MA) in which the discretizations for the generation of the gauge configurations and solution of the quark propagators differ.\cite{Berkowitz:2017opd} Details of tuning the action and its salient features are summarised here.

\subsubsection{Action Details}
To control the continuum limit, infinite volume and physical pion mass extrapolation, a set of LQCD ensembles with multiple lattice spacings, multiple volumes and near physical pion masses must be used.  The only set of publicly-available gauge configurations that satisfy these criteria are the Highly Improved Staggered Quark (HISQ)\protect\citeM{Follana:2006rc} action ensembles with dynamical light, strange and charm quarks ($N_f=2+1+1$) generated by the MILC Collaboration\cite{Bazavov:2012xda}${}^,$\protect\citeM{Bazavov:2010ru}. They were generated with near-physical values of the strange and charm quark masses and three values of the pion mass, $m_\pi \sim \{130, 220, 310\}$~MeV. For this work, we utilize ensembles with three different lattice spacings of $a\sim\{0.15, 0.12, 0.09\}$~fm. Formally the HISQ action has leading discretization errors starting at $\mathrm{O}(\alpha_S a^2, a^4)$, however improved link-smearing greatly suppresses taste-changing interactions leading to numerically smaller discretization errors. The gluons are simulated using the tadpole-improved\protect\citeM{PhysRevD.48.2250}, one-loop Symanzik gauge action\protect\citeM{Alford199587} with leading discretization errors starting at $\mathrm{O}(\alpha_S^2 a^2, a^4)$.

We performed a dedicated volume study at $a\sim0.12$~fm $m_\pi\sim220$~MeV with three volumes. To control the pion mass extrapolation, we generated six new HISQ ensembles with $m_\pi\sim\{350,400\}$~MeV on the same three lattice spacings (these ensembles are available to any interested group upon request). Details of the HISQ ensembles are presented in Extended Data Table~\ref{tab:hisq}.

The valence quark propagators are solved with the M\"{o}bius Domain Wall Fermion (MDWF) action\protect\citeM{Brower:2004xi,Brower:2005qw,Brower:2012vk} after applying the gradient-flow smoothing algorithm\protect\citeM{Luscher:2010iy,Lohmayer:2011si} to the HISQ configurations. The M\"obius kernel with gradient-flow smoothing reduces the residual chiral symmetry breaking such that $m_{\text{res}}<0.1\times m_l^{\text{val.}}$ for moderate values of $L_5$ and thus the valence action has approximate chiral symmetry\protect\citeM{Luscher:1998pqa} as it satisfies the Ginsparg-Wilson relation\protect\citeM{Ginsparg:1981bj} with small corrections. The values of $b_5$ and $c_5$ were chosen (Extended Data Table~\ref{tab:hisq}) such that the M\"{o}bius kernel is a rescaled Shamir kernel\protect\citeM{Shamir:1993zy,Furman:1994ky} of the domain-wall action\protect\citeM{Kaplan:1992bt} ($b_5-c_5=1$). The calculation of these MDWF propagators required the significant majority of computing cycles and were efficiently solved using the QUDA library\protect\citeM{Clark:2009wm} with parallel MPI Support\protect\citeM{Babich:2011np}. After absorbing $m_{\text{res}}$ into the quark mass through the PCAC relation, the MDWF action has discretization errors beginning at $\mathrm{O}(a^2,\alpha_S a^2)$\protect\citeM{Allton:2008pn}.

\subsubsection{Computational Details \label{sec:details}}
The valence quarks are tuned so that the MDWF pion mass matched the taste-5 HISQ pion mass within 2\%, ensuring a unitary theory in the continuum limit. Multiple sources per configuration are used to increase our statistical samplings. On a given configuration, for a series of evenly spaced time-locations, a seeded random origin is chosen, $(x_0,y_0,z_0,t_0)$ along with its antipode, $(x_0,y_0,z_0,t_0) + L/2(1,1,1,0)$ (modulo the periodic spatial boundary conditions). At each point, a smeared source is generated to solve the MDWF quark propagators using the \texttt{SHELL\_SOURCE} with the \texttt{GAUGE\_INV\_GAUSSIAN} routine in \texttt{Chroma}\protect\citeM{Edwards:2004sx}. The correlation functions are constructed with two choices of sink smearing, a \texttt{SHELL\_SINK} with parameters matching the source and a \texttt{POINT\_SINK}. The proton and neutron correlation functions are constructed using the local interpolating operator with the largest overlap on the positive-parity nucleon states\protect\citeM{Basak:2005aq,Basak:2005ir}. We double the statistics by generating analogous correlation functions for the negative parity nucleon. Under a time-reversal transformation the nucleon reverses parity, allowing us to average the forward propagating nucleon correlation functions with the backward propagating negative-parity nucleon correlation functions. Once constructed, all the correlation functions are shifted to $t_0 = 0$ and averaged. We observe no correlation between different sources, resulting in statistical uncertainties inversely proportional to $\sqrt{N_{src}}$ as first observed and studied in detail in previous work\protect\citeM{Beane:2009kya}. All parameter choices for the valence MDWF action are presented in Extended Data Table~\ref{tab:hisq}.

\begin{figure*}{\docfont
	\begin{tabular}{cc}
		\includegraphics[width=0.49\textwidth]{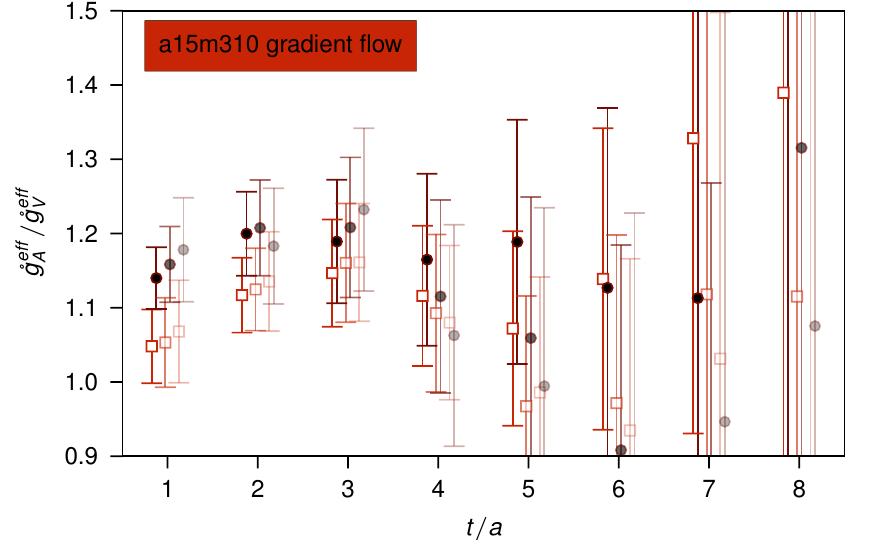}
		&
		\includegraphics[width=0.49\textwidth]{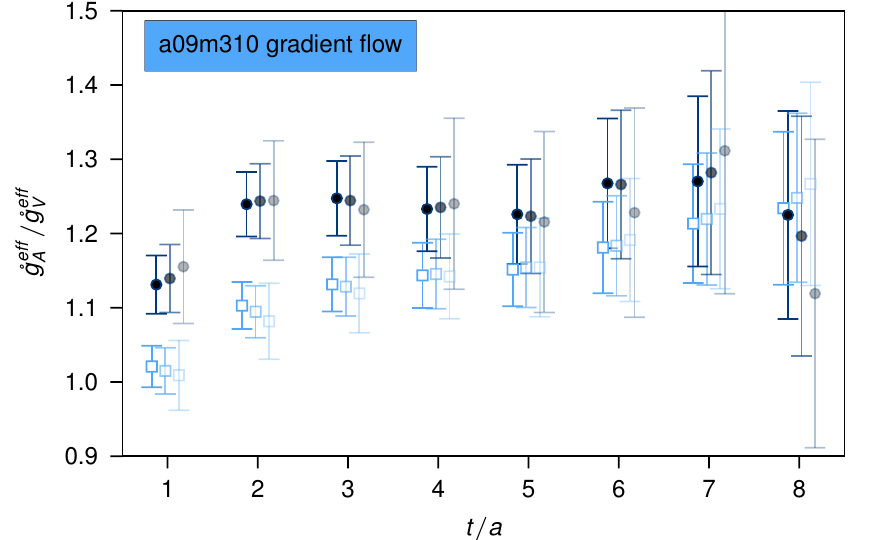}
		\\
		\textbf{a}& \textbf{b}
		\\
		\includegraphics[width=0.49\textwidth]{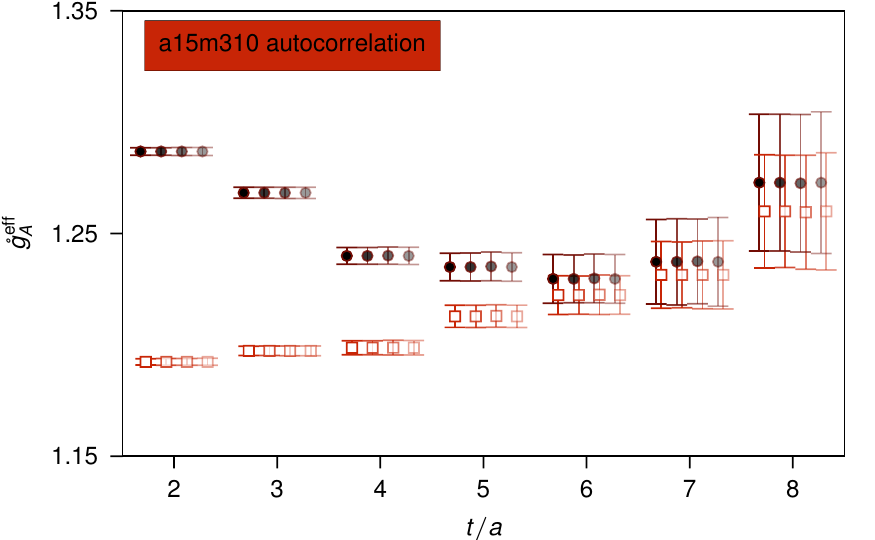}
		&
		\includegraphics[width=0.49\textwidth]{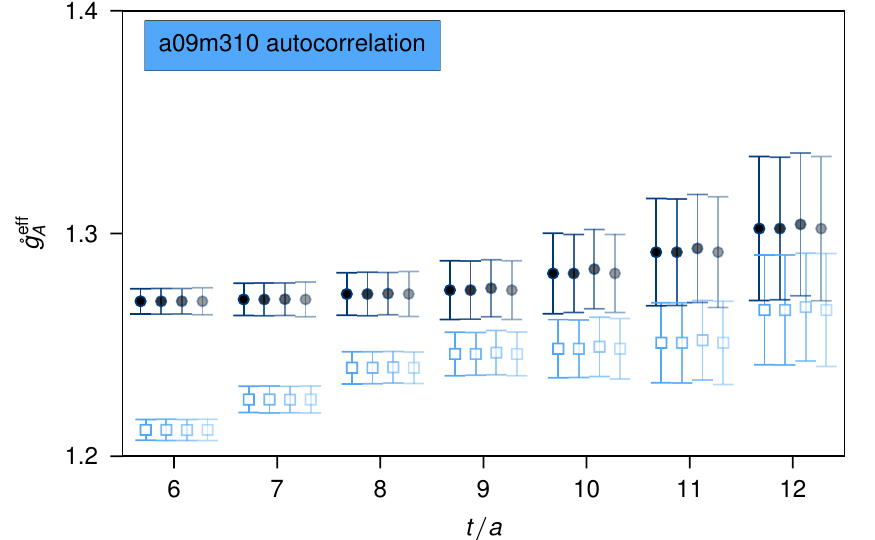}
		\\
		\textbf{c}& \textbf{d}
	\end{tabular}
	\caption{\label{fig:flow_auto_study}
		\textbf{Gradient flow time dependence and Monte Carlo autocorrelation time study.} {\textbf{a, b,}} The flow-time dependence from $t_{gf}=\{1.0, 0.6, 0.2\}$ from left to right (dark to light). We observe flow time independence in the axial to vector correlation function ratios, and smaller stochastic noise at larger flow times. {\textbf{c, d,}} The uncertainty of the mean under bootstrap resampling with successively larger bin sizes of $N_{\text{bins}}=\{1,2,3,4\}$ from left to right (dark to light). We observe the uncertainty of the mean to be unchanged under binning, indicating that there are no autocorrelations present in our data. For all plots, the circle and square data points correspond to the smeared- and point-sink correlation functions.
Uncertainties are one s.e.m.}
}\end{figure*}

We have demonstrated the gradient-flow time ($t_{gf}$) independence of $M_N / F_\pi$ and $F_K / F_\pi$ for our action\cite{Berkowitz:2017opd}. In particular, we demonstrate, with a reduced data set, that the extrapolation of $F_K/F_\pi$ to the physical point in the continuum is independent of flow time and also consistent with the FLAG determination\cite{Aoki:2016frl}. In this work, we also study the flow-time dependence of $g_A$. In Figure~\ref{fig:flow_auto_study}, we show ratios of the axial over the vector FH ratios for the a15m310 and a09m310 ensembles with 196 configurations at a single source. The point-sink (squares) and smeared-sink (circles) FH ratios are plotted with $t_{gf} = \{1.0, 0.6, 0.2\}$ respectively from left to right. In both ensembles, we observe minimal flow time dependence in the ratio of correlators. Additionally the flow time is fixed to $t_{gf}=1.0$ in lattice units on all gauge configurations, ensuring that any quantity extrapolated to the continuum limit will be flow-time independent. For $t_{gf}=1.0$, we find it sufficient to solve the gradient flow diffusion equation with 40 integration steps using the Runge-Kutta algorithm. Furthermore, we observe smaller stochastic uncertainty at increasingly larger values of $t_{gf}$ due to the gradient flow suppressing the ultraviolet noise. These conclusions are consistent with the results observed in previous work\cite{Berkowitz:2017opd} for other hadronic quantities ({\it{e.g.}} Fig.~3 therein).

We also study possible autocorrelations in our data set by binning the FH ratio correlation functions for every ensemble used in this work. Extended Data Figure~\ref{fig:flow_auto_study} shows a representative example of a binning study. We observe that the standard deviation of the raw correlation function is stable under binning for bin sizes up to four, demonstrating that no autocorrelations are present in the data. The complete binning study is presented in the Supplemental Figs.~\ref{fig:autocorrelation_I}--\ref{fig:autocorrelation_III}. We do not bin any of our data in this work.

\subsection{Correlator analysis \label{sec:correlators}}

\begin{table*}{\docfont
	\begin{ruledtabular}
		\begin{tabular}{lcccccccccccc}
			ensemble& $t_{\text{min}}^C$ & $t_{\text{max}}^C$ & $t_{\text{min}}^A$ & $t_{\text{max}}^A$ & $t_{\text{min}}^V$ & $t_{\text{max}}^V$ & $\mathring{g}_A$ & $\mathring{g}_V$ & $\chi^2/\textrm{dof}\, [\textrm{dof}]$& $P$-value & $(Z_A/Z_V-1)\times 10^{5}$\\
			\hline
			a15m400   & 6 & 11 & 3 & 11 & 5 & 11 & 1.213(06) & 0.998(01) & \phantom{0}1.3 [24] & 0.14 & 4.20(17) \\
			a15m350   & 4 & 12 & 2 & 12 & 4 & 12 & 1.195(13) & 0.997(01) & \phantom{0}1.3 [38] & 0.11 & 3.00(19) \\
			a15m310   & 4 & 14 & 4 & 10 & 4 & 15 & 1.216(11) & 1.001(02) & \phantom{0}1.1 [40] & 0.32 & 1.77(24) \\
			a15m220   & 3 & 11 & 4  & 9  & 4  &  9 &  1.275(13) & 1.000(04) & \phantom{0}1.5 [22] & 0.07 &  3.7(1.3) \\
			a15m130   & 2 & 10 & 2 &  6 & 2 & 7 & 1.262(53) & 0.994(35) & \phantom{0}1.6 [20] & 0.05 & 1.87(52) \\
			\hline
			a12m400   & 5 & 10 & 5 & 11 & 7 & 11 & 1.237(10) & 1.016(01) & \phantom{0}1.5 [16]& 0.07 & 4.22(13)\\
			a12m350   & 7 & 14 & 5 & 11 & 5 & 14 & 1.255(14) & 1.016(01) &                  0.93 [30]& 0.57 & 2.75(10)\\
			a12m310   & 5 & 12 & 4 & 12 & 6 & 12 & 1.239(13) & 1.021(02) & \phantom{0}1.5 [28]& 0.06 & 1.86(13)\\
			a12m220S & 4 & 10 & 5 & 10 & 3 & 10 & 1.294(28) & 1.018(03) & \phantom{0}1.1 [22]& 0.35 & 2.02(37)\\
			a12m220   & 3 & 15 & 4 & 10 & 3 & 14 & 1.277(15) & 1.015(02) & \phantom{0}1.1 [44]& 0.34 & 2.02(37)\\
			a12m220L & 4 & 12 & 4 & 12 & 5 & 10 & 1.277(21) & 1.020(05) & \phantom{0}1.4 [28]& 0.09 & 2.02(37)\\
			a12m130   & 2 & 13 & 3 & 13 & 2 & 12 & 1.318(29) & 1.020(08) & \phantom{0}1.1 [48]& 0.24 & 0.45(77)\\
			\hline
			a09m400   & 8 & 18 & 6 & 15 & 7 & 15 & 1.238(08) & 1.023(01) &                  0.98 [40] & 0.50 & 5.10(10)\\
			a09m350   & 7 & 16 & 8 & 16 & 7 & 14 & 1.258(15) & 1.024(02) & \phantom{0}1.3 [34] & 0.09 & 3.40(14)\\
			a09m310   & 9 & 16 & 3 & 12 & 7 & 17 &  1.266(11) &  1.024(01) & \phantom{0}1.0 [38] &   0.40 & 2.09(16)\\
			a09m220   & 9 & 15 & 3 & 14 & 6 & 12 & 1.280(09) & 1.022(02) & \phantom{0}1.3 [32] & 0.09 & 1.86(17)\\
		\end{tabular}
	\end{ruledtabular}
	\caption{\label{tab:complete_corrfit} {\textbf{Correlator fit region and results, and renormalisation coefficients}} Fit regions for the two-point correlation function $t_{\text{min}}^C, t_{\text{max}}^C$, the axial FH ratio $t_{\text{min}}^A, t_{\text{max}}^A$, and the vector FH ratio $t_{\text{min}}^V, t_{\text{max}}^V$ are given in lattice units. The resulting central value and standard deviation of the bare couplings are given in columns $\mathring{g}_A$ and $\mathring{g}_V$ along with the $\chi^2/dof$ and $P$-value of these fits. The last column gives the values of the ratio of renormalisation coefficients $Z_A/Z_V$ at $\mu\approx 2.8$GeV in the $\gamma_\mu$ scheme.
	}
}\end{table*}

The exact wavefunction for the ground state nucleon is unknown, so lattice correlation functions are constructed with interpolating operators for the initial and final states. Therefore, the correlation function describes a superposition of the ground state nucleon of interest and nucleon excited states. Disentangling the ground state from the excited state contamination requires careful analysis of the correlation functions, and has proved to be one of the major challenges for past calculations of $g_A$. As a result of the unique construction of the lattice correlation function though our Feynman-Hellmann strategy, we have access to measurements of the correlation function at both longer and shorter separations between the initial and final states, allowing for a more complete study of excited state contributions compared to previous works. Additionally, nucleon observables suffer from exponentially severe decay in signal-to-noise, posing a serious challenge for high precision calculations. Compared to previous works, exponentially more precise data (at early times) is leveraged in this analysis to combat the severe decay in the signal.

\subsubsection{Analysis Strategy}
For each ensemble, we perform a simultaneous fit to six correlation functions which include the nucleon two-point correlation function, the vector and axial-vector FH ratios [Eq.~(\ref{eq:methods_fhcorrelator})], and for each of these, two different correlation functions corresponding to two different choices of sink smearing for the quark fields. This greatly enhances the amount of correlated data when determining a large subset of shared parameters ({\textit{i.e.}} $E_n$ and $z_n$). Strategies for estimating the unknown parameters were previously discussed\cite{Bouchard:2016heu}. In the present work, we first perform a two-state Bayesian constrained fit to explore the parameter space in $t$, and then take the central value of the resulting posterior distribution as the initial guess to a final two-state unconstrained fit using non-linear $\chi^2$ minimization. Preconditioning the unconstrained fit does not change the final result, but serves as an effective method to explore large parameter spaces, and minimises the iteration count required for convergence. In principle, preconditioning the unconstrained fit from the posterior distribution obtained from Bayes' theorem provides a strategy for avoiding unphysical local minima in the $\chi^2$ manifold. In hindsight however, the data is well-behaved with relatively sharp minima. Bayesian constrained fits with up to eight states were performed resulting in consistent results\cite{Bouchard:2016heu}.

We assess the quality of the candidate fits by first considering only results with $P$-values greater than 0.05 in order to discriminate against fits of poor quality (e.g. Extended Data Fig.~\ref{fig:correlator_fitcurves}e,f). Next, we study the effects of excited state contamination by varying the fit regions over different time separation, and demand that the candidate fit lies in the region of stability (Extended Data Fig.~\ref{fig:correlator_fitcurves}e,f). Finally we quantify the uncertainty of our determination of the matrix element by drawing 5000 bootstrap samples, and accept candidates that are Gaussian distributed, as expected from the distribution of the (path integral) mean as a consequence of the central limit theorem (Extended Data Fig.~\ref{fig:correlator_fitcurves}e,f). The preferred fit is one which satisfies all the above requirements while sampling the largest fit region, such that we maximise the amount of information extracted from these numerically intensive calculations. As a final check, we overlay the preferred fit on top of the data and observe agreement between model and data.

The list of fit regions and preferred results are given in Supplemental Table~\ref{tab:complete_corrfit}. The complete correlator fit study plots are shown in Supplemental Figs.~\ref{fig:a15m400a15m350_curve}--\ref{fig:a09m310a09m220_curve}, with the fit stability plots show in Supplemental Figs.~\ref{fig:stability_I}--\ref{fig:tmax_stability_III} and the bootstrap distributions of the resulting values of $\e_\pi$ in Supplemental Figures~\ref{fig:epi_I}--\ref{fig:epi_II}. All stages of the analysis are implemented using the Python library \texttt{lsqfit}\protect\citeM{lsqfit-9}.

\subsubsection{Discussion}
Studying ground state stability as a function of $t_{\textrm{min}}$ is the most robust way to demonstrate understanding and control of excited state contributions. Such a study is only possible with the data set of the current work because the Feynman-Hellmann strategy makes all possible source-sink separation times accessible. In contrast, all previous calculations use conventional strategies to generate lattice correlation functions, and thus do not generate enough source-sink separation times to perform such a study.

The results of this study are shown in Extended Data Fig.~\ref{fig:correlator_fitcurves}e and Figs.~\ref{fig:stability_I}--\ref{fig:stability_III}. Because excited states are always heavier than the ground state, their contributions are more pronounced at smaller source-sink separations. As a result, we observe that choosing smaller values of $t_{\textrm{min}}$ compared to the preferred fit (in solid black) leads to results that are sometimes in tension, indicating that the ground state signal is being contaminated by excited state artefacts at these times. In contrast, the preferred fits are always consistent with more conservative fits which only include data at larger values of $t_{\textrm{min}}$. To further demonstrate the quality of the ground state determination, we numerically subtract (under bootstrap) the excited state contributions determined in the analysis from the raw correlation functions and plot these processed results, along with the asymptotic value of the ground state (Fig.~\ref{fig:fh_method}, Extended Data Figs.~\ref{fig:correlator_fitcurves}a,b,c and Supplemental Figs.~\ref{fig:a15m400a15m350_curve}--\ref{fig:a09m310a09m220_curve}).

Surveying these figures reveals the existence of correlated statistical fluctuations at $\sim$1~fm separation times and is the manifestation of the signal-to-noise problem suffered by nucleon observables. In all previous work, the nucleon axial and vector correlation functions (\textit{e.g.} Extended Data Fig.~\ref{fig:correlator_fitcurves}b,c and similar Supplemental figures) are constructed in this intermediate time separation region in order to avoid contamination from excited state contributions, and may be affected by these uncontrolled statistical fluctuations. Furthermore, the limited number of source-sink separation times available in previous calculations makes the \textit{a posteriori} identification of correlated fluctuations extremely difficult.

In contrast, this work analyses data at much smaller separation times, where the signal-to-noise has not yet degraded, in order to extract the ground state signal. As a result, this analysis strategy is robust against random statistical fluctuations. To provide more evidence, Extended Data Fig.~\ref{fig:correlator_fitcurves}f and Supplemental Figs.~\ref{fig:tmax_stability_I}--\ref{fig:tmax_stability_III} demonstrate stability of the extracted ground state nucleon couplings under varying $t_{\textrm{max}}$, showing that the result is insensitive to whether the fluctuations are included as part of the fitted data. Additionally, utilizing data at smaller time separations results in an exponentially more precise determination of the nucleon couplings and is the key to obtaining the sub-percent uncertainty presented in this work.

In summary, the Feynman-Hellmann strategy for constructing lattice correlation functions enables leverage over the full source-sink time dependence in the analysis. We observe stability of the ground state nucleon couplings under varying fit regions, demonstrating control over excited-state contamination. The asymptotic values of the ground state are in good agreement with the excited-state subtracted lattice correlation functions. The final bootstrapped results all yield nearly ideal Gaussian distributions. With this preponderance of evidence, we resolve the two major challenges identified from previous works, and demonstrate full control over systematic uncertainties at unprecedented levels of precision.

\subsection{Renormalisation \label{sec:npr}}
\begin{figure}[t]{\docfont
	\begin{tabular}{c}
		\includegraphics[width=0.49\textwidth]{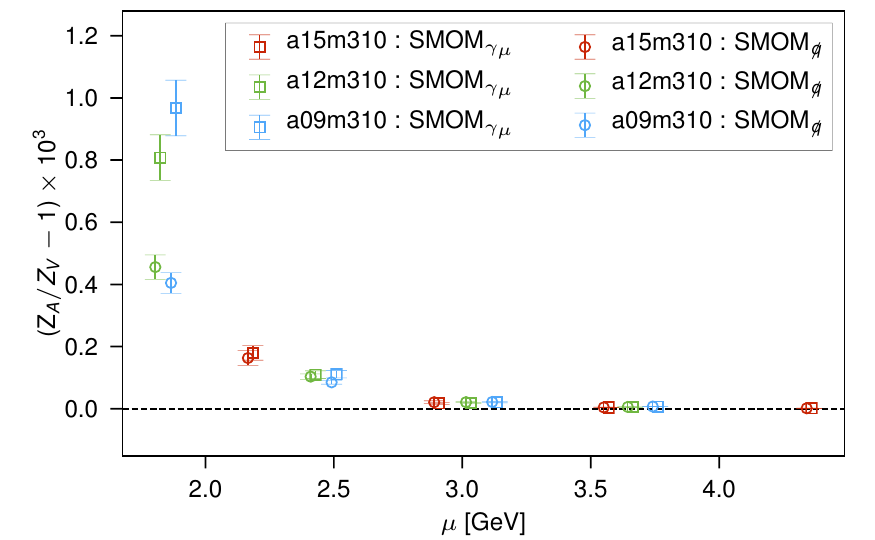}\\
		\textbf{a}\\
		 \includegraphics[width=0.49\textwidth]{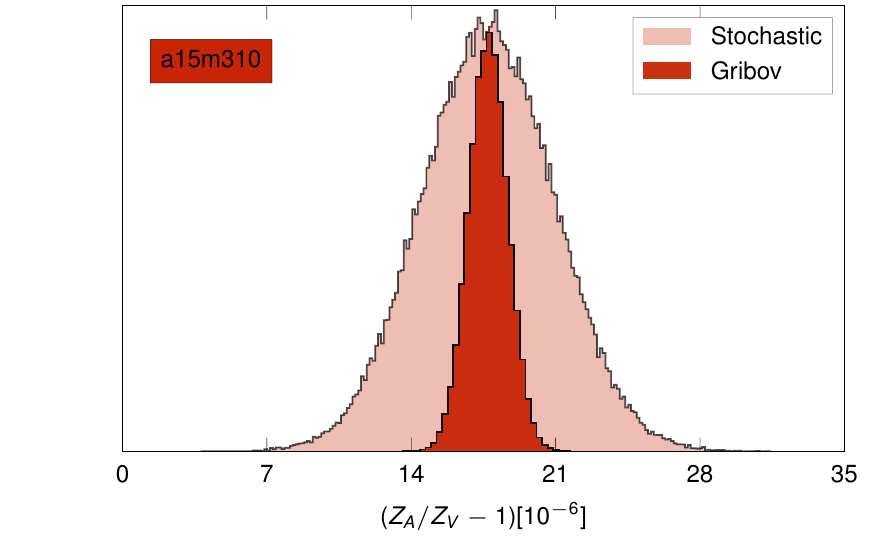}\\
		\textbf{b}
	\end{tabular}
	\caption{\label{fig:renormalisation}
		{\textbf{Renormalisation coefficients versus cut-off scale, and Gribov region study.}} {\textbf{a,}} Renormalisation coefficients from RI-SMOM for the $a\sim\{0.15, 0.12, 0.09\}$~fm $m_\pi\sim310$~MeV ensembles as a function of the renormalisation scale $\mu$ for intermediate $\gamma_\mu$ and $\slashed{q}$ schemes in the Landau gauge. The axial coupling is itself a physical observable, and therefore its value is independent of scale. The scale dependence observed for $\mu \leq 2.5$~GeV comes from the infrared (IR) contamination of a light meson, and in principle at high scales (UV) the coefficient receives large $\mathrm{O}(a\textbf{p})$ corrections. The intermediate region free of IR and UV contaminations is coined the Rome-Southampton window. {\textbf{b,}} The statistical uncertainty of the renormalisation coefficient for the a15m310 ensemble evaluated at $\mu=2.86$~GeV in the $\gamma_\mu$ scheme is shown by the light red histogram. The Landau gauge admits remnant gauge degrees of freedom resulting in the Gribov distribution shown in dark red. Random global gauge transformations are applied to the gauge fields, and the RI-SMOM perscription is repeated to obtain the Gribov distribution. We observe that the systematic uncertainty coming from Landau gauge fixing to be smaller than the statistical uncertainty. Uncertainties are one s.e.m.}
}\end{figure}

Discretization of the Dirac action leads to differences between the local current used in the calculation and the conserved current. We correct for this difference using the non-perturbative Rome-Southampton renormalisation procedure\protect\citeM{Martinelli:1994ty}, with non-exceptional kinematics\protect\citeM{Aoki:2007xm,Sturm:2009kb}. Explicitly we compute
\begin{equation}
\label{eq:ren_ga}
g_A = \frac{Z_A}{Z_V} \frac{\mathring{g}_A}{\mathring{g}_V} \;,
\end{equation}
where $Z_A$ and $Z_V$ are the renormalisation factors of the axial and vector current, while $\mathring{g}_A$ and $\mathring{g}_V$ correspond to the bare (un-renormalized) couplings. In Eq.~\eqref{eq:ren_ga}, we take advantage of the fact that $Z_V \mathring{g}_V = 1$. Furthermore, because of the good chiral property of our lattice discretization, we expect $Z_A=Z_V$ up to small artefacts. We have computed these factors in RI-SMOM schemes and with  momentum sources, as proposed in\protect\citeM{Gockeler:1998ye}, resulting in high statistical precision. We observe the ratio of the renormalisation coefficients $Z_A/Z_V$ to be commensurate with unity at one part in 10,000, indicating that the lattice action we use preserves chiral symmetry to very good approximation, see Supplemental Data Fig.~\ref{fig:renormalisation}. In this procedure, the renormalisation scale is given by $\mu=\sqrt{q^2}$ ($q^2 \geq0$) and $q$ is the vertex momentum transfer. This result, together with the improved stochastic uncertainty gained from simultaneously fitting the vector FH ratio, further reduces the final uncertainty of our result.

Since the quark bi-linear matrix elements used to determine the renormalisation coefficients are not gauge invariant, we perform these calculations in Landau gauge. Landau gauge fixing however, is {\it{incomplete}} and the resulting coefficients will be evaluated at one of many Gribov regions\protect\citeM{Giusti:2001xf}. We sample the distribution of renormalisation coefficients over different Gribov regions by repeating the calculation after performing random global gauge transformations to the gauge fields. We observe that the systematic uncertainty from this effect is subdominant to the statistical uncertainty of the renormalisation coefficients (Supplemental Data Fig.~\ref{fig:renormalisation}). We performed a dedicated flow-time study of $Z_A / Z_V$ on a subset of the ensembles and find this ratio is also flow-time independent.

\subsection{Parameterization of the chiral, continuum and infinite volume extrapolations  \label{sec:extrap}}
The results of these calculations (Extended Data Table~\ref{tab:ga_extrap}) must be extrapolated to the physical point. $g_A$ is a dimensionless quantity and therefore the entire extrapolation can be performed by using ratios of physical quantities that form dimensionless variables without the need for performing a scale setting. On each ensemble, we determine the three quantities
\begin{equation}
\e^2_a = \frac{1}{4\pi}\frac{a^2}{w_0^2}, \quad
m_\pi L, \quad
\e_\pi = \frac{m_\pi}{4\pi F_\pi}\, ,
\end{equation}
which are used to parameterise the continuum, infinite volume and physical pion mass extrapolations. $w_0$ is a gradient-flow scale that can be precisely and accurately determined\protect\citeM{Borsanyi:2012zs} and $F_\pi$ is the pion decay constant (with $F_\pi\sim92$~MeV normalization). EFT methods can be used to parameterise the dependence upon these variables.

\subsubsection{\label{sec:chiral_extrapolation}$\chi$PT through N3LO}
For a static quantity such as $g_A$,  Heavy Baryon $\chi$PT (HB$\chi$PT)\cite{Jenkins:1990jv} can be used to parameterise the pion mass dependence. The convergence issues of $SU(3)$ HB$\chi$PT\protect\citeM{WalkerLoud:2008bp,Torok:2009dg,Ishikawa:2009vc,Jenkins:2009wv,WalkerLoud:2011ab} require two-flavor HB$\chi$PT\protect\citeM{Bernard:1992qa} to be used for a controlled extrapolation. The complete pion mass dependence of $g_A$ is known through $\mathrm{O}(m_\pi^3)$\protect\citeM{Jenkins:1991es,Bernard:1992qa,Kambor:1998pi}, which is next-to-next-to-leading order (NNLO) in the chiral expansion. In terms of $\e_\pi$, the pion mass dependence is given by
\begin{align}\label{eq:ga_xpt}
g_A =&\phantom{+} g_0
	+c_2\e_\pi^2
	-\e_\pi^2(g_0 + 2g_0^3) \ln(\e_\pi^2)+g_0 c_3 \e_\pi^3\, ,
\end{align}
where $g_0$, $c_2$ and $c_3$ are low-energy constants (LECs) that must be determined in the analysis. In this expression, we have set the $\chi$PT renormalisation scale to $\mu=4\pi F_\pi$. The corrections to using a fixed renormalisation scale enter at $\mathrm{O}(\e_\pi^4)$ which can be seen by expanding $F_\pi / F$ in the above expression, where $F = \lim_{m_\pi\rightarrow0} F_\pi$.

The complete next-to-next-to-next-to-leading order (N3LO) calculation of $g_A$ has not been determined, however, the $\ln^2(\e_\pi)$ corrections have been determined with a renormalisation group analysis\protect\citeM{Bernard:2006te}. Even though the complete calculation has not been performed, the full parameterisation of these corrections is given by
\begin{align}\label{eq:ga_xpt_4}
\d_\chi^{(4)} g_A = \e_\pi^4 \Big[&
	c_4
	+\tilde{\g}_4 \ln(\e_\pi^2)
\nonumber\\&
	+\left(\frac{2}{3}g_0 + \frac{37}{12}g_0^3 +4g_0^5\right) \ln^2(\e_\pi^2)
	\bigg]\, .
\end{align}
The extrapolation formula was provided formula in terms of $m_\pi / F$ and so the difference in the coefficient of the $\ln^2(\e_\pi^2)$ term given here and previously\protect\citeM{Bernard:2006te} is attributed to using $F_\pi$ rather than $F$ in the expression. The LEC $\tilde{\g}_4$ differs from $\g_4$\protect\citeM{Bernard:2006te}. Our data set is not sufficient to use the full N3LO expression as it contains a total of 5 unknown LECs, and we have results at 5 different values of $m_\pi$. We do, however, include partial corrections from N3LO, like the $c_4$ counter term (\emph{i.e.} NNLO+ct), to check the stability of the analysis, and the Bayesian Framework allows us to use the full expression.

\subsubsection{\label{sec:delta}Including explicit delta degrees of freedom}
The $\e_\pi$ dependence described above stems from the chiral Lagrangian with only pions and nucleons as explicit degrees of freedom. There are many publications in the literature advocating for the explicit inclusion of the delta resonances in the theory in order to accurately describe properties of the nucleon. While the delta states are strong resonances, in the large-$N_c$ limit\protect\citeM{'tHooft:1973jz,Witten:1979kh}, the splitting between them and the nucleons vanishes. Further, the deltas are strongly coupled to the nucleons and the mass gap between them ($\D\equiv M_\D-M_N$) is comparable to the pion mass, such that contributions from the delta states to nucleon quantities can be poorly captured without explicitly including them as dynamical states in the EFT\protect\citeM{Butler:1992ci,Griesshammer:2012we}. In lattice QCD calculations of nucleon quantities, the pion masses are still generally heavier than in nature and for $m_\pi \gtrsim 290$~MeV, the deltas become stable, asymptotic states. Finally, it has been observed that including explicit deltas in the EFT leads to a milder pion mass dependence for $g_A$\protect\citeM{Hemmert:2003cb}.

This observation follows straightforwardly from the large-$N_c$ formalism\protect\citeM{Dashen:1993as,Dashen:1993ac,Jenkins:1993zu,Dashen:1993jt,Dashen:1994qi,Jenkins:1995gc,Jenkins:1998wy,Fettes:2000gb}. Combining the large-$N_c$ expansion with the chiral expansion leads to an improved perturbative expansion for many quantities, including the baryon spectrum\protect\citeM{Jenkins:1995td}, which has been observed numerically with lattice QCD calculations\protect\citeM{Jenkins:2009wv,WalkerLoud:2011ab}. It has been shown that there are cancellations between nucleon and delta virtual corrections for $g_A$ as well, which lead to the milder pion mass dependence\protect\citeM{CalleCordon:2012xz}.

In the present work, an extrapolation including the deltas explicitly is a \textit{phenomenological} extrapolation as there are three new quantities that are required to perform the chiral extrapolation, which we have not determined in our calculation, and therefore, some knowledge from experiment must be used to constrain them. One must know the delta-nucleon mass splitting, $\D$, as well as two additional axial couplings, the $\D\rightarrow\D$ coupling and the $\D\rightarrow N\pi$ transition coupling which we denote $\mathring{g}_{\D\D}$ and $\mathring{g}_{N\D}$, respectively (the mathrings denote the chiral limit value of these couplings, just as $g_0$ is the chiral limit value of $g_A$). These quantities are particularly challenging to compute due to the resonant nature of the delta (for sufficiently light pion masses), and require calculations of not only the external states, but also the $\pi N$ scattering phase shifts\protect\citeM{Lellouch:2000pv,Agadjanov:2014kha,Briceno:2014uqa}. The first lattice QCD calculation of such $1\rightarrow2$ transitions has only recently been performed for mesons\protect\citeM{Owen:2015fra,Briceno:2015dca,Briceno:2016kkp,Gerardin:2016cqj}. For our mixed-action calculation, this problem is further exacerbated by the non-unitary nature of the theory as these non-unitary effects can go on-shell in the $\pi N$ scattering system, thus precluding the use of the known formalism\protect\citeM{Briceno:2015axa}.

The continuum, infinite volume extrapolation function including deltas in the $SU(2)$ chiral expansion was first determined at NLO, and is given by\protect\citeM{Hemmert:2003cb}
\begin{align}\label{eq:gA_delta}
g_A =&\phantom{+} g_0
	+c_2^{\D} \e_\pi^2
	+\frac{32\pi}{27}g_0 g_{N\D}^2 \frac{\e_\pi^3}{\e_\D}
\nonumber\\&
	-\e_\pi^2 \ln(\e_\pi^2) \left[
		g_0 + 2g_0^3
		+\frac{2}{9}g_0 g_{N\D}^2 +\frac{50}{81}g_{\D\D} g_{N\D}^2
		\right]
\nonumber\\&
	-R\left(\frac{\e_\pi^2}{\e_\D^2}\right) g_{N\D}^2 \left[
		\e_\pi^2 \frac{32}{27}g_0
		+\e_\D^2 \left( \frac{76}{27}g_0 + \frac{100}{81}g_\D \right)
	\right]
\nonumber\\&
	-\e_\D^2 \ln \left(\frac{4\e_\D^2}{\e_\pi^2}\right) \left[
		\frac{76}{27}g_0 g_{N\D}^2 + \frac{100}{81}g_\D g_{N\D}^2
	\right]\, ,
\end{align}
where we have defined
\begin{equation}\label{eq:e_delta}
\e_\D \equiv \frac{\D}{4\pi F_\pi}\, ,
\end{equation}
and the new non-analytic function is given by
\begin{equation}\label{eq:R}
R\left(z\right) \equiv \left\{ \begin{matrix}
	\sqrt{1-z} \ln \left( \frac{1 - \sqrt{1 - z}}{1 + \sqrt{1 - z}} \right) +\ln(4/z),
		& z \leq 1 \\
	2\sqrt{z-1} \arctan z +\ln(4/z),
		& z > 1
	\end{matrix}
	\right. \, .
\end{equation}

In order to use this extrapolation formula, we take the value of $\D\simeq293$~MeV from the experimental splitting of the nucleons and deltas. As the pion mass is increased above the physical value, this splitting is known to reduce\protect\citeM{WalkerLoud:2008bp}, but we do not account for this change in the extrapolation. At this order in the expansion, it is sufficient to pick a single value, with the difference appearing at higher orders in the expansion. In order to constrain the two new axial couplings, we set the central values to those predicted from the large-$N_c$ expansion\protect\citeM{Dashen:1994qi,Jenkins:1995gc}
\begin{align}\label{eq:gND_Nc}
&\mathring{g}_{N\D} = -\frac{6}{5}g_0\, ,&
&\mathring{g}_{\D\D} = -\frac{9}{5}g_0\, .&
\end{align}
Our numerical results are insufficient to constrain these couplings, so to study the sensitivity to them, we include them in the analysis under the Bayesian Framework with prior widths varying between 5\% and 40\%.

The sign difference between these two axial couplings and $g_0$ is what leads to the milder pion mass dependence as this sign difference results in axial coefficients of the non-analytic terms in Eq.~\eqref{eq:gA_delta} being $\sim2$ times smaller. Another interesting feature of the delta-full extrapolation is the presence of $m_\pi^3$ dependence with a fixed coefficient. This does not appear until NNLO in the delta-less expansion, and with an unknown LEC.

\subsubsection{\label{sec:taylor_extrapolation}Taylor expansion}
In addition to these two $\chi$PT-derived extrapolation functions, a simple Taylor expansion can also be considered. In particular, for a quantity with mild pion mass dependence, such as $g_A$, the Taylor expansion should provide an adequate description of the quark mass dependence\cite{Durr:2008zz}. The most natural parameter for performing the Taylor expansion is a reference light quark mass.  Since the squared pion mass scales approximately as the quark mass, a corresponding Taylor expansion can be performed about $\e_\pi^2$.
This is equivalent to dropping the non-analytic contributions to Eq.~\eqref{eq:ga_xpt}.

One can also consider a Taylor expansion in the parameter $\e_\pi$. This is not a natural expansion parameter as it scales approximately as the the square root of the input quark masses.  However, it has been observed that the nucleon mass displays a remarkably linear dependence upon the pion mass, such that phenomenological estimate of the nucleon mass in lattice calculations is given by $m_N \simeq 800 +m_\pi$~MeV\protect\citeM{WalkerLoud:2008bp,WalkerLoud:2008pj,Walker-Loud:2013yua}. This observation motivates us to consider this Taylor expansion as an alternative model for describing the pion mass dependence. As with the Taylor expansion in $\e_\pi^2$, we observe that the choice of the reference point to perform the expansion has insignificant impact on extrapolated value of $g_A$ when the NNLO ($\e_\pi^2$ in this case) expansion is considered.

\subsubsection{\label{sec:volume_extrapolation}Dependence upon $m_\pi L$}
The finite-volume corrections can be incorporated into the EFT through an infrared modification of the pion propagators\cite{Gasser:1987zq}. In the asymptotically large volume limit, these corrections vanish at least as fast as $e^{-m_\pi L}$. The leading volume corrections to $g_A$ can be parameterised as\protect\citeM{Beane:2004rf}
\begin{align}\label{eq:gA_FV}
\d_L &\equiv g_A(L) - g_A(\infty)
\nonumber\\&
    = \frac{8}{3} \e_\pi^2 \left[
 g_{0}^3 F_1(m_\pi L)
 +g_{0} F_3(m_\pi L)
 \right]
\end{align}
where
\begin{align}
F_1(x) &= \sum_{\mathbf{n} \neq 0} \left[
 K_0(x|\mathbf{n}|) - \frac{K_1(x|\mathbf{n}|)}{x|\mathbf{n}|}
 \right]\, ,
\nonumber\\
F_3(x) &= -\frac{3}{2} \sum_{\mathbf{n} \neq 0} \frac{K_1(x|\mathbf{n}|)}{x|\mathbf{n}|}\, .
\end{align}
$K_\nu(z)$ are modified Bessel functions of the second kind and $g_0$ is the leading order (LO) contribution to $g_A$ in the chiral expansion.
In the large $m_\pi L$ limit,
\begin{equation}
\d_L = 8 g_0^3 \e_\pi^2 \sqrt{2\pi} \frac{e^{-m_\pi L}}{\sqrt{m_\pi L}}
	+\mathrm{O}\left( e^{-\sqrt{2}m_\pi L}, \frac{1}{(m_\pi L)^{3/2}} \right) .
\end{equation}

In order to asses the uncertainty arising from the FV corrections, we can also model higher order contributions. The NNLO contribution to $g_A$ also arises from single loop diagrams (rather than two loops). It is therefore reasonable to model the finite volume corrections from these terms as similar to those arising from the NLO contributions, particularly the graph that gives rise to the $F_1$ correction. Therefore, we add an additional FV correction with the following form,
\begin{equation}
\label{eq:f3_fv}
\d_{L_3} \equiv f_3 \e_\pi^3 F_1(m_\pi L)
\end{equation}
where $f_3$ is an unknown LEC.

We do not currently have any prior knowledge for what the value of $f_3$ is. Therefore, the prior central value for $f_3$ is set to zero, while the width is determined by an empirical Bayes study shown in Supplemental Data Fig.~\ref{fig:fv_bayes}. For the six models that enter the final result, we vary the prior width from 0.5 to 40 and for each model, choose the value with the largest Bayes factor. The priors for $f_3$ are listed in Supplemental Data Table~\ref{tab:priors}. When compared to the coefficient of the leading finite volume discretization correction $8g_0/3$, the width of $f_3$ is approximately 3 to 5 times wider when determined by the empirical Bayes analysis, and provides a conservative estimate for the finite volume uncertainty. The complete finite volume correction considered is defined to be
\begin{equation}
\label{eq:full_fv_correction}
\delta^\prime_L\equiv \delta_L + \delta_{L_3}.
\end{equation}

\begin{figure*}{\docfont
	\includegraphics[width=0.32\textwidth]{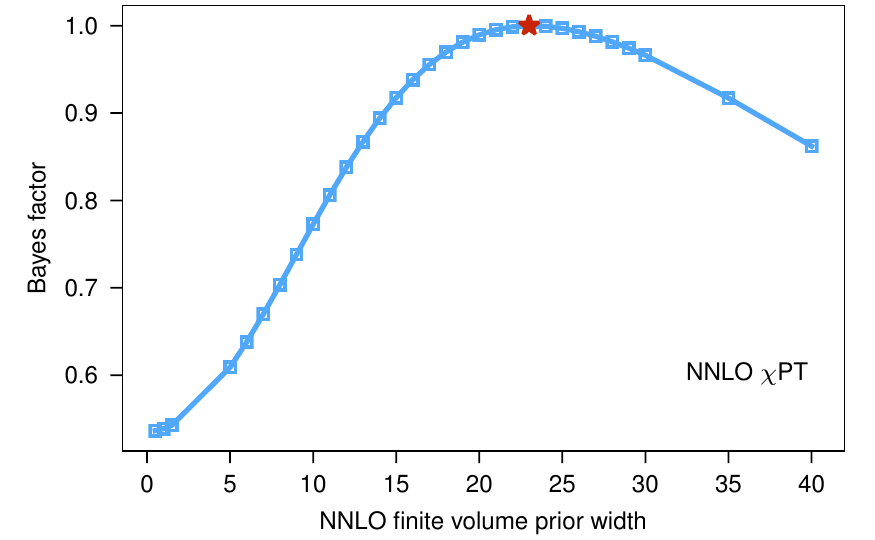}
	\includegraphics[width=0.32\textwidth]{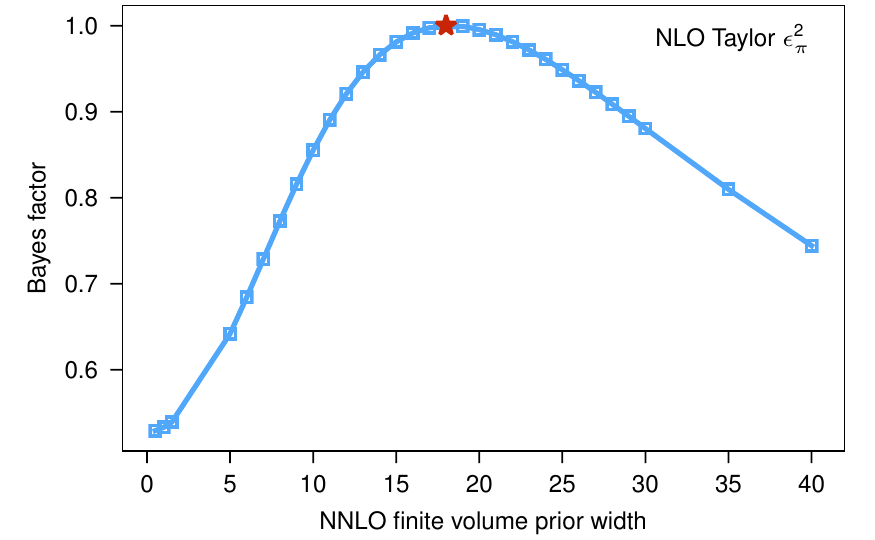}
	\includegraphics[width=0.32\textwidth]{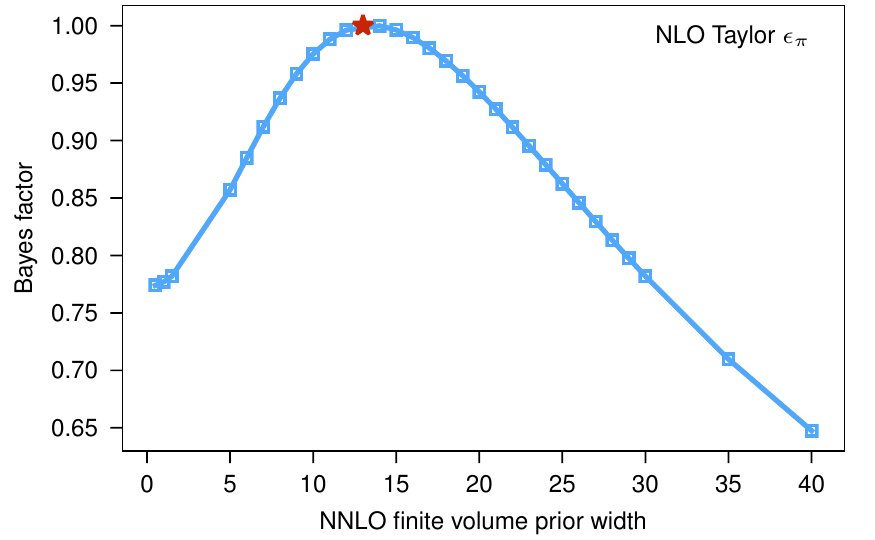}
	\includegraphics[width=0.32\textwidth]{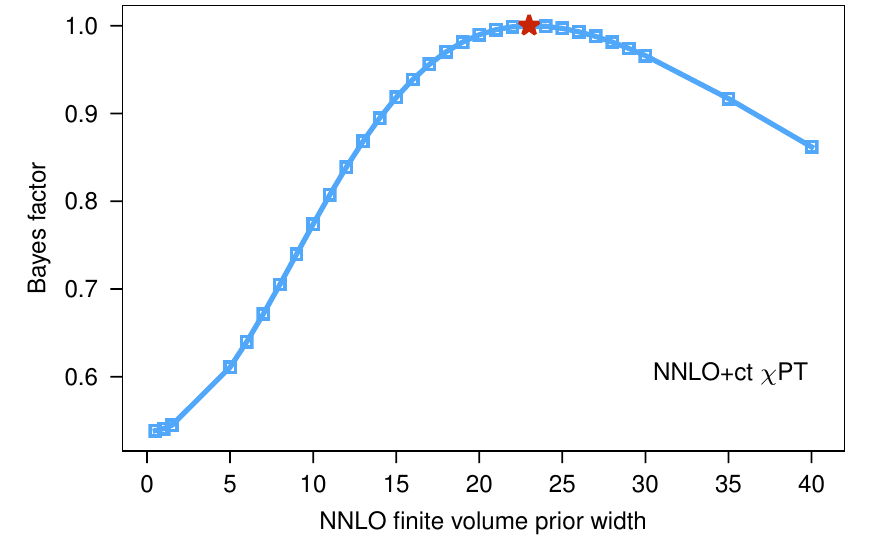}
	\includegraphics[width=0.32\textwidth]{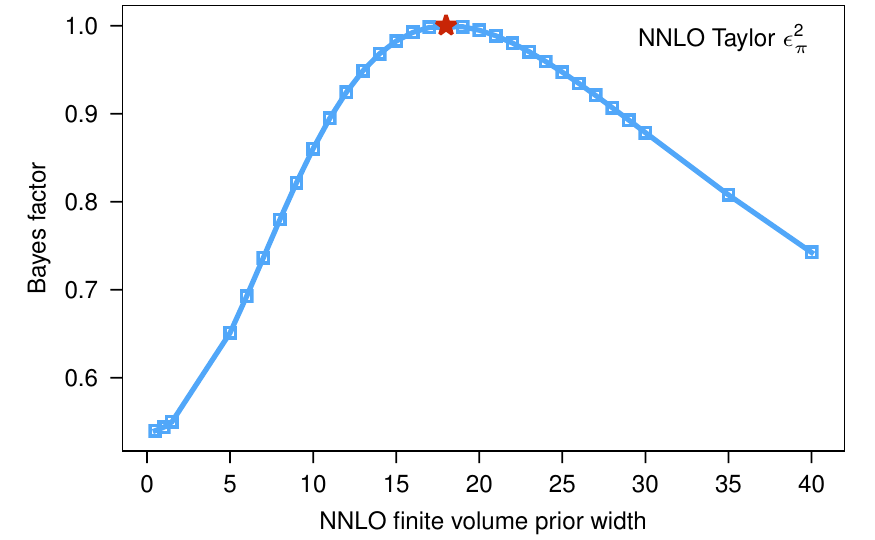}
	\includegraphics[width=0.32\textwidth]{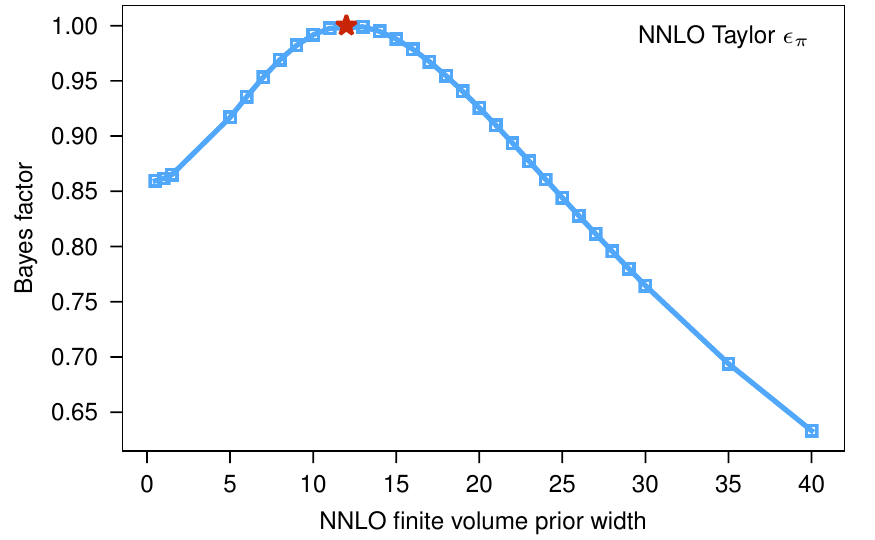}
	\caption{\label{fig:fv_bayes}
		\textbf{Empirical Bayes analysis for $f_3$ prior width}. The prior width of the dimensionless LEC $f_3$ is plotted with the resulting Bayes factors for the six models that enter in the final result. The red star marks the maximum Bayes factor and sets the prior width for $f_3$ in the final analysis.
	}
}\end{figure*}

There has been some discussion in the literature that $g_A$ may be particularly susceptible to finite-volume corrections such that the leading $\chi$PT prediction for the volume dependence is grossly insufficient to explain the observed volume dependence\protect\citeM{Jaffe:2001eb,Cohen:2001bg,Yamazaki:2008py,Yamazaki:2009zq}. In Supplemental Data Fig.~\ref{fig:nlo_FV}, we plot the resulting NLO $\chi$PT prediction of the volume dependence (determined in a NNLO fit to all 16 ensembles) as well as the estimated NNLO corrections, Eq.~\eqref{eq:full_fv_correction}, alongside the three a12m220 ensembles, which are in perfect accord.

\begin{figure}{\docfont
\includegraphics[width=\columnwidth]{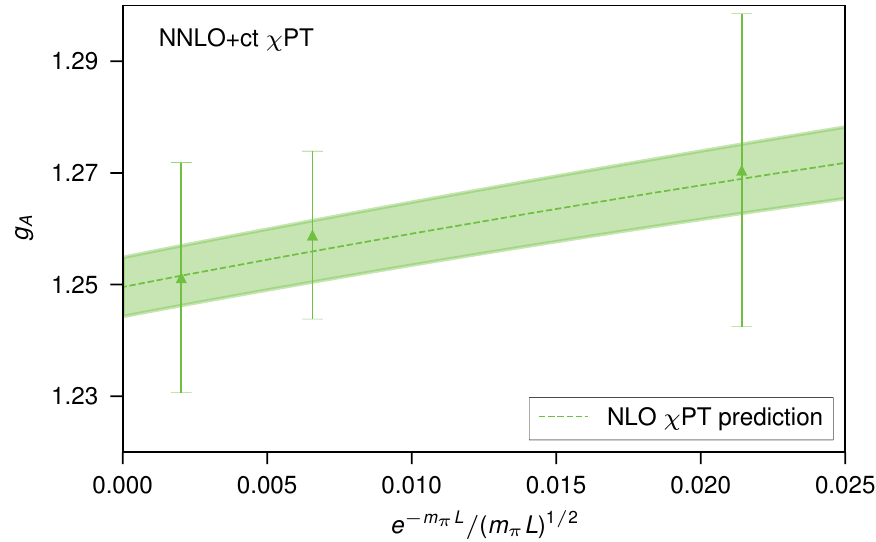}
\includegraphics[width=\columnwidth]{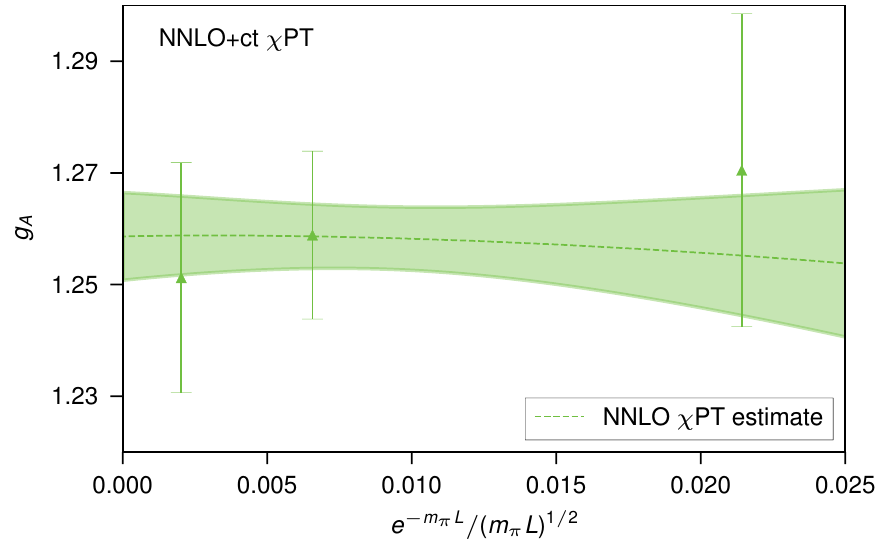}
\caption{\label{fig:nlo_FV}
\textbf{Finite volume dependence}
We plot the predicted NLO finite volume corrections (top) for the $m_\pi\sim220$~MeV ensembles at $a\sim0.12$~fm, with coefficients determined from an NNLO $\chi$PT analysis of all 16 ensembles, along with the numerical results on the a12m220S, a12m220 and a12m220L ensembles.
In the bottom figure, we plot the same result except using the analysis with the estimated NNLO FV correction as well.
Uncertainties are one s.e.m.}
}\end{figure}

\subsubsection{\label{sec:continuum_extrapolation}Dependence upon $\e_a$}
The discretization corrections can be incorporated into \textit{mixed-action} EFT (MAEFT)\protect\citeM{Bar:2003mh},\protect\citeM{Sharpe:1998xm} which is known for our MDWF on HISQ action\protect\citeM{Bar:2005tu,Tiburzi:2005is,Chen:2005ab,Chen:2006wf,Orginos:2007tw,Jiang:2007sn,Chen:2007ug,Chen:2009su} through next-to-leading order (NLO) in the chiral and continuum expansion. Unfortunately, the MAEFT introduces new unknown coefficients that are not well constrained by our results. However, we observe our results are well described by a simple Taylor expansion in the discretization scale, with no discernible pion-mass-dependent discretization effects. We are therefore able to supplement the continuum HB$\chi$PT formula with corrections that parameterise the possible discretization effects to NNLO in the Symanzik exapansion\cite{Symanzik:1983dc}\protect\citeM{Symanzik:1983gh},
\begin{equation}
	\d_a = a_2\e_a^2 + b_4 \e_a^2\e_\pi^2 + a_4 \e_a^4\, ,
\end{equation}
where the first term is an NLO correction and the second and third term arise at NNLO in a power counting where $\e_\pi^2 \sim \e_a^2$. The coefficients are unknown constants which must be determined in the extrapolation analysis.

We also consider discretization corrections of the form
\begin{equation}
	\d^\prime_a = a_1 \sqrt{4\pi} \e_a + s_2 \alpha_s \e_a^2\, ,
\end{equation}
where $a_1 = \mathrm{O}(m_{\text{res}})$ and $s_2 = \mathrm{O}(1)$.
The first term could arise from residual chiral symmetry breaking corrections arising from our use of local axial and vector currents and the second term originates from generic one-loop radiative gluon corrections at finite lattice spacing.

\subsection{Extrapolation analysis}
In the following sections, we discuss how the physical-point extrapolation is performed in order to obtain Eq.~(\ref{eq:ga_breakdown}), the concluding result of this work, followed by a discussion on all sources of uncertainty, and end by studying the sensitivity of our final result under changes to different inputs of the analysis.

\subsubsection{Model averaging \label{sec:uncertainty}}
The extrapolation analysis uses the various Ans\"atze described in Sec.~\ref{sec:extrap}. While $\chi$PT provides our best hope for a model-independent extrapolation in the pion mass, it is not known \textit{a priori} whether the chiral expansion converges for a given quantity near the physical pion mass. For this reason, we include all models equally in our model average. Convergence of the chiral expansion will be discussed in Sec.~\ref{sec:chipt_convergence}.

Our main result comes from the Bayesian model averaging\protect\citeM{BMA} of a set of six different models. This procedure accounts for model selection uncertainty, and avoids over-confident inferences resulting from trusting any single model. The six extrapolation models used are:
\begin{subequations}
\begin{align}
\label{eq:nnlo_xpt}
\textrm{NNLO $\chi$PT}:\quad &
	\textrm{Eq.~\eqref{eq:ga_xpt}} +\d_a + \d_L^\prime
\\
\label{eq:nnloct_xpt}
\textrm{NNLO+ct $\chi$PT}:\quad &
	\textrm{Eq.~\eqref{eq:ga_xpt}} +c_4 \e_\pi^4 +\d_a + \d_L^\prime
\\
\label{eq:nlo_Tesq}
\textrm{NLO Taylor $\e_\pi^2$}:\quad &
	c_0 + c_2 \e_\pi^2 + \d_a + \d_L^\prime
\\
\label{eq:nnlo_Tesq}
\textrm{NNLO Taylor $\e_\pi^2$}:\quad &
	c_0 + c_2 \e_\pi^2 +c_4 \e_\pi^4 +\d_a + \d_L^\prime
\\
\label{eq:nlo_Te}
\textrm{NLO Taylor $\e_\pi$}:\quad &
	c_0 + c_1 \e_\pi + \d_a + \d_L^\prime
\\
\label{eq:nnlo_Te}
\textrm{NNLO Taylor $\e_\pi$}:\quad &
	c_0 + c_1 \e_\pi + c_2 \e_\pi^2 + \d_a + \d_L^\prime
\end{align}
\end{subequations}
In Table~\ref{tab:priors}, we list all the priors used in the analysis of each of the extrapolations.
\begin{table}{\docfont
\begin{ruledtabular}
\begin{tabular}{ccccccccc}
\multicolumn{9}{c}{main parameters}\\
Model& $\tilde{g}_0$& $\tilde{c}_0$& $\tilde{c}_2$& $\tilde{c}_3$& $\tilde{c}_4$& $\tilde{a}_2$& $\tilde{a}_4$& $\tilde{b}_4$\\
\hline
$\chi$PT & 1(50)& --& 0(50)& 0(50)& 0(1)     & 0(50)& 0(1)& 0(1) \\
Taylor $\e_\pi^2$      & 1.2(1.0)& 1(50)& 0(50)& --& 0(1)& 0(50)& 0(1)& 0(1)\\
Taylor $\e_\pi$      & 1.2(1.0)& 1(50)& 0(50)& --& 0(1)& 0(50)& 0(1)& 0(1)\\
\end{tabular}
\begin{tabular}{ccccccc}
\multicolumn{6}{c}{alternate parameters}\\
Model & $\tilde{g}_{N\D}$& $\tilde{g}_{\D\D}$& $\tilde{\g}_4$& $\tilde{a}_1$& $\tilde{s}_2$&$\tilde{f}_3$\\
\hline
$\chi$PT & $-1.44(35)$& $-2.16(52)$& 0(50)& $0(10^{-3})$& 0(1)& 0(23)  \\
Taylor $\e_\pi^2$& --& --& --& $0(10^{-3})$& 0(1) & 0(18)\\
Taylor $\e_\pi$& --& --& --& $0(10^{-3})$& 0(1) & 0(12.5)\\
\end{tabular}
\end{ruledtabular}
\caption{\label{tab:priors}
\textbf{Priors used in extrapolation analysis.} The priors used in the extrapolation analysis for all unknown constants. For the lower order coefficients, we use unconstraining priors. For the NNLO coefficients, we use $\mathrm{O}(1)$ priors. For the Taylor fits, the coefficient $g_0$ is used to parameterize the leading finite volume corrections. The impact of varying the prior widths is discussed in Sec.~\ref{sec:priors_cuts} and show in Extended Data Fig.~\ref{fig:final_stability}. The choice of priors for the alternate parameters are discussed in Secs.~\ref{sec:error_budget} and \ref{sec:delta_fits}.
}
}\end{table}

Under the Bayesian framework, the model averaged posterior distribution of $g_A$ is determined by marginalising over the set of models $\{M_k\}$,
\begin{equation}
P(g_A|D) = \sum_{k} P(g_A|M_k,D)P(M_k|D),
\label{eq:model_weight_avg}
\end{equation}
where $P(M_k|D)$ is the posterior distribution of model $k$ given data $D$ and is related to the likelihood of the model $P(D|M_k)$ through Bayes' Theorem,
\begin{equation}
\label{eq:bayes_weights}
P(M_k|D)=\frac{P(D|M_k)P(M_k)}{\sum_l P(D|M_l)P(M_l)}.
\end{equation}
In particular, the likelihood that the data is produced under model $k$ is given by the marginalising over all (continuous) parameters $\theta_k$
\begin{equation}
P(D|M_k) = \int P(D|\theta_k, M_k) P(\theta_k |M_k) d\theta_k
\end{equation}
and as a result $P(D|M_k)$ is now explicitly a number and not a distribution. The act of marginalising over all parameters naturally penalises over-parameterised models. In our average, we choose agnostic priors $P(M_k)$ for all models listed in Eq.~(\ref{eq:nnlo_xpt}--\ref{eq:nnlo_Te}). Consequently, the posterior mean $\textrm{E}[g_A]$ and variance $\textrm{Var}[g_A]$ for the model weighted average follows
\begin{align}
\textrm{E}[g_A] & = \sum_k \textrm{E}[g_A|M_k] P(M_k|D),\\
\textrm{Var}[g_A] & = \sum_k \textrm{Var}[g_A|M_k] P(M_k|D) \nonumber \\
& + \left\{\sum_k\textrm{E}^2[g_A|M_k]P(M_k|D)\right\} - \textrm{E}^2[g_A|D],
\label{eq:model_avg_var}
\end{align}
where $\textrm{Var}(g_A)$ is a direct consequence of the \textit{law of total variance}. The first line of Eq.~(\ref{eq:model_avg_var}) yields the \textit{expected value of the process variance} which we refer to as the \textit{model averaged variance} while the second line gives the \textit{variance of the hypothetical means} which we refer to as the \textit{model uncertainty}. The weighted average is performed with \texttt{lsqfit}\protect\citeM{lsqfit-9}.

The resulting physical point extrapolations are provided in Supplemental Data Tab.~\ref{tab:model_selection} and plotted in Extended Data Fig.~\ref{fig:final_stability}, and the model average extrapolation is presented in Extended Data Fig.~\ref{fig:model_selection}. The convergence of each model, the model average continuum, and infinite volume extrapolations are presented in Extended Data Figs.~\ref{fig:chipt_convergence}, \ref{fig:continuum_extrapolation}, and \ref{fig:IV_extrapolation} respectively.

\begin{table}{\docfont
\begin{tabular}{rcccc}
\hline\hline
Fit& $\chi^2/\textrm{dof}$& $\mathcal{L}(D|M_k)$& $P(M_k|D)$& $P(g_A|M_k)$\\
\hline
NNLO $\chi$PT               & 0.727 & 22.734& 0.033& 1.273(19) \\
NNLO+ct $\chi$PT          & 0.726 & 22.729& 0.033& 1.273(19) \\
NLO Taylor $\e_\pi^2$     & 0.792 & 24.887& 0.287& 1.266(09) \\
NNLO Taylor $\e_\pi^2$  & 0.787 & 24.897& 0.284& 1.267(10) \\
NLO Taylor $\e_\pi$        & 0.700 & 24.855& 0.191& 1.276(10) \\
NNLO Taylor $\e_\pi$     & 0.674 & 24.848& 0.172& 1.280(14) \\
\hline
\textbf{average}&&&& \gaM \\
\hline\hline
\end{tabular}
\caption{\label{tab:model_selection}
\textbf{Model selection analysis.} We explore model uncertainties by choosing six different models and studying the variation in the extrapolation to the physical point. With the inclusion of priors, the \textit{augmented} $\chi^2/\textrm{dof}$ listed assumes there to be 16 degrees of freedom (equal to the number of data points) for all six models. $\mathcal{L}(D|M_k)$ lists the log-likelihood distribution $\log P(D|M_k)$. The Taylor expansion fits are strongly favored over the $\chi$PT fits as measured by the posterior of the model. In \texttt{lsqfit}, $\mathcal{L}(D|M_k)$ is called the \textit{log Gaussian Bayes Factor} (logGBF). The final result is given with two uncertainties: the first is the averaged variance (line one of Eq.~(\ref{eq:model_avg_var})) and the second is the model uncertainty (line two of Eq.~(\ref{eq:model_avg_var})).
}
}\end{table}

We present our final physical point extrapolation as a function of pion mass in Fig.~\ref{fig:results}a. A comparison with other LQCD results\cite{Edwards:2005ym,Capitani:2012gj,Horsley:2013ayv,Bali:2014nma,Abdel-Rehim:2015owa,Bhattacharya:2016zcn,Bhattacharya:2016zcn,Alexandrou:2017hac,Capitani:2017qpc} is presented in Fig.~\ref{fig:results}b

\subsubsection{Uncertainty analysis \label{sec:error_budget}}
The final uncertainty budget receives contributions from statistical uncertainty, extrapolation to the chiral, continuum and infinite volume limits, the model selection uncertainty, and isospin symmetry breaking.

\bigskip\noindent\textit{\scriptsize Statistical uncertainty}\smallskip

Statistical uncertainty incorporates the correlated uncertainties of $\mathring{g}_A$, $\mathring{g}_V$, $m_\pi$, and $F_\pi$, as well as the uncorrelated uncertainty of $m_\pi$ and $F_\pi$ obtained from the PDG\cite{Olive:2016xmw}, which is used to evaluate the chiral-continuum extrapolation at the physical point.

\bigskip\noindent\textit{\scriptsize Chiral and continuum extrapolation uncertainty}\smallskip

The chiral extrapolation uncertainty is determined from the uncertainty on the resulting LECs that control the $\e_\pi$ dependence and do not vanish in the continuum limit.
Uncertainty from the continuum extrapolation includes statistical uncertainty from $\e_a$ and the resulting uncertainty on all LECs associated with corrections that vanish in the continuum limit.
Additional generic one-loop and chirally-suppressed tree-level discretization errors are investigated in Fig.~\ref{fig:final_stability}, labeled as `$\plus\mathrm{O}(\alpha_s a^2)~\textrm{disc.}$' and `$\plus\mathrm{O}(a)~\textrm{disc.}$' respectively, yielding insignificant changes to the result and are therefore omitted from the final extrapolation.

\bigskip\noindent\textit{\scriptsize Infinite volume extrapolation uncertainty}\smallskip

We include the finite volume correction given by Eq.~(\ref{eq:full_fv_correction}). Specifically, the leading volume correction to $g_A$ derived from $\chi$PT\protect\citeM{Beane:2004rf} is used and higher-order finite volume corrections are estimated by  Eq.~(\ref{eq:f3_fv}) with an unkown LEC $f_3$. The FV uncertainty is derived through the uncertainties on the LECs that determine the infinite volume dependence. In Extended Data Fig.~\ref{fig:IV_extrapolation}\textbf{a}, we display the model averaged FV correction along with the raw a12m220 ensembles on the three volumes used. In panel \textbf{b}, we add to the model average extrapolation, black horizontal ticks to denote the central value of the renormalized values of $g_A$ from Extended Data Table~\ref{tab:ga_extrap}.  In all but two ensembles, the FV shift is significantly less than one sigma, with one data point shifting about one sigma and the other closer to two sigma. As a cross check of our FV uncertainty, we compare two additional analyses: one without FV corrections and one including only NLO corrections, which are displayed in Extended Data Fig.~\ref{fig:final_stability} as ``omit FV'' and ``NLO FV'' respectively. These both result in a relative difference with our final analysis consistent with our estimated FV uncertainty.

\bigskip\noindent\textit{\scriptsize Isospin breaking uncertainty}\smallskip

Finally, isospin breaking estimates both strong and electromagnetic isospin breaking added in quadrature. Experimental results of $g_A$ include radiative corrections up to one-loop\protect\citeM{Czarnecki:2004cw, Marciano:2005ec}, therefore, we estimate the two-loop radiative corrections to be $(\alpha_{\text{EM}}/\pi)^2\sim0.0005\%$. Strong isospin breaking corrections can enter at $\mathrm{O}(m_d-m_u)$ with the leading correction to the axial current in the chiral Lagrangian appearing at NLO\protect\citeM{Fettes:2000gb,Arndt:2001ye,Beane:2002vq}
\begin{equation}
\d^{\rm NLO} j^\mu_{5,a} = \frac{ b_1^{\d} }{(4\pi F)^2}
	 \bar{N} \{ \t_a^\xi, \chi_\d \} S^\mu N\, ,
\end{equation}
where $S^\mu$ is the spin operator, the spurion fields are
\begin{align}
\t_a^\xi &= \frac{1}{2} \left( \xi \t_a \xi^\dagger + \xi^\dagger \t_a \xi \right)\, ,
\\
\chi_\d &= \frac{1}{2} \left( \xi 2B\d\t_3 \xi + \xi^\dagger 2B\d\t_3 \xi^\dagger \right)\, ,
\end{align}
with $\xi^2 = \S = \exp\{\sqrt{2}i \phi / F\}$, $2\d = m_d-m_u$ and $B$ is the chiral condensate (scaled by $F^2$) related to the pion mass by the Gell-Mann--Oakes--Renner relation\protect\citeM{GellMann:1968rz}. However, this contribution vanishes for the $n\rightarrow p$ transition used to determine $g_A$ as $\{ \t_+, \t_3\} = 0$. In order to contribute to $g_A$, there needs to be two insertions of isospin breaking corrections, such that the anti-commutator of $\t_+$ and the isospin breaking contributions are non-vanishing. In QCD, the only isospin breaking parameter is the quark mass operator and so pure strong isospin breaking corrections enter as $\mathrm{O}\left(\frac{(m_d-m_u)^2}{(m_d+m_u)^2} \e_\pi^4\right)\sim 0.002\%$. There can also be mixed QED + QCD isospin breaking corrections which will scale as $\mathrm{O}\left(\a_\textrm{EM} \frac{m_d-m_u}{m_d+m_u} \e_\pi^2\right) \sim 0.004\%$.

Finally, QED corrections modify the values of $m^{\text{nature}}_\pi$ and $F^{\text{nature}}_\pi$. The $\pi^0$ mass provides a good estimate of the pion mass in the isospin limit\cite{Aoki:2016frl} and the QED corrections to $F_\pi^-$ are given by
\begin{equation}
F_{\pi^-} = F_{\pi^-}^{\rm LQCD} \left(1 + \frac{\d R_{\pi^-}}{2}\right)\, ,
\end{equation}
where the correction has been estimated to be\protect\citeM{Lubicz:2016mpj}
\begin{equation}
\d R_{\pi^-} = 0.0169(15)\, ,
\end{equation}
in good agreement with the $\chi$PT estimate\protect\citeM{Cirigliano:2011tm,Rosner:2015wva}. We can make a conservative estimate of the uncertainty from these corrections by splitting the difference of our extrapolated answers by using
\begin{align}
\e_{\pi^-} &= \frac{m_{\pi^-}}{4\pi F_{\pi^{-}}}\, ,
\\
\e_{\pi^0} &= \frac{m_{\pi^0}(1+\d R_{\pi^-}/2)}{4\pi F_{\pi^{-}}}\, ,
\end{align}
resulting in an uncertainty estimate of
\begin{equation}
	\left| \frac{g_A(\e_{\pi^0}) - g_A(\e_{\pi^-})}{2} \right| = 0.00038(14)\, ,
\end{equation}
which is a $0.03\%$ uncertainty.

\bigskip\noindent\textit{\scriptsize Model uncertainty}\smallskip

The model selection uncertainty is determined as described in Sec.~\ref{sec:uncertainty}.

\bigskip\noindent\textit{\scriptsize Final uncertainty}\smallskip

The final model averaged uncertainty breakdown, including the uncertainty arising from the different extrapolation functions, is presented in the main text, Eq.~\eqref{eq:ga_breakdown}. Broken down to the different contributions of statistical (s), chiral ($\chi$), continuum (a), infinite volume (v), isospin breaking (I) and model (M), we have
\begin{equation*}
g_A = \gaindividual\, .
\end{equation*}
The total uncertainty arises from adding these uncertainties in quadrature. More precise values at the physical pion mass will have the largest impact in simultaneously reducing the extrapolation and model-selection uncertainty. This demonstrates a straightforward path towards a sub-percent precision, which may be able to offer insight to the upward trending values of the measurements of $g_A$, including the most recent determination\protect\citeM{Brown:2017mhw}.

\subsubsection{Sensitivity analysis \label{sec:priors_cuts}}
Robustness of the final result is tested under changes in the initial prior distributions, and data. Specifically, subsets of data are explored to quantify sensitivity for different regions of pion mass and lattice spacing.

\begin{figure*}{\docfont
\includegraphics[width=0.325\textwidth]{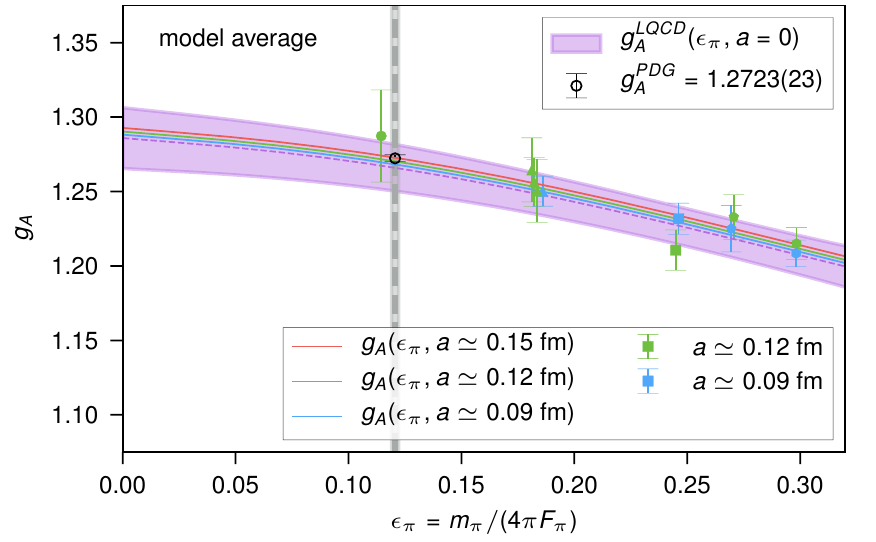}
\includegraphics[width=0.325\textwidth]{figs_main/chiral_modelavg}
\includegraphics[width=0.325\textwidth]{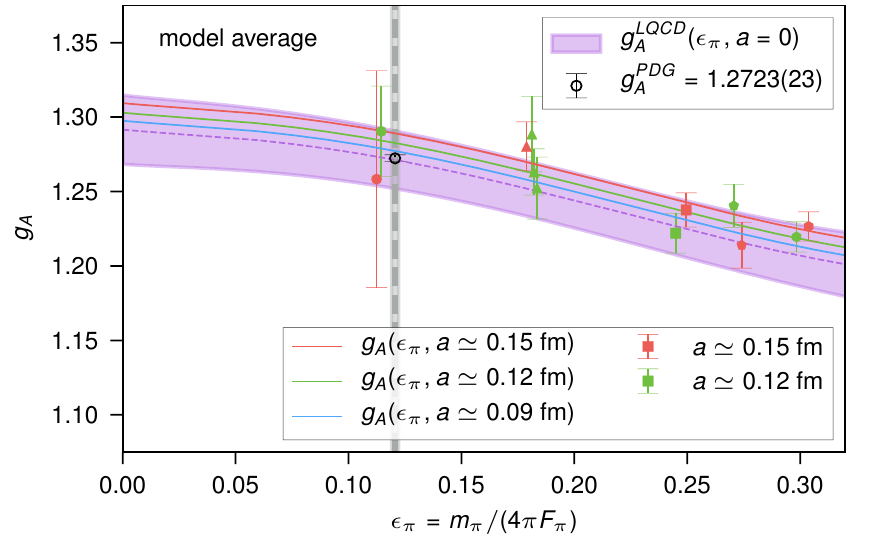}
\caption{\label{fig:disc_cut}
\textbf{Sensitivity to cutting lattice spacings.} We plot the resulting model average analysis when either the coarsest or finest ensembles are cut from those considered, with the full data set in the middle for comparison.  Uncertainties are one s.e.m.
}
}\end{figure*}

\bigskip\noindent\textit{\scriptsize Prior sensitivity}\smallskip

Our final result, displayed as the black square in Extended Data Fig.~\ref{fig:final_stability}, is robust and extrapolated with unconstraining priors.  In this figure, the set of results labeled `$2\times$~LO~width' and `$2\times$~all~widths' explores doubling the prior widths for the leading-order and all LECs respectively. We observe indiscernible change to the final result when doubling only the widths on the priors for the leading-order LECs, demonstrating these priors to be unconstraining. Doubling the prior widths of all LECs leads to insignificant changes in the final result, demonstrating that none of the priors are biasing the fit.

\bigskip\noindent\textit{\scriptsize Pion mass sensitivity and posterior correlations}\smallskip

Next, we demonstrate that the heavy pion masses do not skew the results by analysing subsets of our full data. We study the variation of our final analysis if the two points close to the physical pion mass are removed. The results of these extrapolation analyses are presented in Extended Data Fig.~\ref{fig:final_stability}. We observe that, as expected, when some of the results are removed, the uncertainty grows, but the resulting extrapolation is consistent with our main result within the 1-$\sigma$ level. In Supplemental Table~\ref{tab:pion_cut}, we show the impact of cutting these heavy pion mass points for each of the models which enter our model average. We conclude there is no statistical justification for truncating the heavy pion mass points. The impact of including or excluding these heavier pion mass points on our final extrapolated uncertainty is already included through the use of varying orders in the various model extrapolation functions. We also observe that the growth in the uncertainty from the full data set to the $m_\pi \lesssim310$~MeV data set scales as one would expect when going from 5 to 3 different pion masses.

Additionally, to assess the influence of the heavy pion mass region on our extrapolated result, we use the resulting analyses to compute the correlation between between the extrapolated result at the physical pion mass and at $m_\pi=400$~MeV. For simplicity, we compute the correlation in both cases in the continuum and infinite volume limits. This correlation allows for the determination of the conditional mean between these two points, which provides a measure of how much the extrapolated answer at one point (physical pion mass) would shift, given a fluctuation at the other point (heavy pion mass)
\begin{equation}
g_A(\e_\pi^{(1)}, \e_\pi^{(2)}) = \hat{g}_A(\e_\pi^{(1)})
	+ C_{1,2}\ \s_1 \frac{g_A(\e_\pi^{(2)}) - \hat{g}_A(\e_\pi^{(2)})}{\s_2}\, ,
\end{equation}
where $\hat{g}_A(\e_\pi)$ is the expected value of $g_A$ at $\e_\pi$ given the analysis, $g_A(\e_\pi^{(2)}) - \hat{g}_A(\e_\pi^{(2)})$ is the hypothetical fluctuation at point 2, $\s_{i}$ are the continuum and infinite volume extrapolated uncertainties at the two points and the coefficient $C_{1,2}$ is the correlation coefficient between $g_A$ at $\e_\pi^{(1)}$ and at $\e_\pi^{(2)}$.

The uncertainty of the continuum and infinite volume extrapolated result at the heavy pion mass, $\s_{400}$ is approximately the same as the uncertainty on the input data points at that pion mass ($\sim0.009$ compared to 0.006, 0.010 and 0.008 on the a15m400, a12m400 and a09m400 ensembles respectively, see Extended Data Table~\ref{tab:ga_extrap}). Therefore, if we consider a hypothetical 1-$\s$ fluctuation of the data at $m_\pi\sim400$~MeV, the expected shift in our extrapolated value of $g_A$ at the physical point is well approximated by
\begin{equation}
\d g_A^{phys.} = C_{phys., 400}\ \s_{phys.}\, .
\end{equation}
The model-averaged correlation coefficient, with weights given by the model-selection analysis (Supplemental Data Table~\ref{tab:model_selection}), is given by
\begin{equation}
C_{phys., 400}^\textrm{model avg.} = 0.37\, .
\end{equation}
The shift in the extrapolated value of $g_A$ at the physical pion mass, due to a hypothetical 1-$\s$ fluctuation at $m_\pi\simeq400$~MeV is given by the weighted average of $C_{phys., 400}^{(i)} \s_{phys.}^{(i)}$ over the models, $i$, resulting in
\begin{equation}
\d g_A^{phys.} = 0.0030 \times \hat{g}_A(\e_\pi^{phys.})\, ,
\end{equation}
where $\hat{g}_A(\e_\pi^{phys.})$ is our final result, Eq.~\eqref{eq:ga_breakdown}, or a 0.3\% change. We conclude that the physical-point extrapolation from the model-averaged fit ansatz is relatively insensitive to fluctuations at larger pion masses, leading further evidence for the robustness of our final result.

\begin{table}
{\scriptsize
\begin{ruledtabular}
\begin{tabular}{r|ccc}
$m_\pi$ range& $\lesssim400$~MeV& $\lesssim350$~MeV& $\lesssim310$~MeV\\
Fit&weight, $g_A$& weight, $g_A$& weight, $g_A$ \\
\hline
NNLO $\chi$PT              & 0.033, 1.273(19)& 0.044, 1.281(25)& 0.076, 1.275(27) \\
NNLO+ct $\chi$PT         & 0.033, 1.273(19)& 0.044, 1.280(25)& 0.077, 1.275(27) \\
NLO Taylor $\e_\pi^2$    & 0.287, 1.266(09)& 0.305, 1.278(12)& 0.300, 1.282(15) \\
NNLO Taylor $\e_\pi^2$ & 0.284, 1.267(10)& 0.306, 1.278(13)& 0.300, 1.282(16) \\
NLO Taylor $\e_\pi$       & 0.191, 1.276(10)& 0.156, 1.288(14)& 0.125, 1.290(17) \\
NNLO Taylor $\e_\pi$     & 0.172, 1.280(14)& 0.146, 1.288(16)& 0.121, 1.290(18) \\
\hline
\textbf{model average}& 1.271(11)(06)& 1.281(15)(04)& 1.283(18)(05) \\
\end{tabular}
\end{ruledtabular}
\caption{\label{tab:pion_cut}
\textbf{Effect of cutting heavy pion mass points.} The weights are determined from the log Bayes Factors as described in the text. The $m_\pi \lesssim400$~MeV analysis is described in the text and listed in Extended Data Fig. ~\ref{fig:final_stability} as \textbf{model avg}.  The model average results here correspond to the same points in the figure.
}}
\end{table}

\bigskip\noindent\textit{\small Lattice spacing sensitivity}\smallskip

To further check the sensitivity of our results on the continuum limit, we perform the full analysis discarding, one at a time, each of the individual $a\sim0.15$~fm and $a\sim0.09$~fm ensembles. In Fig.~\ref{fig:disc_cut}, we display the resulting model average extrapolations with these two data cuts side-by-side with the extrapolation of the full data set. The extrapolated final results are also shown in Extended Data Fig.~\ref{fig:final_stability}. Similar to the pion mass cuts, removing results from one of the discretization scales leads to a larger, but consistent result.

\subsection{$\chi$PT convergence and inclusion of the $\D$s \label{sec:chipt_convergence}}

For sufficiently light pion masses, $\chi$PT provides a model independent description of low-energy QCD. What is not known \textit{a priori} is the range of pion masses for which $\chi$PT is a converging, perturbative expansion about the chiral limit. As discussed in Sec.~\ref{sec:delta}, there is also theoretical evidence that the explicit inclusion of the delta degrees of freedom in the chiral Lagrangian will improve the convergence of $g_A$. We explore these extrapolations in more detail.

\subsubsection{Convergence of the $\chi$PT expansion}

We discuss the convergence of the $\chi$PT expansion without explicit delta degrees of freedom. The first fit that results in an acceptable $\chi^2_{\textrm{aug}}/\textrm{dof}$ is the NNLO $\chi$PT fit. This fit has 3 LECs determined from our 5 different pion mass points. The convergence of the fit is displayed in Extended Data Fig.~\ref{fig:chipt_convergence}a. Each curve is the sum of all contributions up to the order listed. One observes large cancellations between the NLO and NNLO contributions already at pion masses lighter than nature (grey vertical line). The strong curvature of the NLO curve is driven by competition between the counter term, $c_2$ and the $\ln(\e_\pi^2)$ contribution.  Because of this competition, it is more difficult to assess the convergence of the theory.

We are not able to perform a meaningful fit with the full N3LO formula as there are 5 unknown LECs, and we have only 5 pion mass points. However, we can check the convergence by adding just the local counter term contribution, $c_4 \e_\pi^4$, the NNLO+ct $\chi$PT fit. The convergence of this fit is depicted in Extended Data Fig.~\ref{fig:chipt_convergence}b. We see that the addition of this term has negligible impact on the resulting extrapolated value of $g_A$ for all pion masses depicted. The resulting order-by-order contributions of this analysis and the resulting correlation matrix for the LECs are given in Supplemental Data Table~\ref{tab:xpt4_order}.
\begin{table}{\docfont
\begin{ruledtabular}
\begin{tabular}{ccrcr}
\multicolumn{5}{c}{Order-by-order contribution}\\
order $n$ & $\d^{(n)} g_A$& $\%$ of total& LEC& value \\
\hline
0& $+1.236(34)$ & $97.1(2.7)$                     & $g_0$& 1.236(34) \\
2& $-0.026(30)$ & $-\phantom{9}2.0(2.4)$   & $c_2$& -23.0(3.5) \\
3& $+0.062(14)$ & $+\phantom{9}4.9(1.1)$ & $c_3$& 28.7(5.5) \\
4& $+0.0000(2)$& $+\phantom{9}0.0(0.0)$  & $c_4$& 0.007(1.000) \\
\hline
Total& \phantom{+}1.273(19)
\end{tabular}
\begin{tabular}{ccccc}
\multicolumn{5}{c}{LEC correlation matrix}\\
\hline
LEC& $g_0$ & $c_2$ & $c_3$ & $c_4$ \\
$g_0$ & 1 & -0.02010 & -0.09365 & 0.03797 \\
$c_2$ & -0.02010 & 1 & 0.97231 & -0.99050 \\
$c_3$ & -0.09365 & 0.97231 & 1 & -0.99401 \\
$c_4$ & 0.03797 & -0.99050 & -0.99401 & 1
\end{tabular}
\end{ruledtabular}
\caption{\label{tab:xpt4_order} \textbf{NNLO+ct $\chi$PT analysis results}, We provide the order-by-order contribution to $g_A$ and the resulting LEC correlation matrix from the NNLO+ct $\chi$PT analysis.}
}\end{table}

Under the Bayesian Framework, we can perform the full N3LO fit. In Supplemental Data Fig.~\ref{fig:xpt_n3lo_full}, we show the resulting fit as well as the convergence of the expansion. We first observe the uncertainty of the extrapolation begins to grow significantly outside the region of constraining data, which is a clear sign of overfitting. One also observes that the convergence of the expansion changes markedly from the NNLO+ct $\chi$PT extrapolation, where now, the NNLO result is also quickly dropping as a function of the pion mass for pion masses near and above the physical pion mass. The full extrapolation has a mild pion mass dependence, which also demonstrates there is a large cancellation between different terms in the expansion. This is not surprising given the large and positive coefficient of the $\ln^2(\e_\pi)$ contribution appearing at N3LO, Eq.~\eqref{eq:ga_xpt_4}. These results suggest that the $\chi$PT expansion may be particularly poor for $g_A$. However, a strong conclusion can not be drawn without having results at more pion mass points. In particular, having precise results in the lighter pion mass region is desirable. The resulting extrapolation is compared with our final result in Extended Data Fig.~\ref{fig:final_stability}.

\begin{figure}{\docfont
\includegraphics[width=\columnwidth]{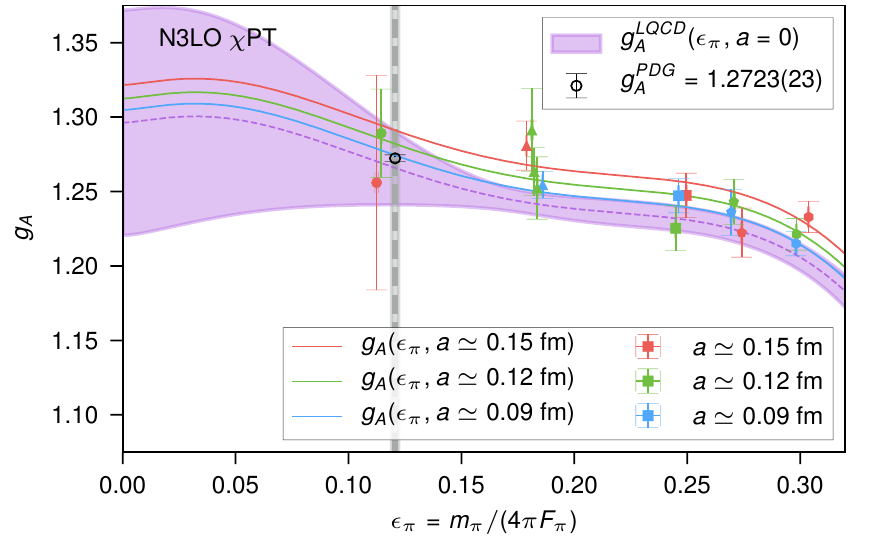}
\includegraphics[width=\columnwidth]{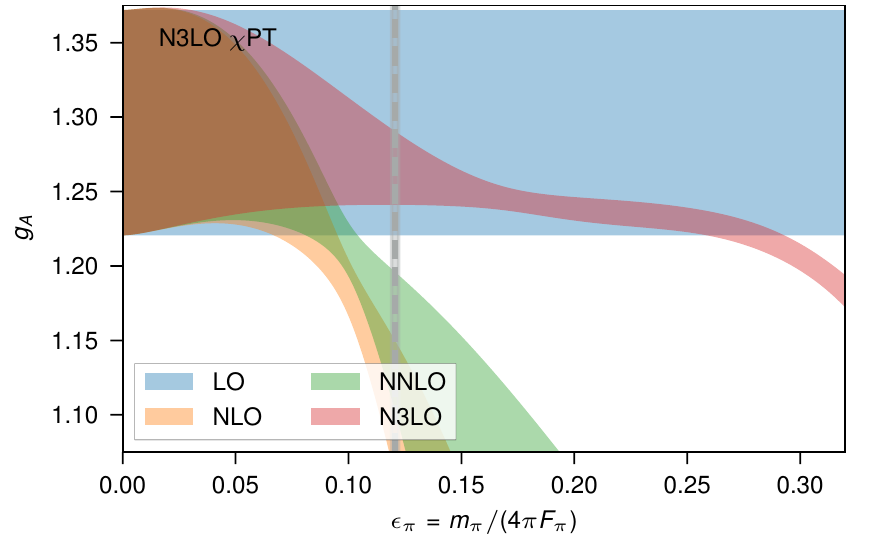}
\caption{\label{fig:xpt_n3lo_full}
\textbf{The N3LO $\chi$PT extrapolation}. We plot the N3LO extrapolation of our results and resulting convergence of the expansion.
Uncertainties are one s.e.m.}
}\end{figure}

\subsubsection{Including the $\D$s \label{sec:delta_fits}}

We finally turn to an extrapolation including explicit $\D$ degrees of freedom. As discussed in Sec.~\ref{sec:delta}, it is expected the pion mass dependence of $g_A$ predicted from $\chi$PT will be milder with the inclusion of these states, due to cancellations between the nucleon and delta virtual loops imposed by the large-$N_c$ expansion. Including the deltas at NLO introduces two new axial coupling LECs and the extra mass splitting parameter, for a total of 4 LECs. An input parameter to the analysis is $\e_\D$, defined in Eq.~\eqref{eq:e_delta}. Our results do not directly constrain these two new axial couplings, and so the resulting fits have larger uncertainties. We do not include the analysis with explicit deltas in our set that are averaged as they require phenomenological input, unlike all the fits which go in the final analysis. Nevertheless, it is interesting to explore how the delta degrees of freedom impact the analysis.

We begin by exploring how varying the prior width on the new axial couplings impacts the analysis. We take the central values in Eq.~\eqref{eq:gND_Nc} and vary the prior width of these couplings from 1\% to 40\%. In Supplemental Data Fig.~\ref{fig:delta_BF}, we plot the resulting Bayes factors normalized to the maximum Bayes factor which occurs with a 18\% prior width on $\mathring{g}_{N\D}$ and $\mathring{g}_{\D\D}$. It is interesting to note that the prior width motivated by Empirical Bayes is comparable to the nominal $1/N_c$ correction which would be $\sim33\%$.

\begin{figure}
\includegraphics[width=0.49\textwidth]{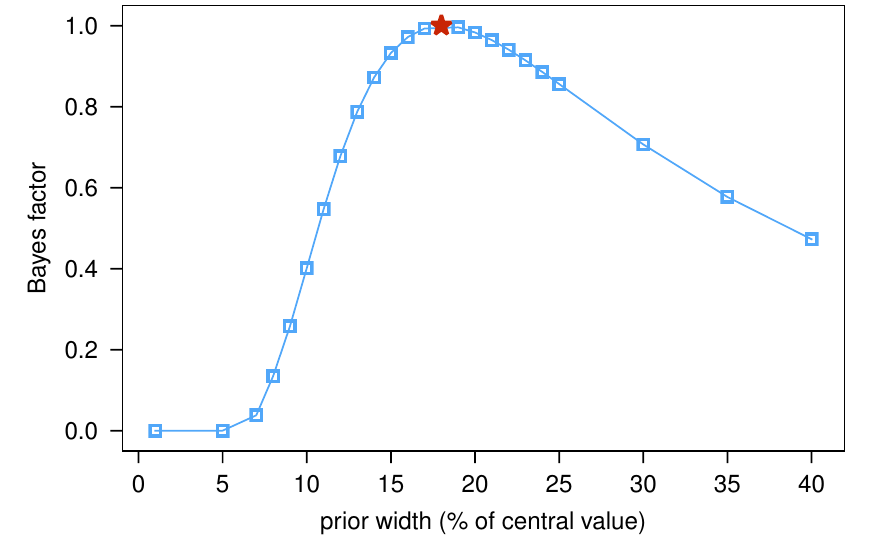}
\caption{\label{fig:delta_BF}
\textbf{Bayes factors versus prior width}. We plot Bayes factors of the resulting NLO $\chi$PT($\D$) fits versus the prior width (in \%) given to the two new axial couplings.
An Empirical Bayes analysis selects the 18\% as the optimal prior width.
}
\end{figure}

Next, we explore the convergence of the expansion when the delta degrees of freedom are included. While we are taking $\D\simeq293$~MeV, $\e_\D$ depends upon $\e_a$ and $\e_\pi$ through the denominator of $F_\pi$. We have not parameterised this dependence, precluding our ability to plot the resulting fit, but we can observe the values of the LECs that are determined with and without explicit delta degrees of freedom to compare the size of the NLO contributions. This study can only be performed for the NLO LEC $c_2$, as we do not have the extrapolation function with deltas beyond NLO. In Supplemental Data Table~\ref{tab:delta_fit}, we list the resulting value of $c_2$ and the size of the NLO contribution for the analysis with and without the delta. The NLO $\chi$PT($\slashed{\D}$) results in a very poor fit, so we do not report the values of $c_2$ or $\d g_A^{\rm NLO}$.  The NLO $\chi$PT($\D$) does result in a good fit with $\chi^2_{\textrm{aug}}/\textrm{dof} = 0.49$. The resulting fit is displayed in the bottom entry of Extended Data Fig.~\ref{fig:final_stability}. Comparing the NNLO fits, which means either NLO $\chi$PT($\D$) and NLO $\chi$PT($\slashed{\D}$) plus NNLO $\chi$PT($\slashed{\D}$), we observe the value of $c_2(\D)$ is approximately half as big as $c_2(\slashed{\D})$, indicating that the non-analytic terms are smaller when the deltas are included. This suggests that the convergence will be improved with the explicit inclusion of $\D$ degrees of freedom.

In order to fully constrain this fit and test the convergence of the expansion with and without delta degrees of freedom, first, a calculation including the $N\rightarrow\D$ and $\D\rightarrow\D$ axial matrix elements is needed, and second, the EFT including the deltas must be worked out to at least one higher order in the expansion. This is a particularly interesting question to resolve as the $\chi$PT($\slashed{\D}$) seems to be poorly converging, if at all, as observed in the prior section.

\begin{table}{\docfont
\begin{ruledtabular}
\begin{tabular}{rcccc}
Order& \multicolumn{2}{c}{with delta}& \multicolumn{2}{c}{without delta}\\
         & $c_2$& $\d g_A^{\rm NLO}$& $c_2$& $\d g_A^{\rm NLO}$\\
\hline
NLO & -4.8(1.8)& 0.28(13)& --& --\\
NNLO & 12.1(8.2)& 0.75(40)& -23.0(3.5)&-0.026(30) \\
\end{tabular}
\end{ruledtabular}
\caption{\label{tab:delta_fit}
\textbf{Effect of including delta degrees of freedom.} We list the resulting value of $c_2$ and the NLO contribution to $g_A$ at the physical point for two fits with and without the delta degrees of freedom. The NLO $\chi$PT($\slashed{\Delta}$) fit has a sufficiently poor $\chi^2_{\textrm{aug}}/\textrm{dof}$ we do not report the resulting LEC and $\d g_A^{\rm NLO}$ values.
}
}\end{table}

\newpage

\begin{figure*}[h]{\docfont
\includegraphics[width=0.49\textwidth]{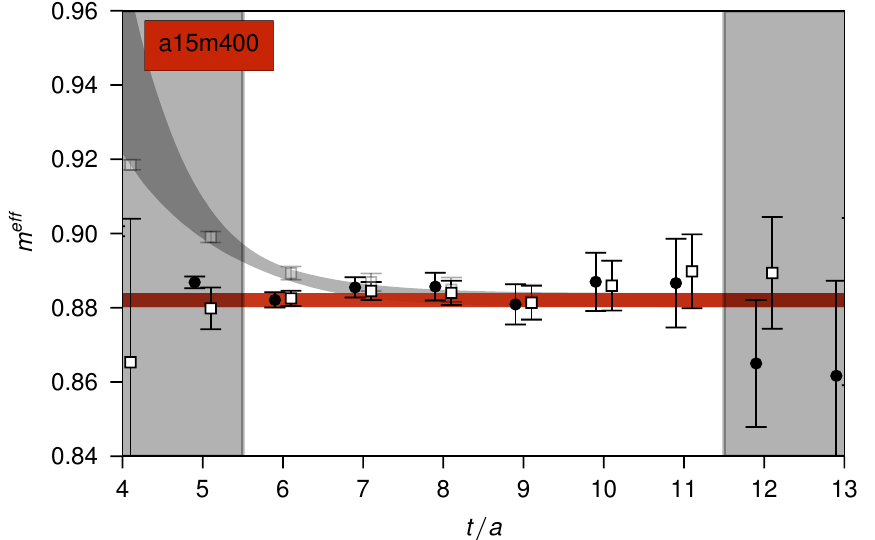}
\includegraphics[width=0.49\textwidth]{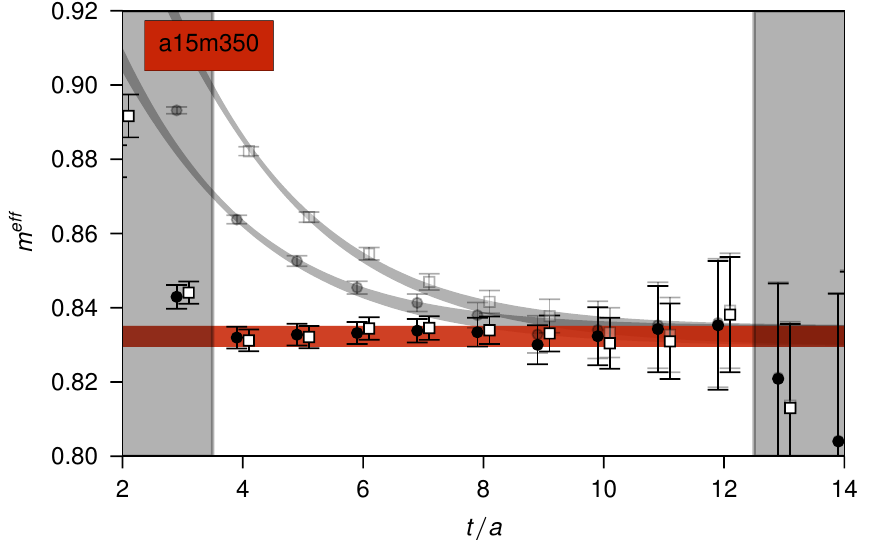}
\includegraphics[width=0.49\textwidth]{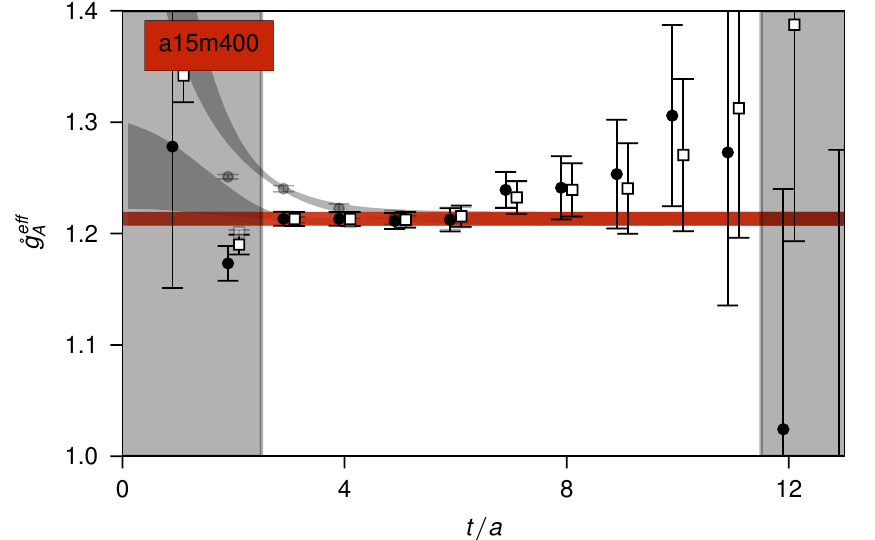}
\includegraphics[width=0.49\textwidth]{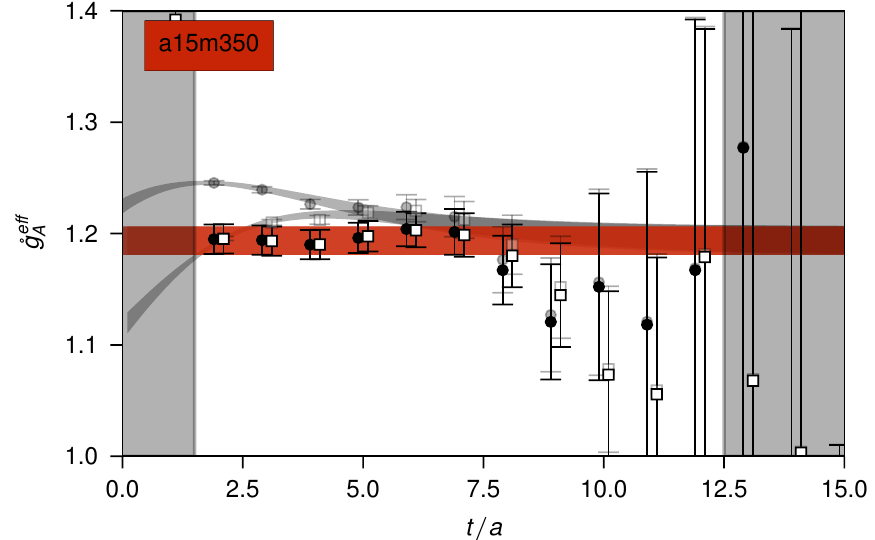}
\includegraphics[width=0.49\textwidth]{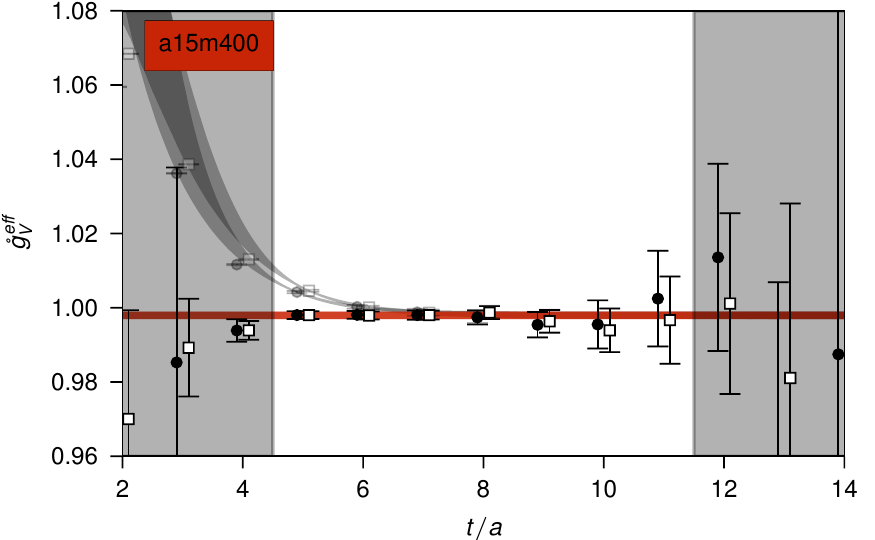}
\includegraphics[width=0.49\textwidth]{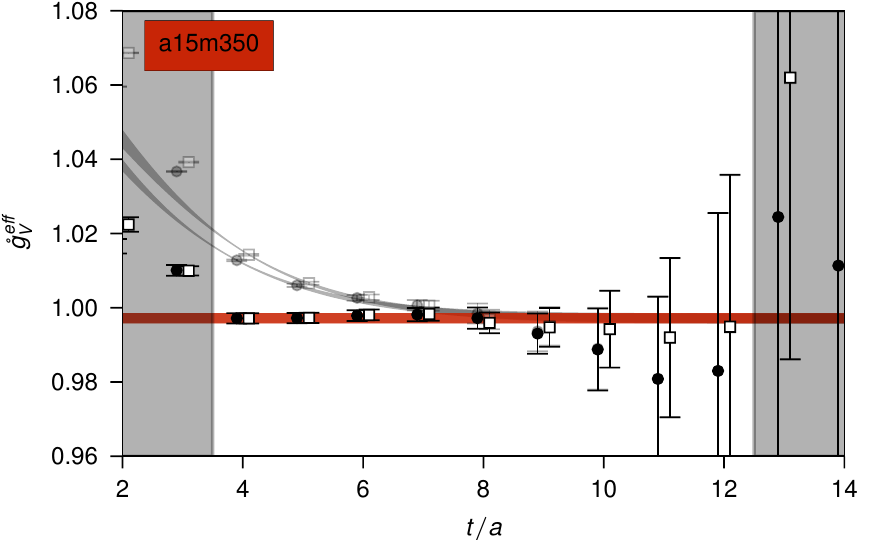}
\includegraphics[width=0.49\textwidth]{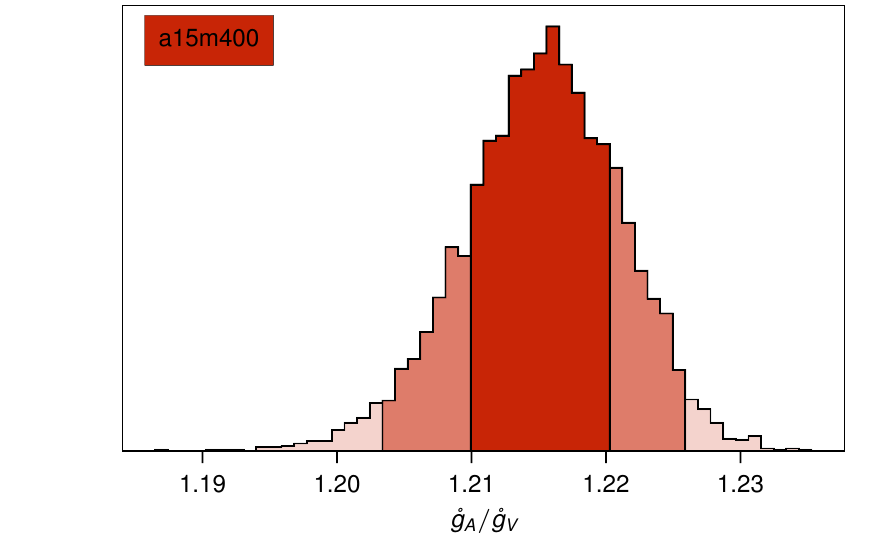}
\includegraphics[width=0.49\textwidth]{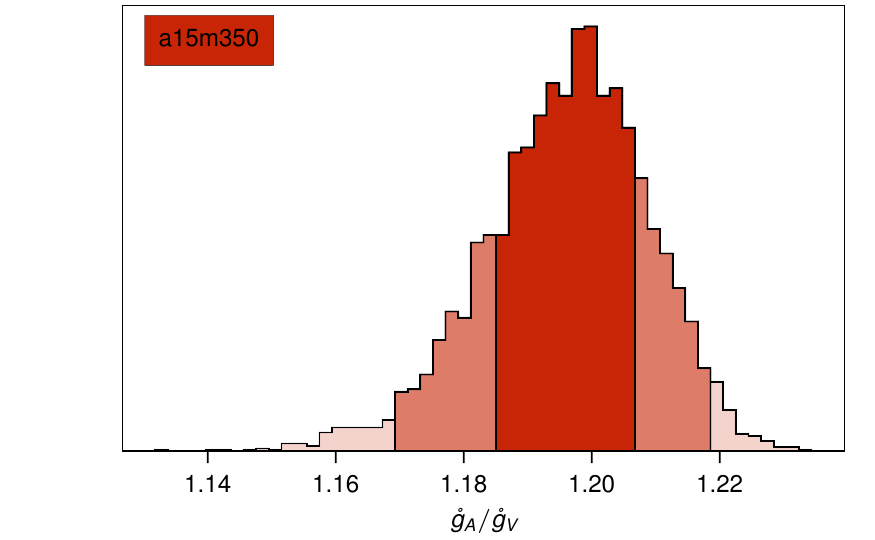}
\caption{\label{fig:a15m400a15m350_curve}
{\textbf{Correlator fit study I.}} Analogous to Extended Data Fig.~\ref{fig:correlator_fitcurves}a,  b, c and d for the a15m400 and a15m350 ensembles. Unbiased bootstrap fit curves with 68\% confidence intervals. Results from one simultaneous fit are represented in each column. The resulting biased bootstrap histograms for $\mathring{g}_{A}/\mathring{g}_{V}$ follow at the bottom. In the histograms, regions mark the 68\% and 95\% confidence interval. Uncertainties are one s.e.m.}
}\end{figure*}

\begin{figure*}[h]{\docfont
\includegraphics[width=0.49\textwidth]{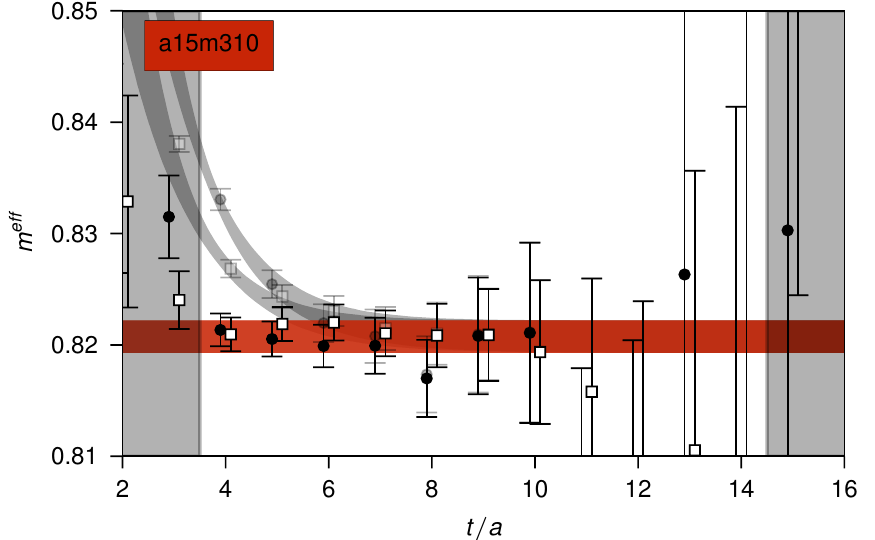}
\includegraphics[width=0.49\textwidth]{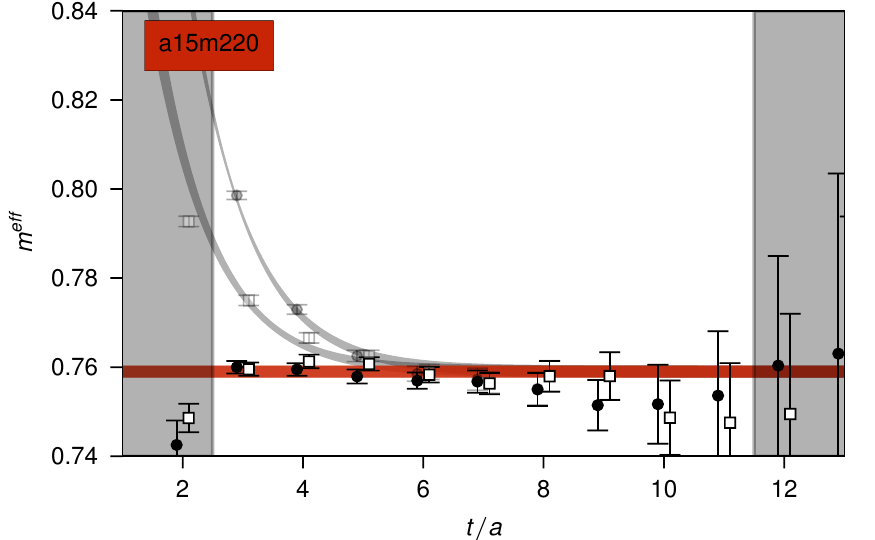}
\includegraphics[width=0.49\textwidth]{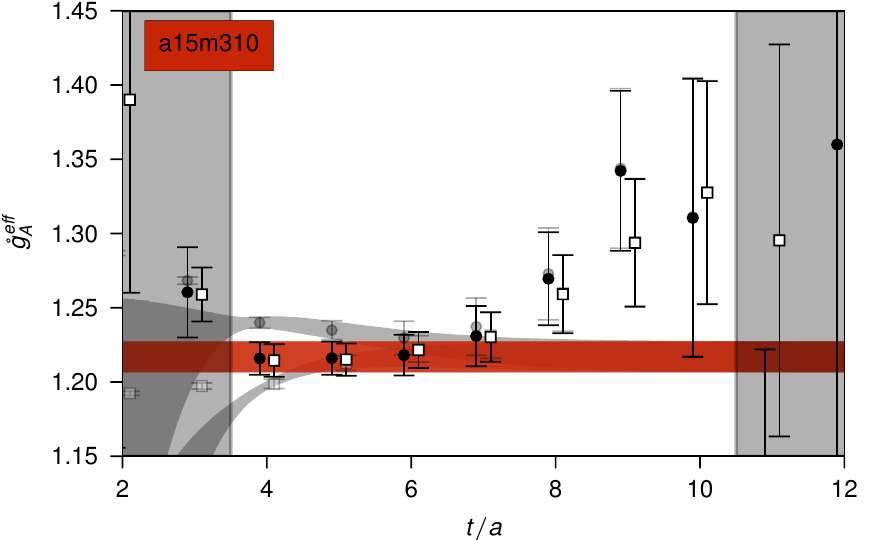}
\includegraphics[width=0.49\textwidth]{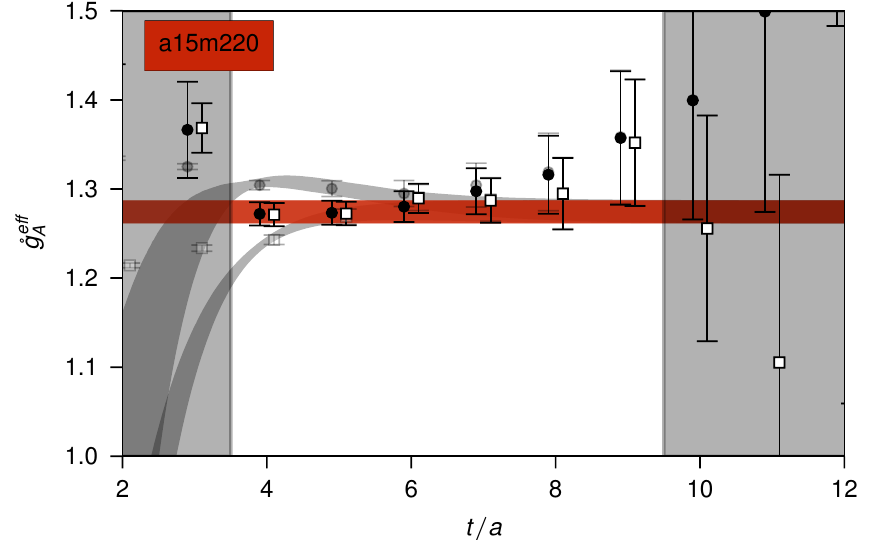}
\includegraphics[width=0.49\textwidth]{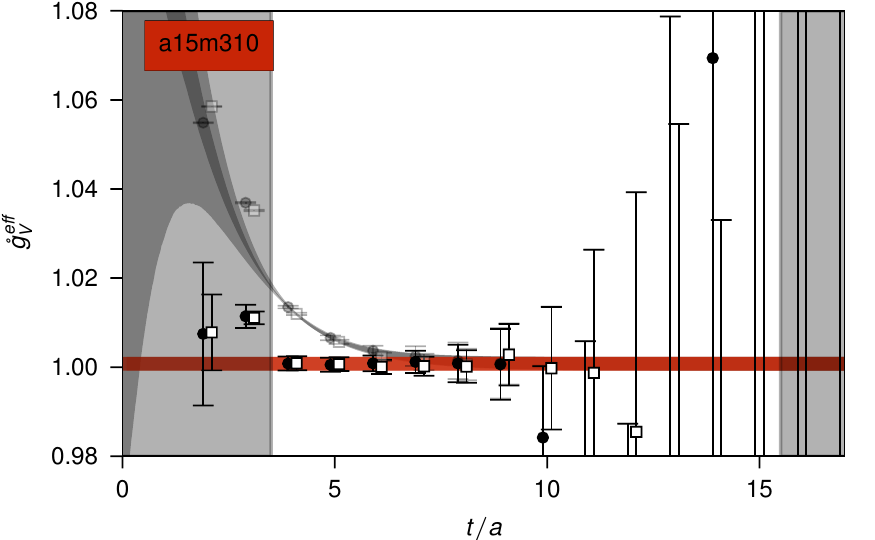}
\includegraphics[width=0.49\textwidth]{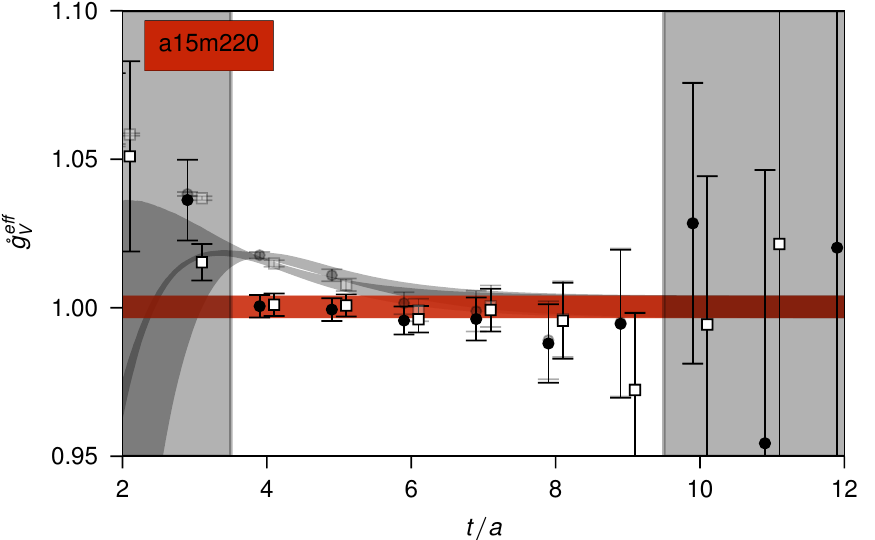}
\includegraphics[width=0.49\textwidth]{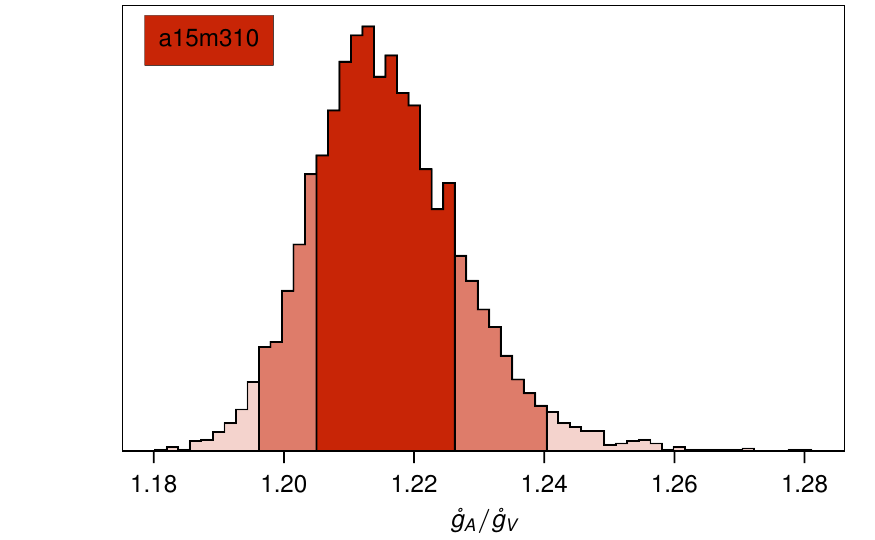}
\includegraphics[width=0.49\textwidth]{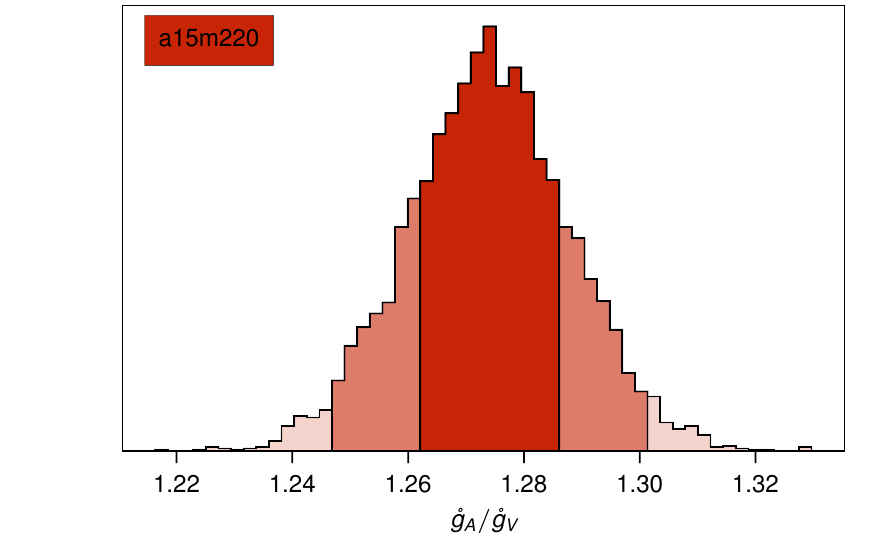}
\caption{\label{fig:a15m310a15m220_curve}
{\textbf{Correlator fit study II.}} Analogous to Extended Data Fig.~\ref{fig:correlator_fitcurves}a,  b, c and d for the a15m130 and a12m400 ensembles.
Uncertainties are one s.e.m.}
}\end{figure*}

\begin{figure*}[h]{\docfont
\includegraphics[width=0.49\textwidth]{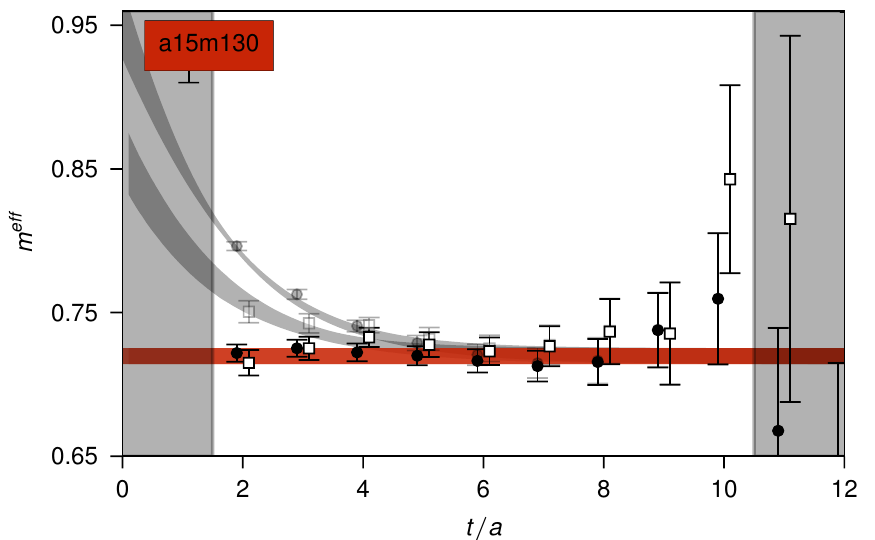}
\includegraphics[width=0.49\textwidth]{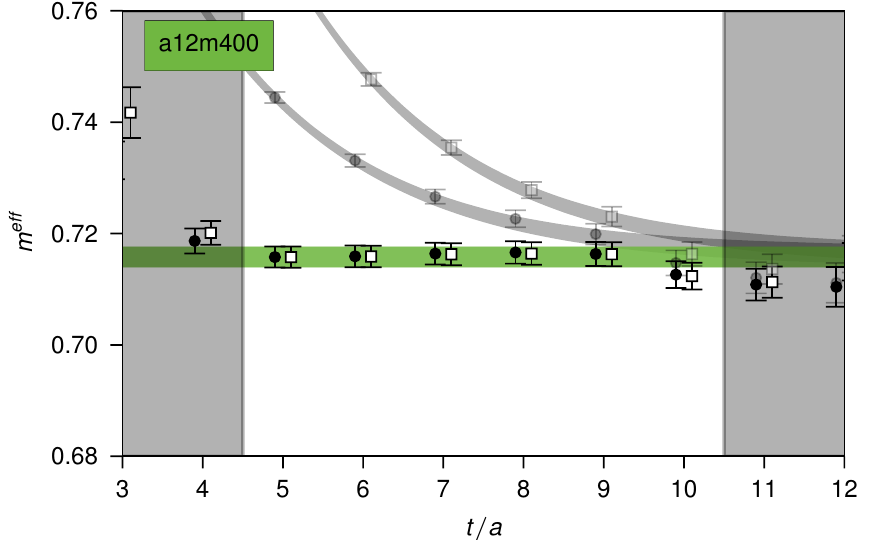}
\includegraphics[width=0.49\textwidth]{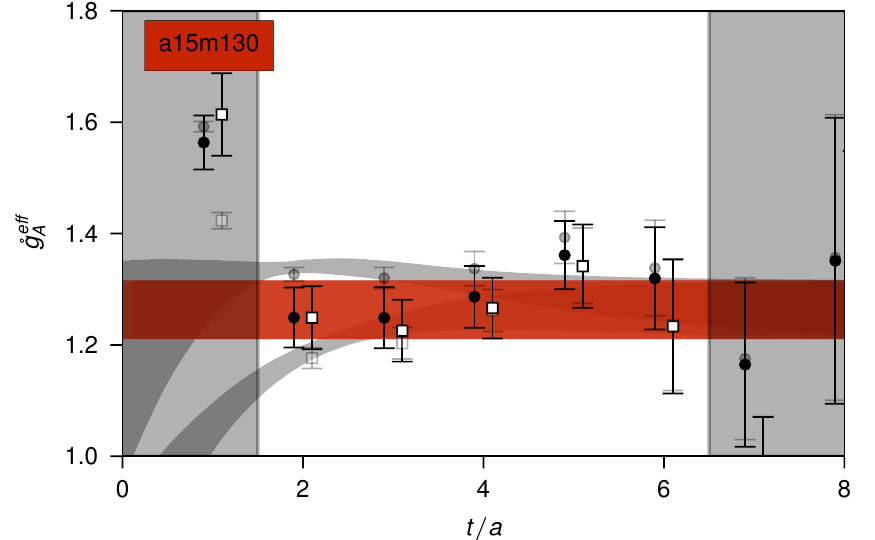}
\includegraphics[width=0.49\textwidth]{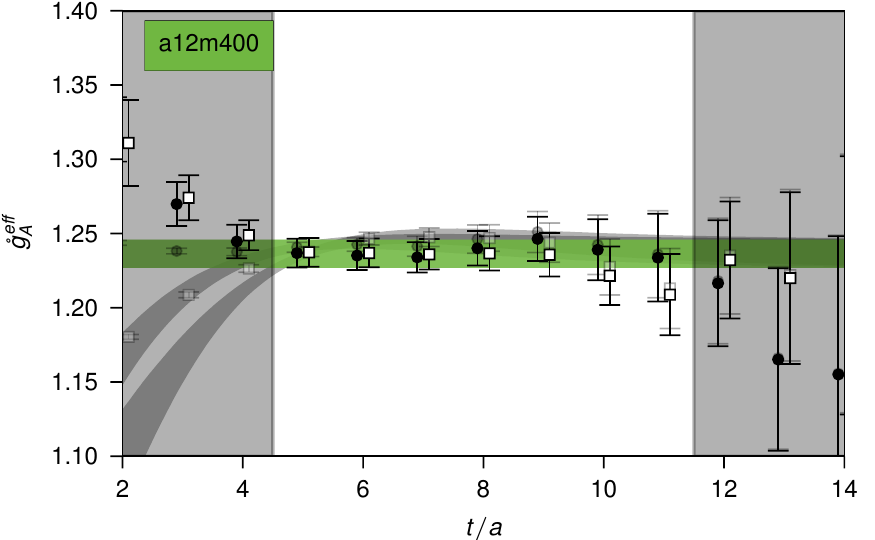}
\includegraphics[width=0.49\textwidth]{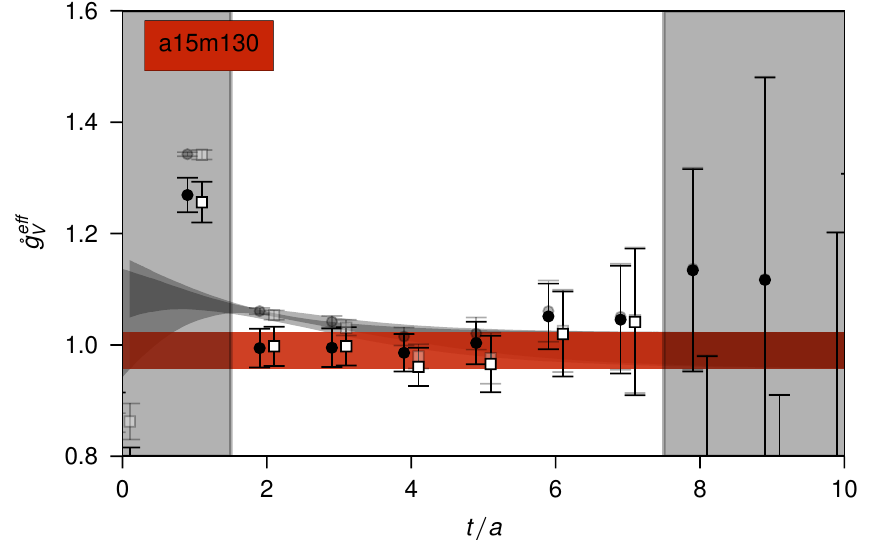}
\includegraphics[width=0.49\textwidth]{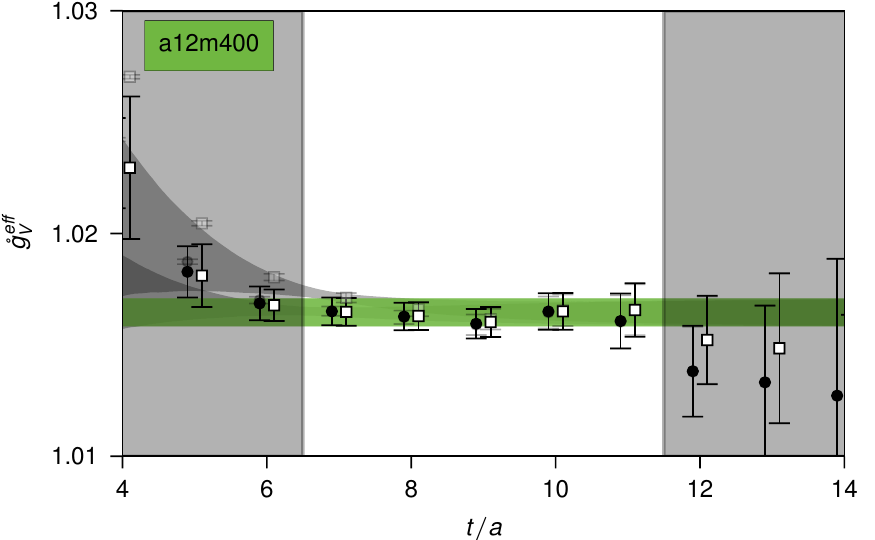}
\includegraphics[width=0.49\textwidth]{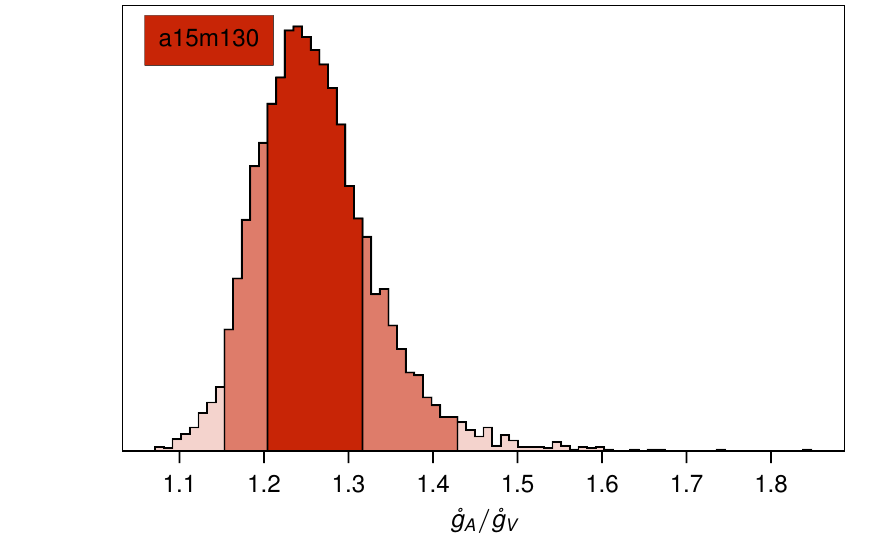}
\includegraphics[width=0.49\textwidth]{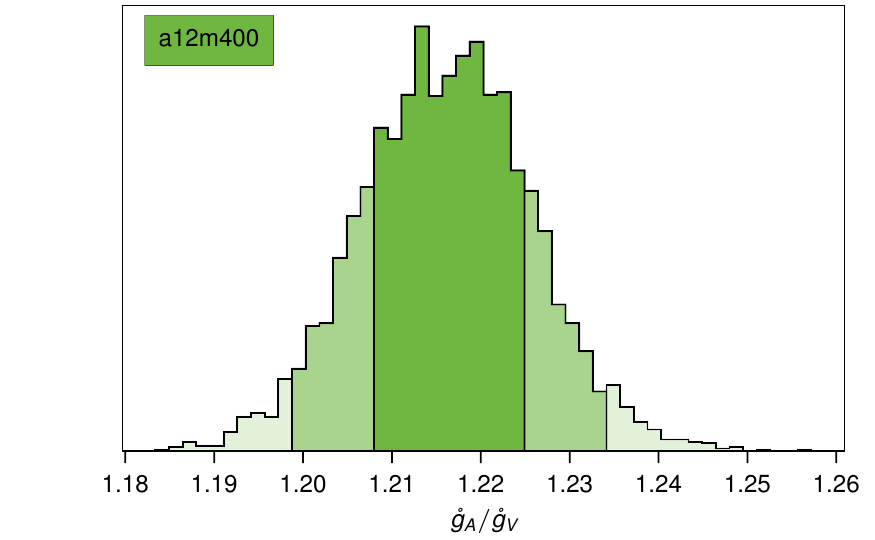}
\caption{ \label{fig:a15m130a12m400_curve}
{\textbf{Correlator fit study III.}} Analogous to Extended Data Fig.~\ref{fig:correlator_fitcurves}a,  b, c and d for the a15m130 and a12m400 ensembles.
Uncertainties are one s.e.m.}
}\end{figure*}

\begin{figure*}[h]{\docfont
\includegraphics[width=0.49\textwidth]{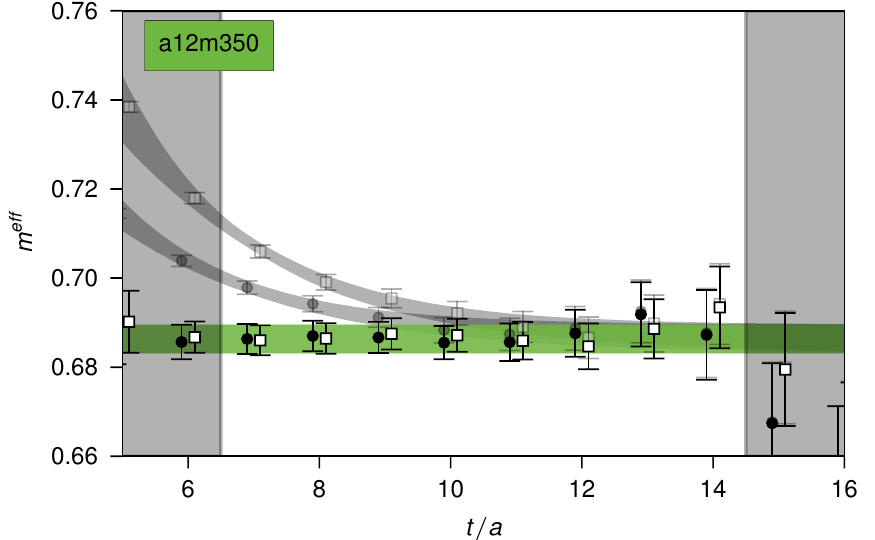}
\includegraphics[width=0.49\textwidth]{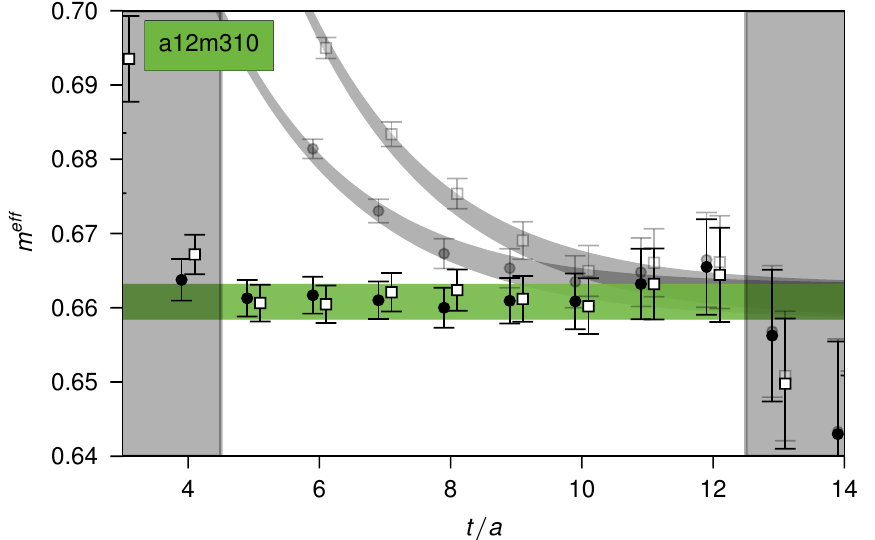}
\includegraphics[width=0.49\textwidth]{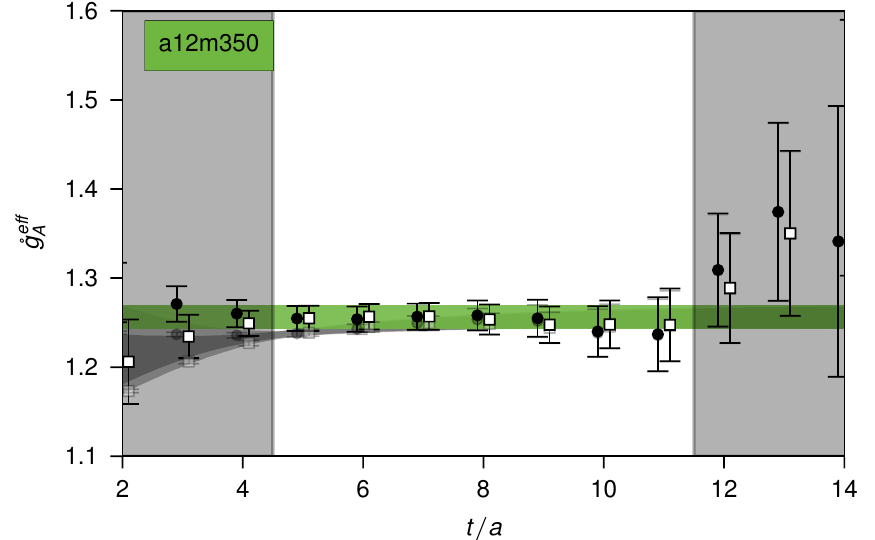}
\includegraphics[width=0.49\textwidth]{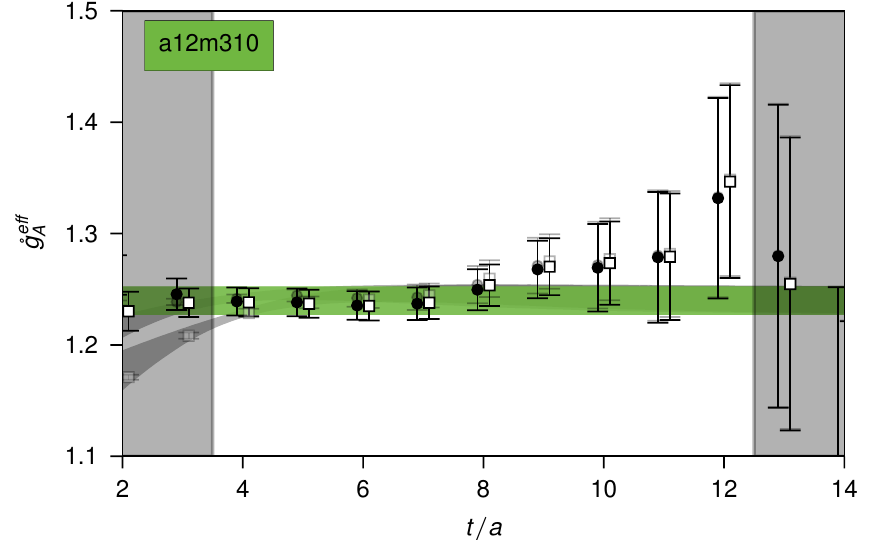}
\includegraphics[width=0.49\textwidth]{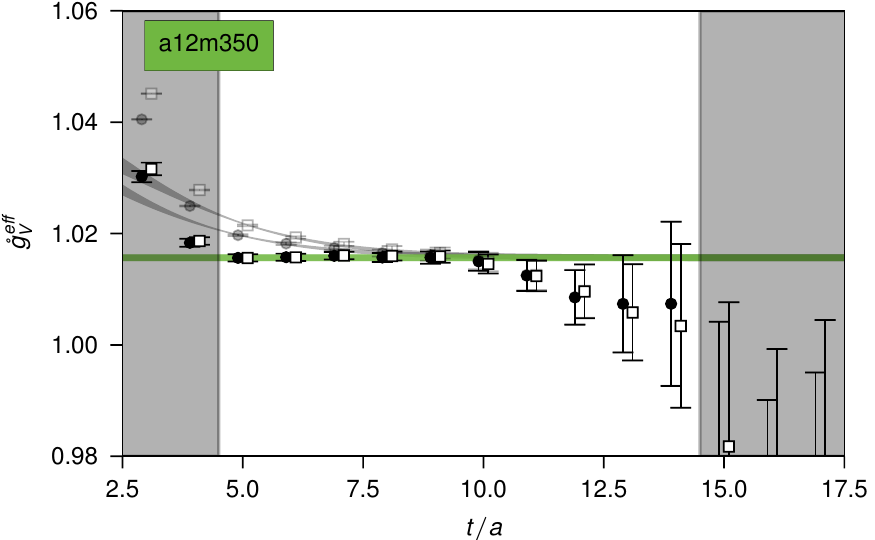}
\includegraphics[width=0.49\textwidth]{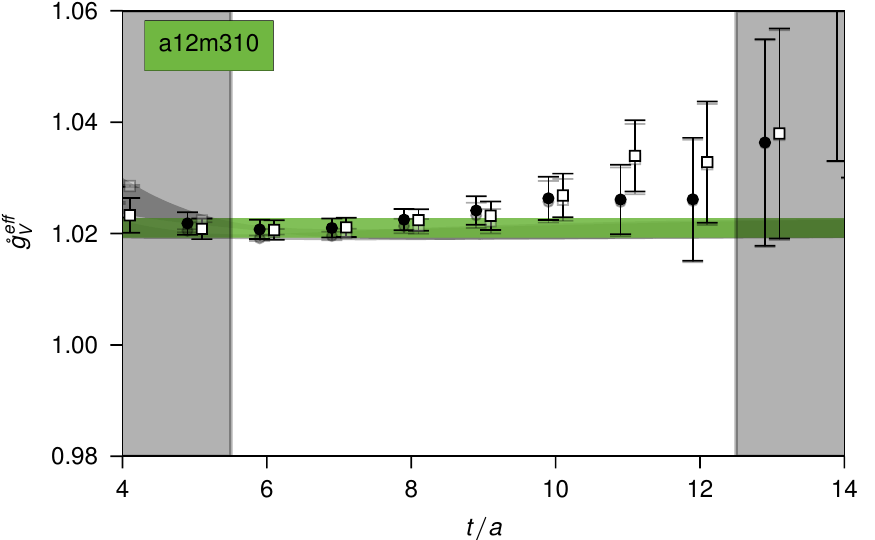}
\includegraphics[width=0.49\textwidth]{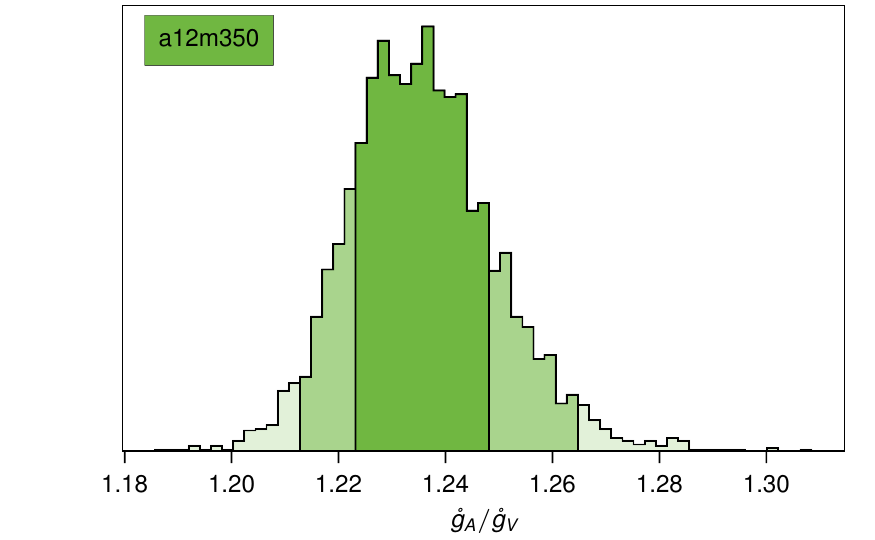}
\includegraphics[width=0.49\textwidth]{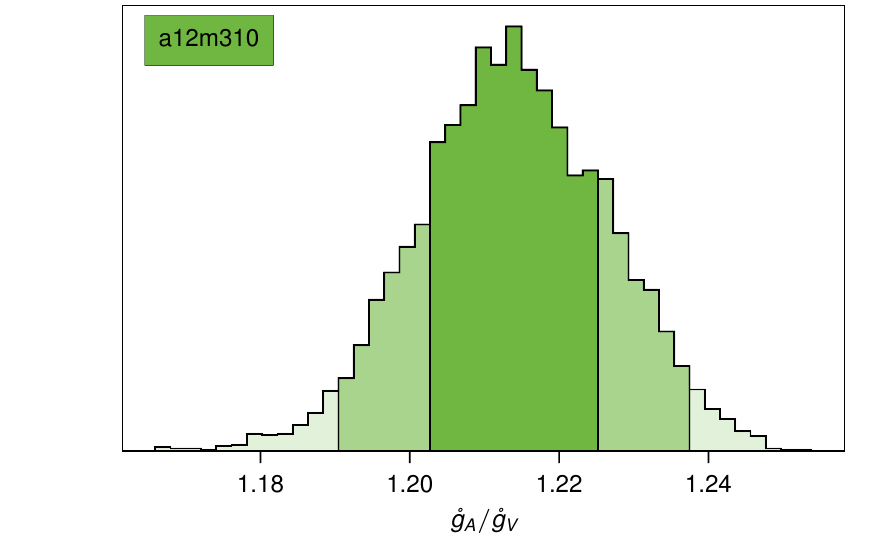}
\caption{\label{fig:a12m350a12m310_curve}
{\textbf{Correlator fit study IV.}} Analogous to Extended Data Fig.~\ref{fig:correlator_fitcurves}a,  b, c and d for the a12m350 and a12m310 ensembles.
Uncertainties are one s.e.m.}
}\end{figure*}

\begin{figure*}[h]{\docfont
\includegraphics[width=0.49\textwidth]{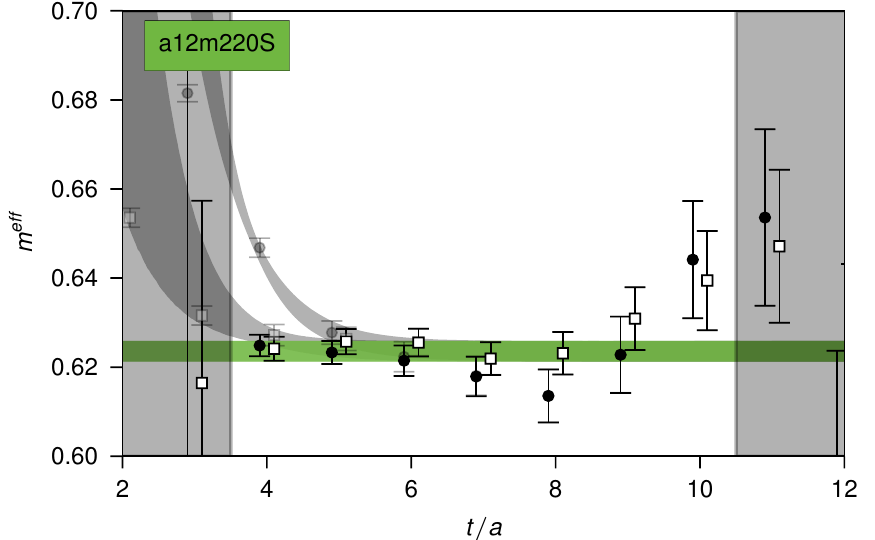}
\includegraphics[width=0.49\textwidth]{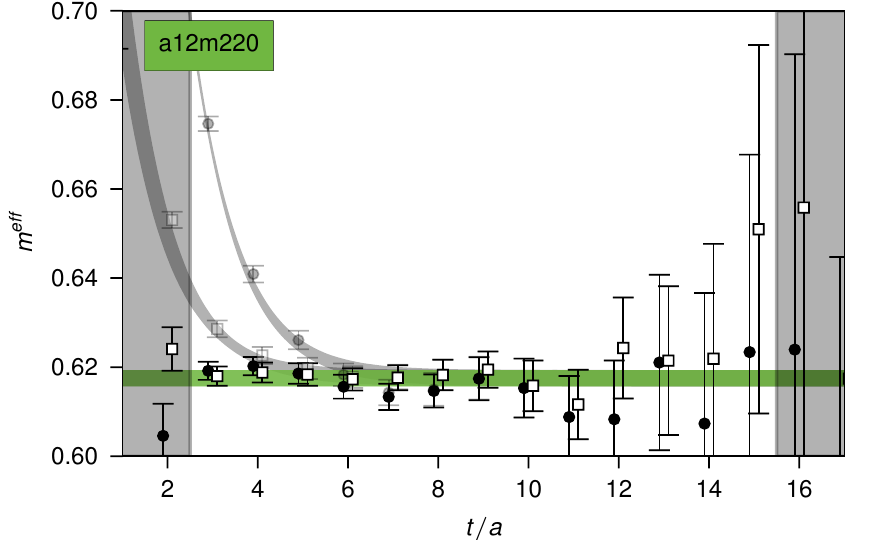}
\includegraphics[width=0.49\textwidth]{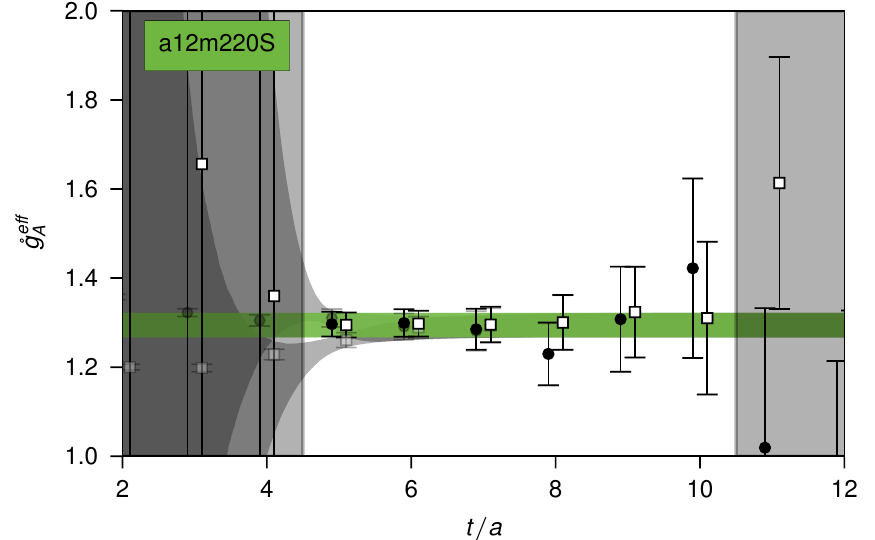}
\includegraphics[width=0.49\textwidth]{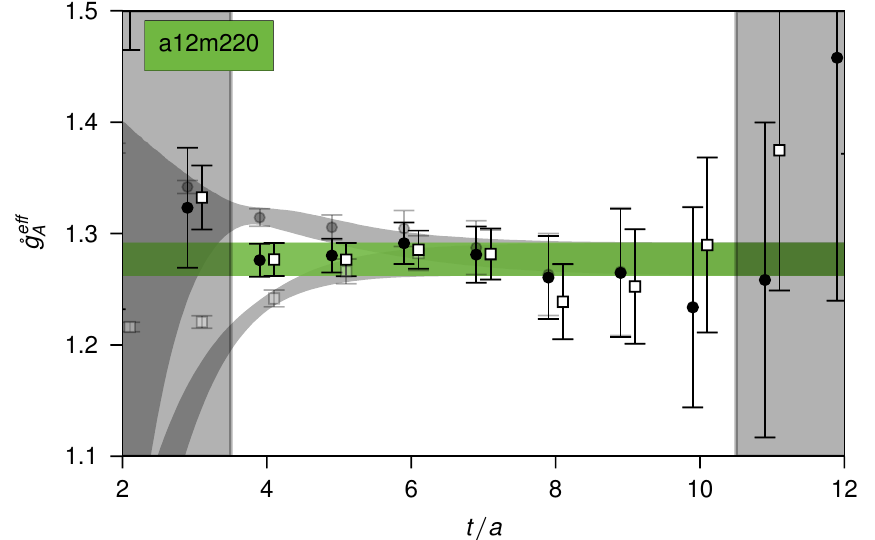}
\includegraphics[width=0.49\textwidth]{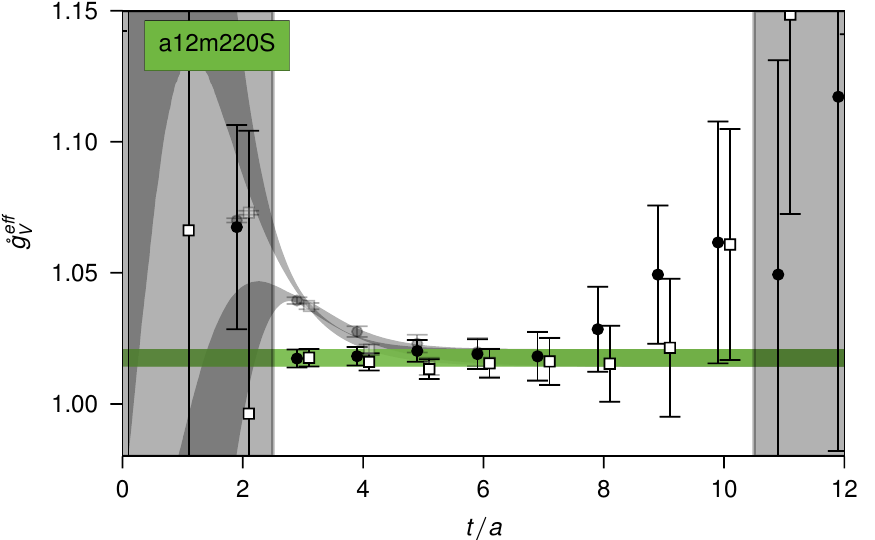}
\includegraphics[width=0.49\textwidth]{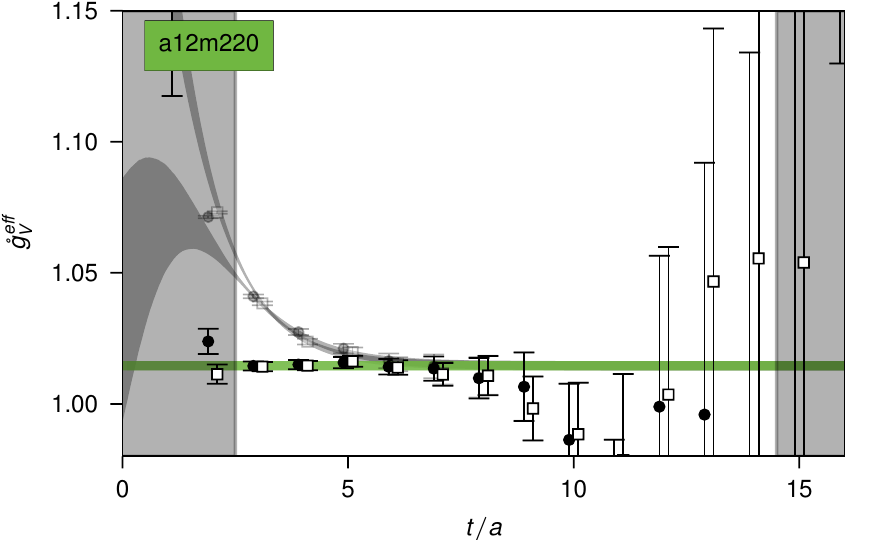}
\includegraphics[width=0.49\textwidth]{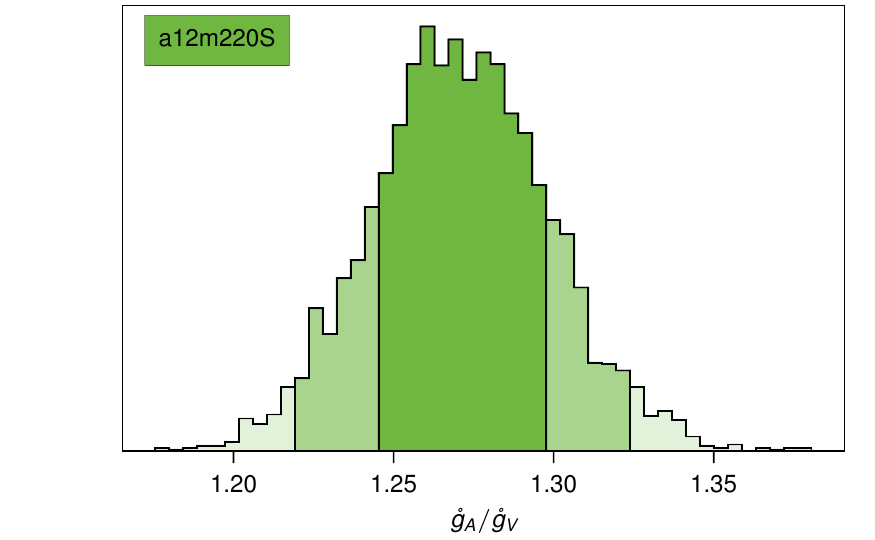}
\includegraphics[width=0.49\textwidth]{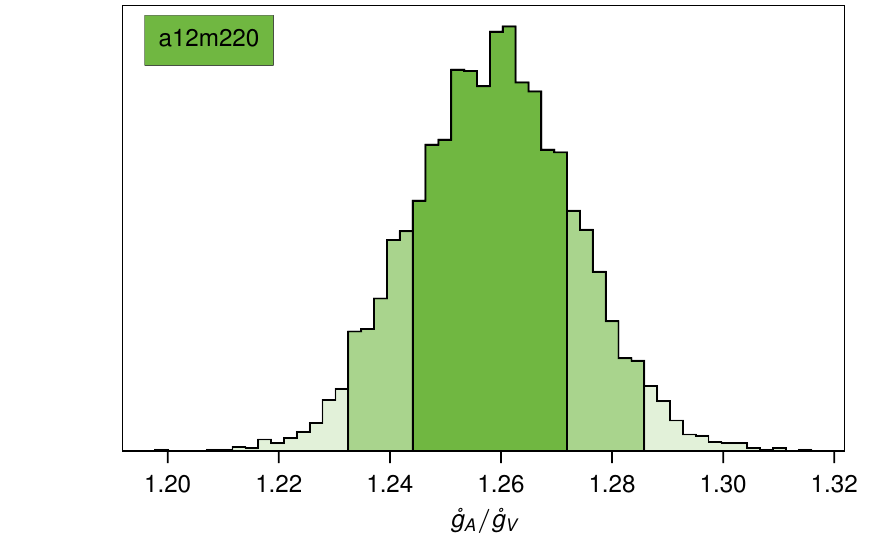}
\caption{\label{fig:a12m220a12m220S_curve}
{\textbf{Correlator fit study V.}} Analogous to Extended Data Fig.~\ref{fig:correlator_fitcurves}a,  b, c and d for the a12m220 and a12m220S ensembles.
Uncertainties are one s.e.m.}
}\end{figure*}

\begin{figure*}[h]{\docfont
\includegraphics[width=0.49\textwidth]{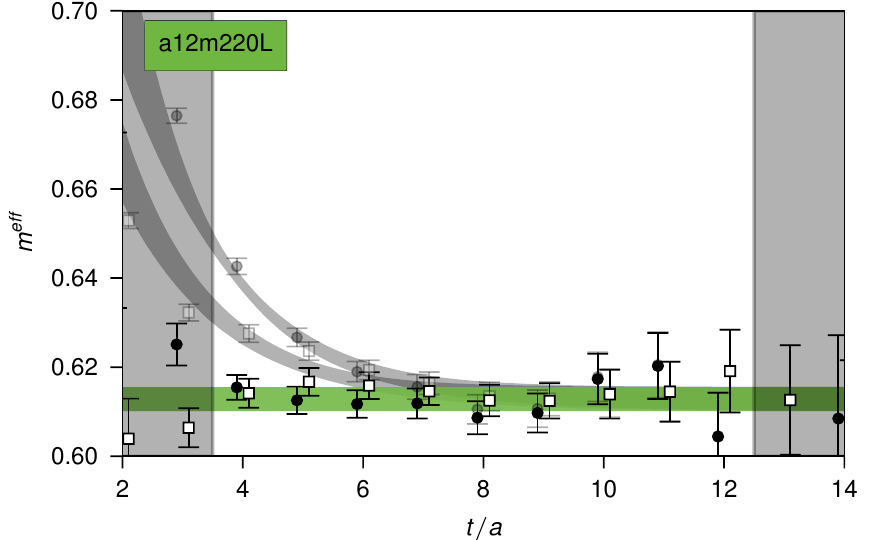}
\includegraphics[width=0.49\textwidth]{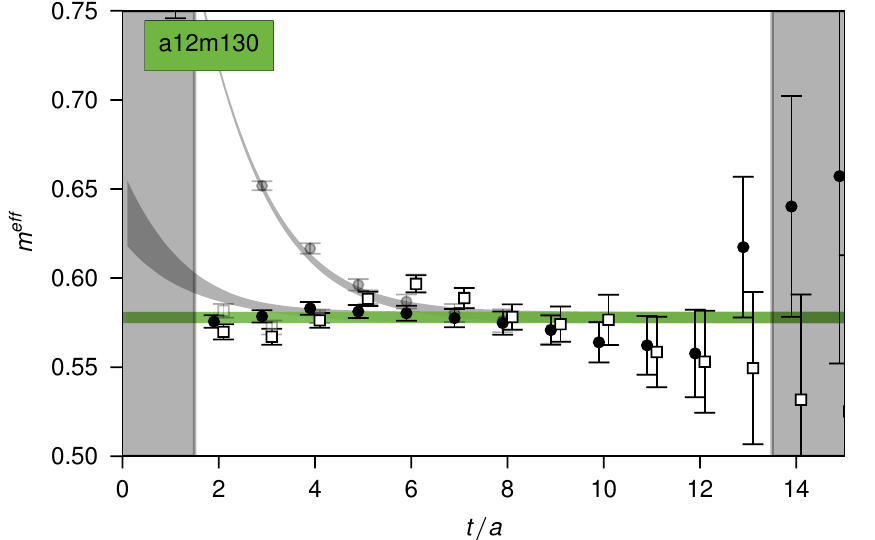}
\includegraphics[width=0.49\textwidth]{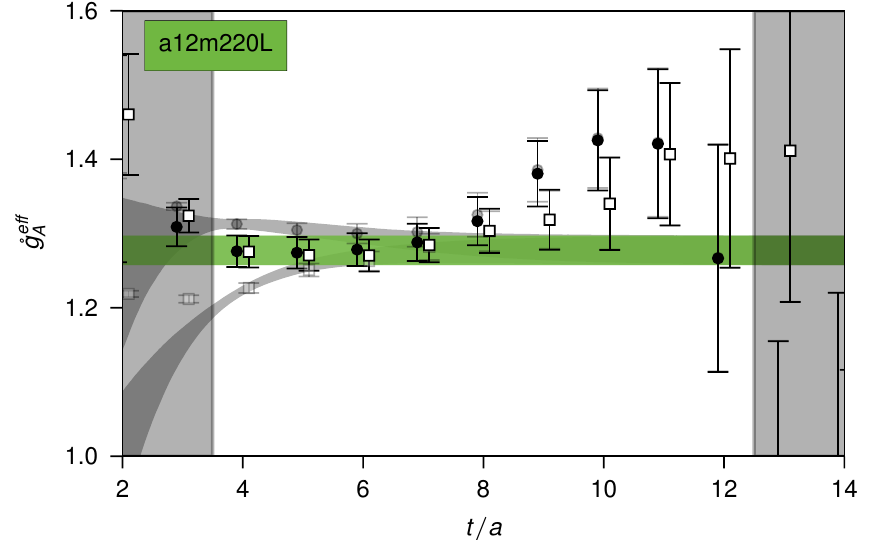}
\includegraphics[width=0.49\textwidth]{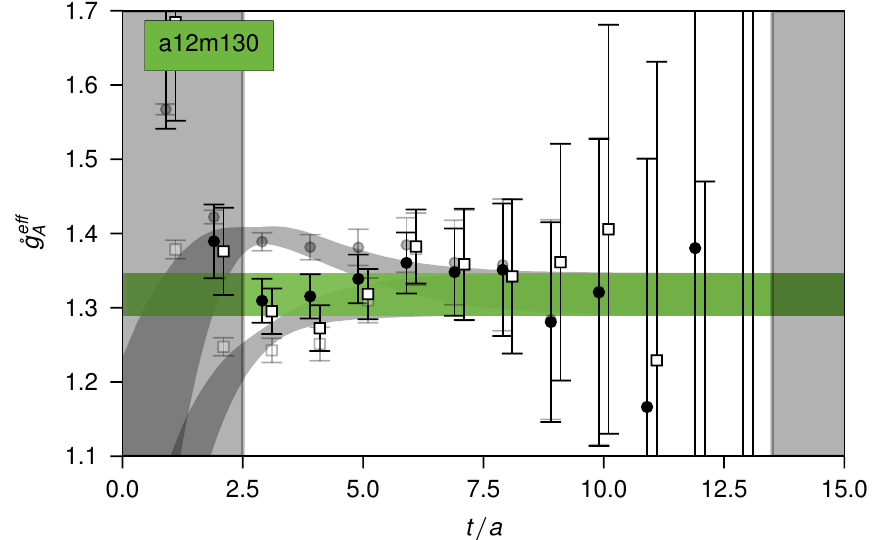}
\includegraphics[width=0.49\textwidth]{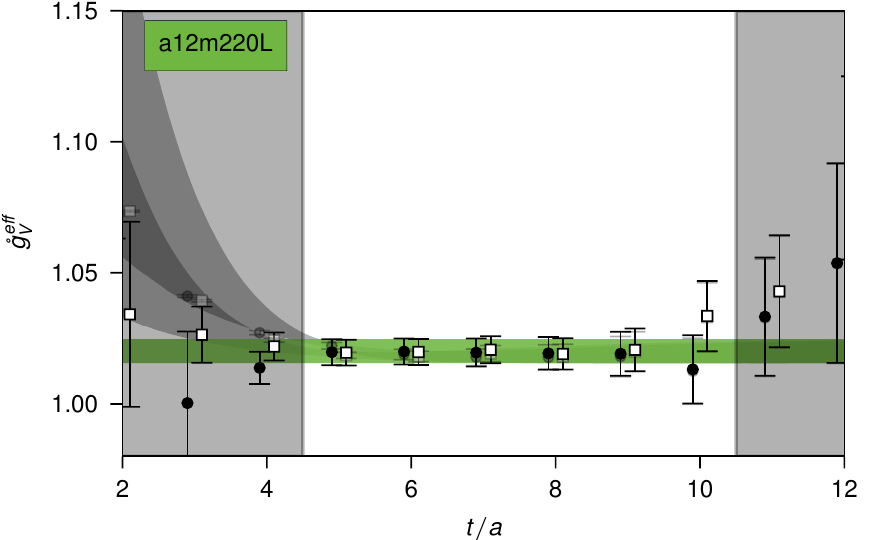}
\includegraphics[width=0.49\textwidth]{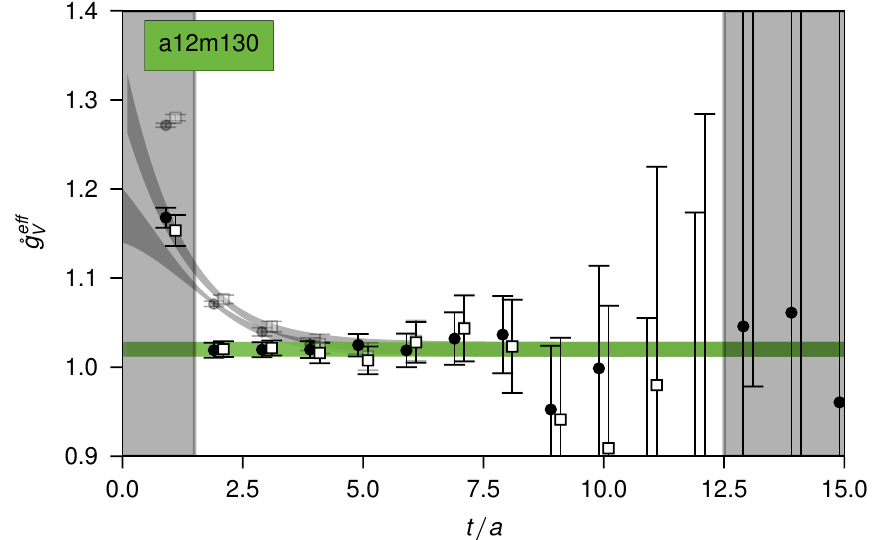}
\includegraphics[width=0.49\textwidth]{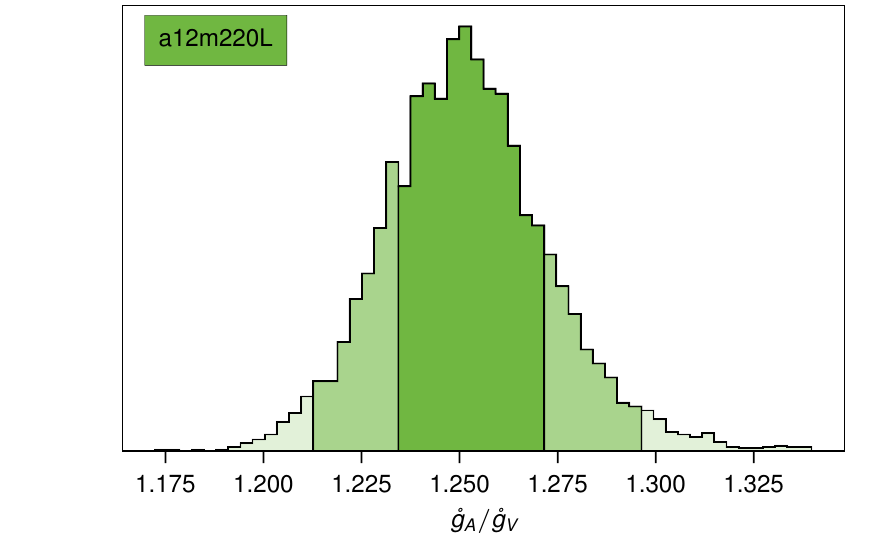}
\includegraphics[width=0.49\textwidth]{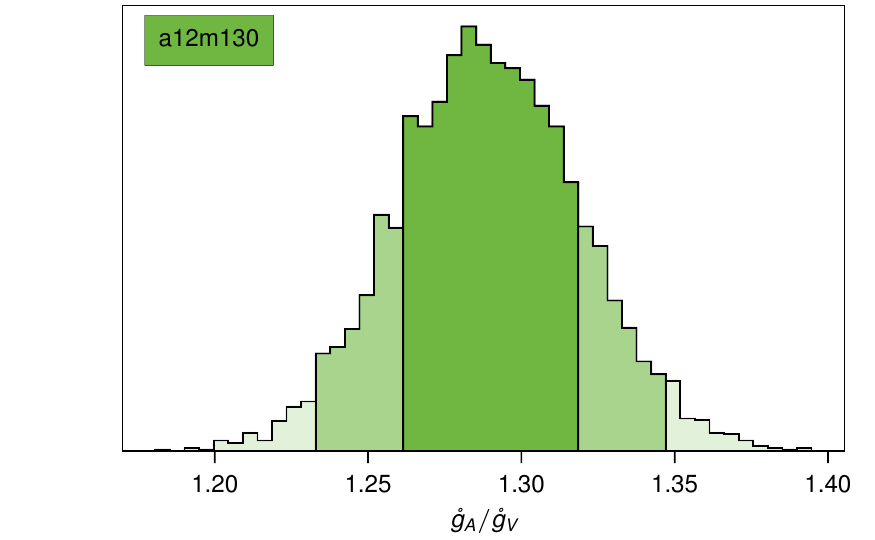}
\caption{\label{fig:a12m220La12m130_curve}
{\textbf{Correlator fit study VI.}} Analogous to Extended Data Fig.~\ref{fig:correlator_fitcurves}a,  b, c and d for the a12m220L and a12m130 ensembles.
Uncertainties are one s.e.m.}
}\end{figure*}

\begin{figure*}[h]{\docfont
\includegraphics[width=0.49\textwidth]{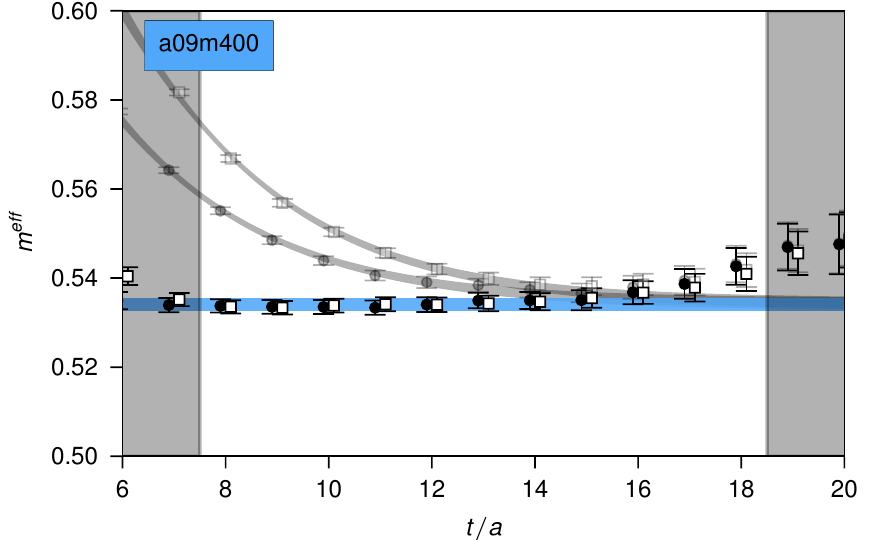}
\includegraphics[width=0.49\textwidth]{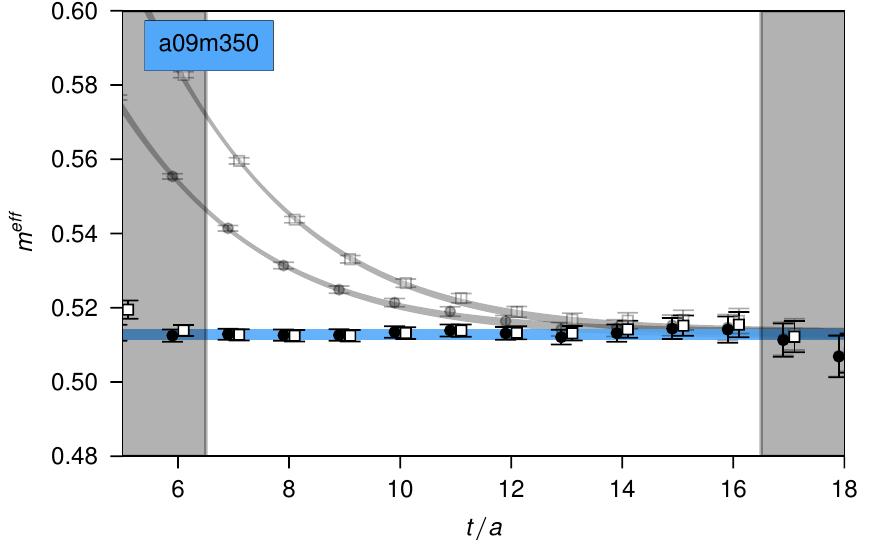}
\includegraphics[width=0.49\textwidth]{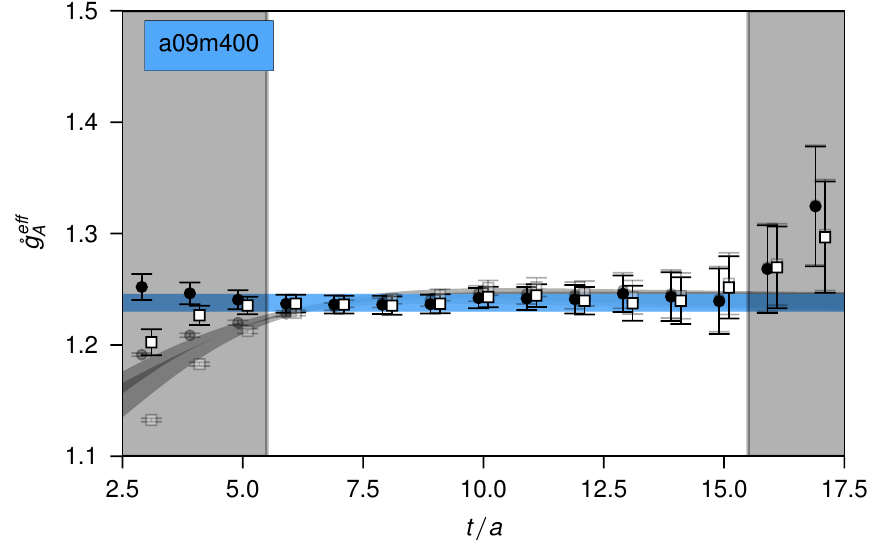}
\includegraphics[width=0.49\textwidth]{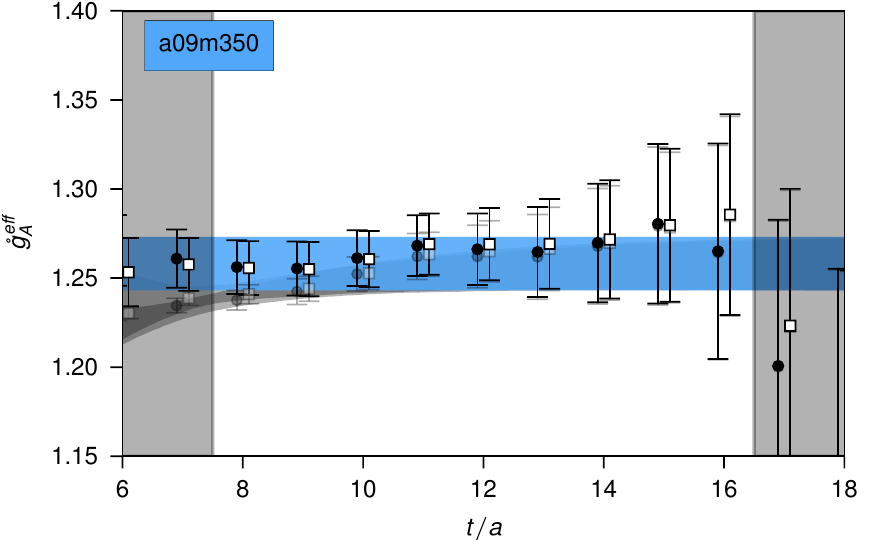}
\includegraphics[width=0.49\textwidth]{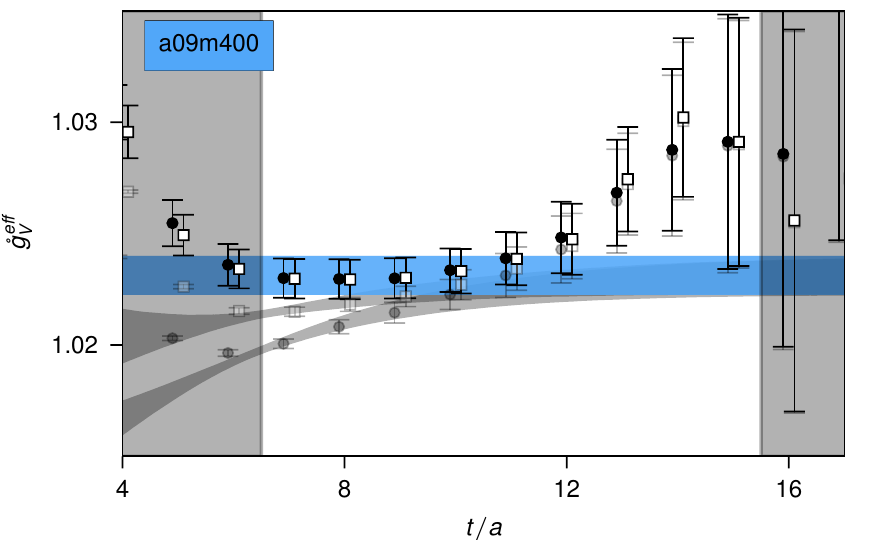}
\includegraphics[width=0.49\textwidth]{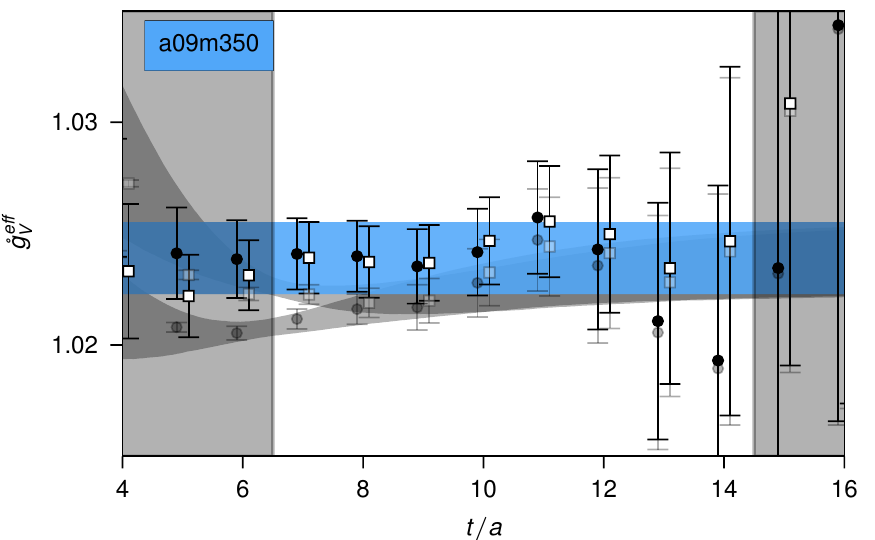}
\includegraphics[width=0.49\textwidth]{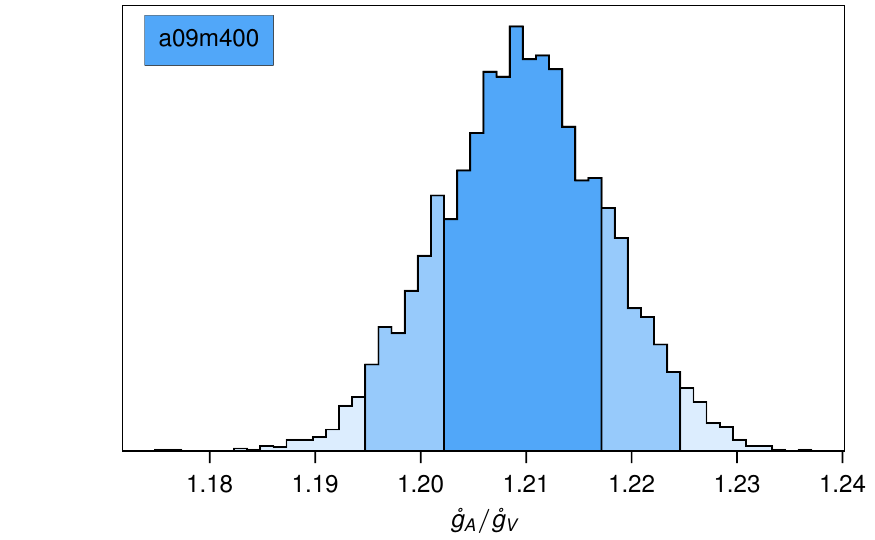}
\includegraphics[width=0.49\textwidth]{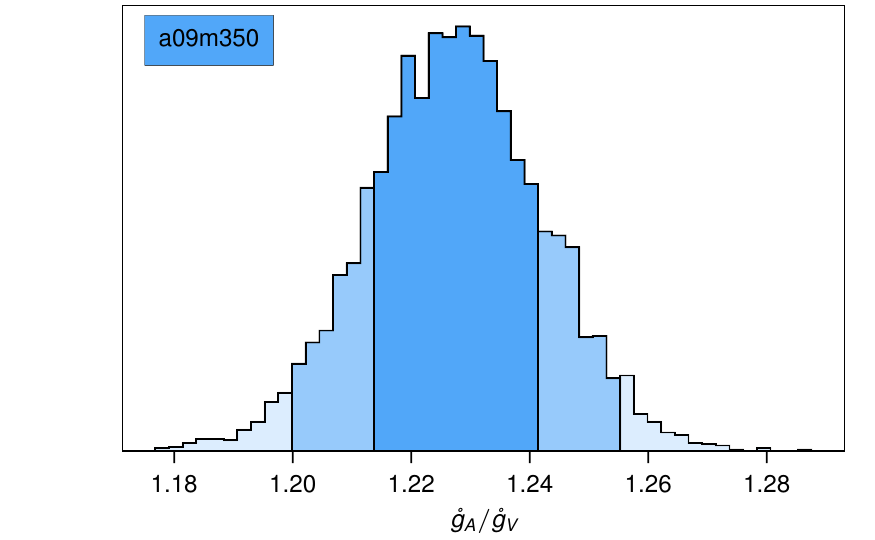}
\caption{\label{fig:a09m400a09m350_curve}
{\textbf{Correlator fit study VII.}} Analogous to Extended Data Fig.~\ref{fig:correlator_fitcurves}a,  b, c and d for the a09m400 and a09m350 ensembles.
Uncertainties are one s.e.m.}
}\end{figure*}

\begin{figure*}[h]{\docfont
\includegraphics[width=0.49\textwidth]{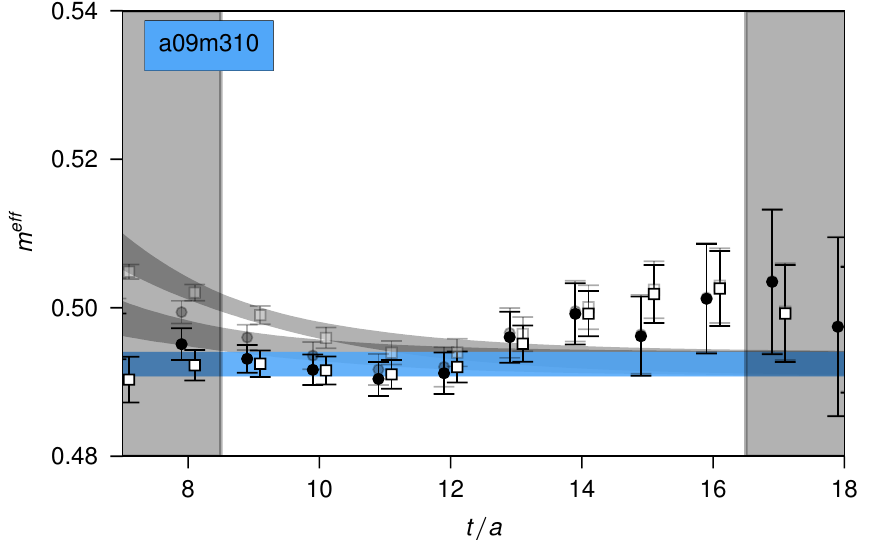}
\includegraphics[width=0.49\textwidth]{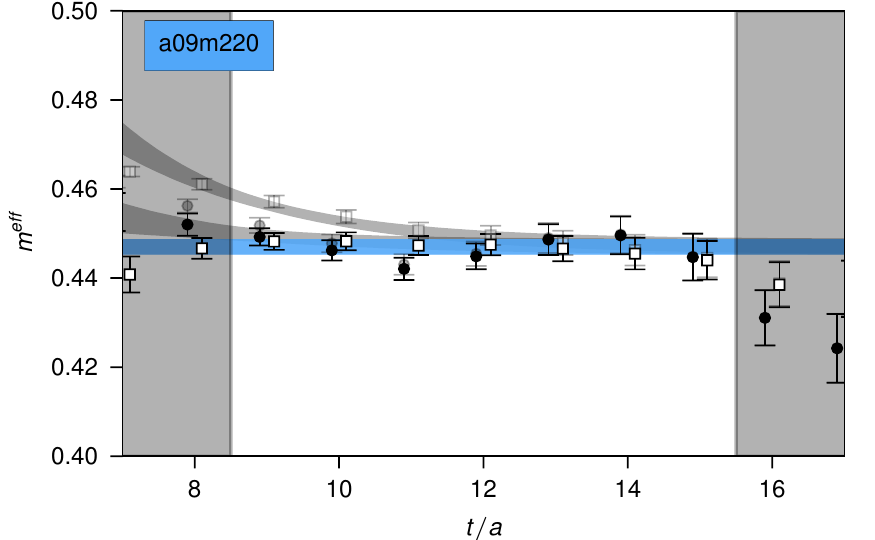}
\includegraphics[width=0.49\textwidth]{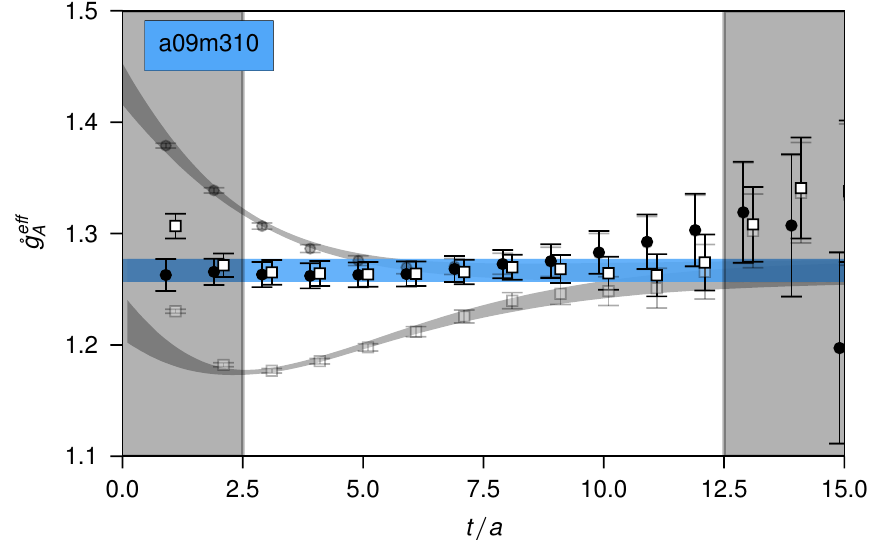}
\includegraphics[width=0.49\textwidth]{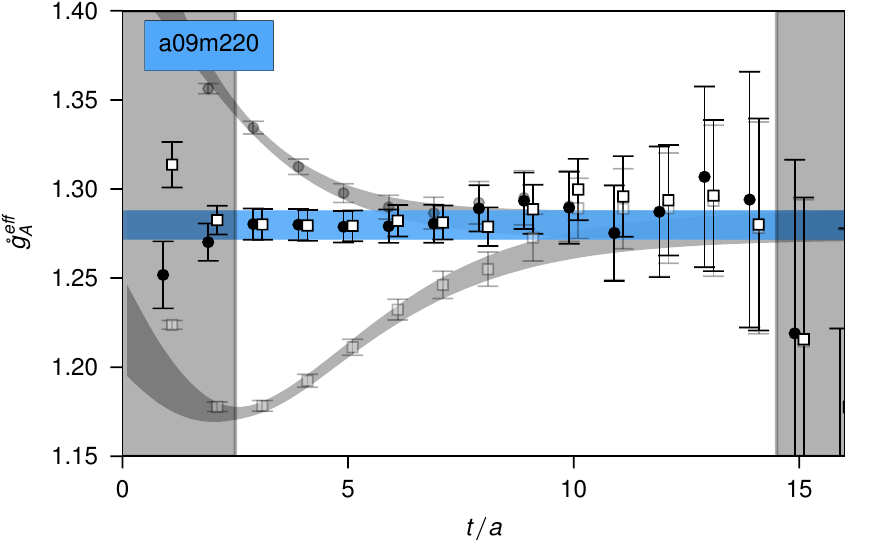}
\includegraphics[width=0.49\textwidth]{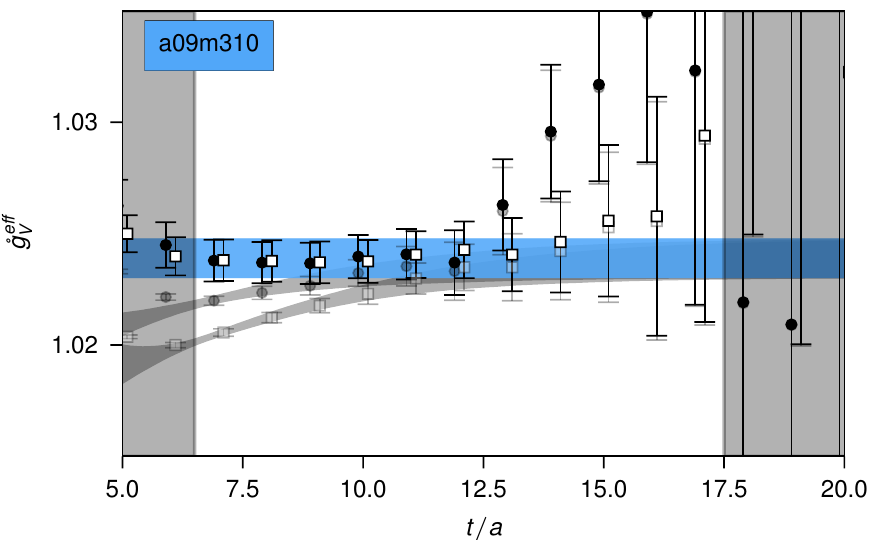}
\includegraphics[width=0.49\textwidth]{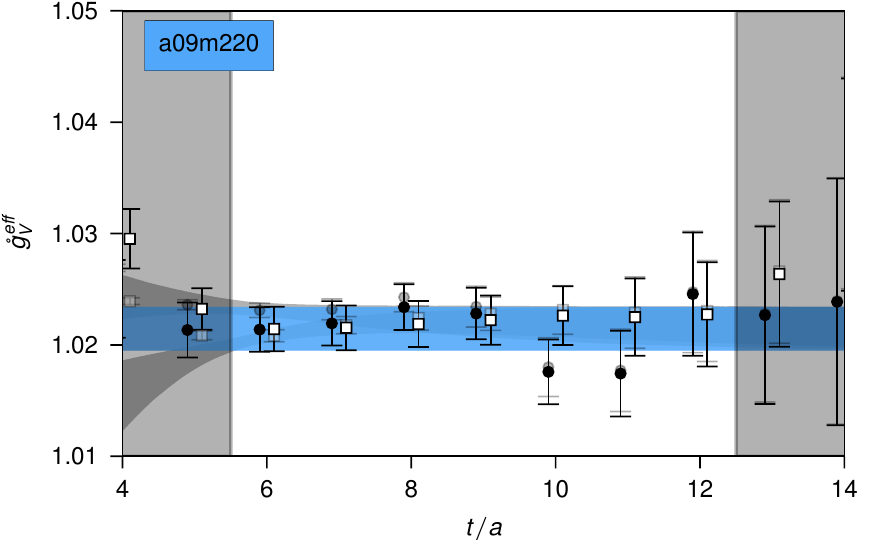}
\includegraphics[width=0.49\textwidth]{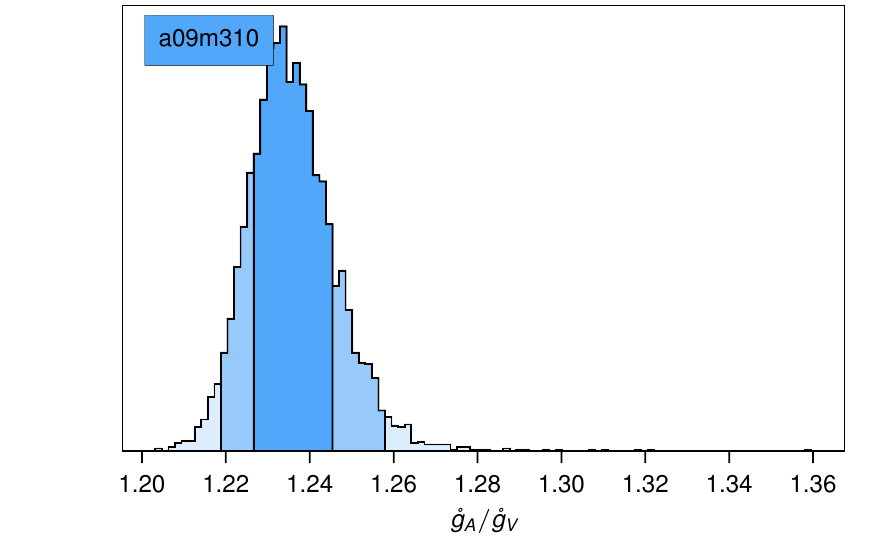}
\includegraphics[width=0.49\textwidth]{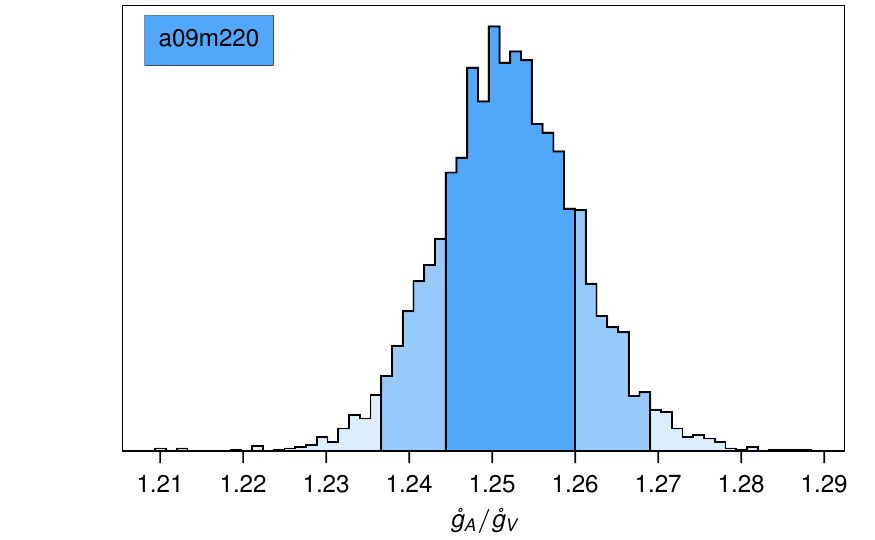}
\caption{\label{fig:a09m310a09m220_curve}
{\textbf{Correlator fit study VIII.}} Analogous to Extended Data Fig.~\ref{fig:correlator_fitcurves}a,  b, c and d for the a09m310 and a09m220 ensembles.
Uncertainties are one s.e.m.}
}\end{figure*}

\begin{figure*}[h]{\docfont
\includegraphics[width=0.49\textwidth]{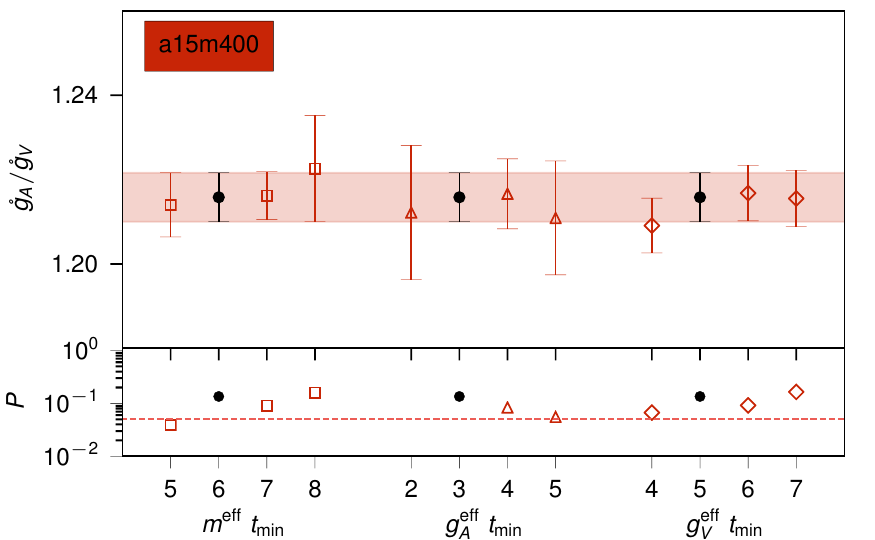}
\includegraphics[width=0.49\textwidth]{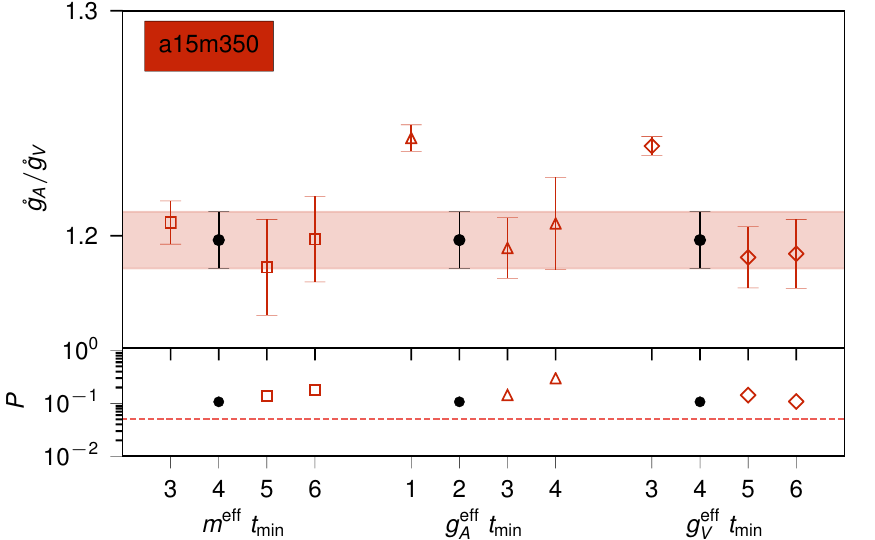}
\includegraphics[width=0.49\textwidth]{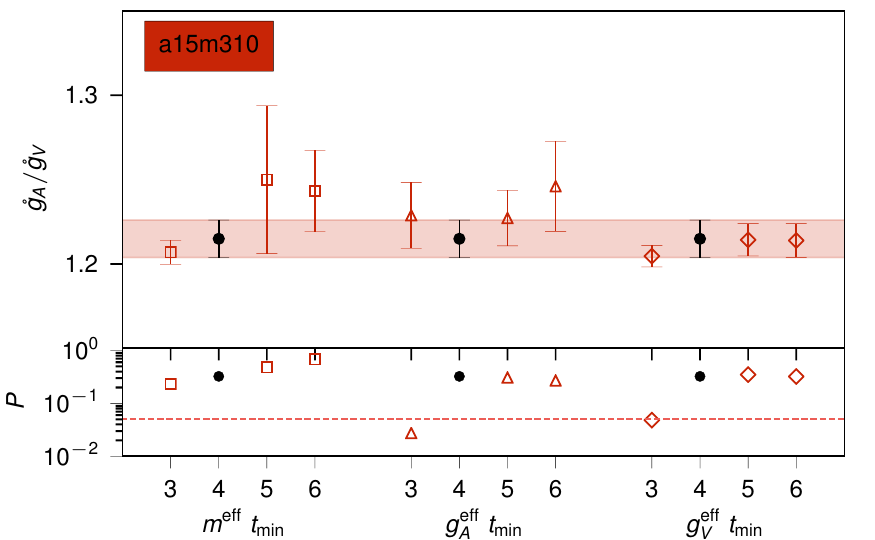}
\includegraphics[width=0.49\textwidth]{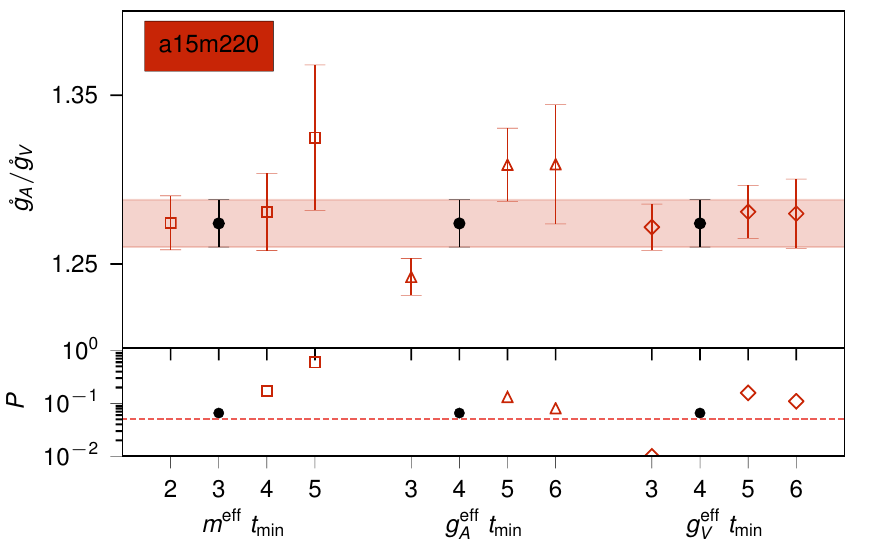}
\includegraphics[width=0.49\textwidth]{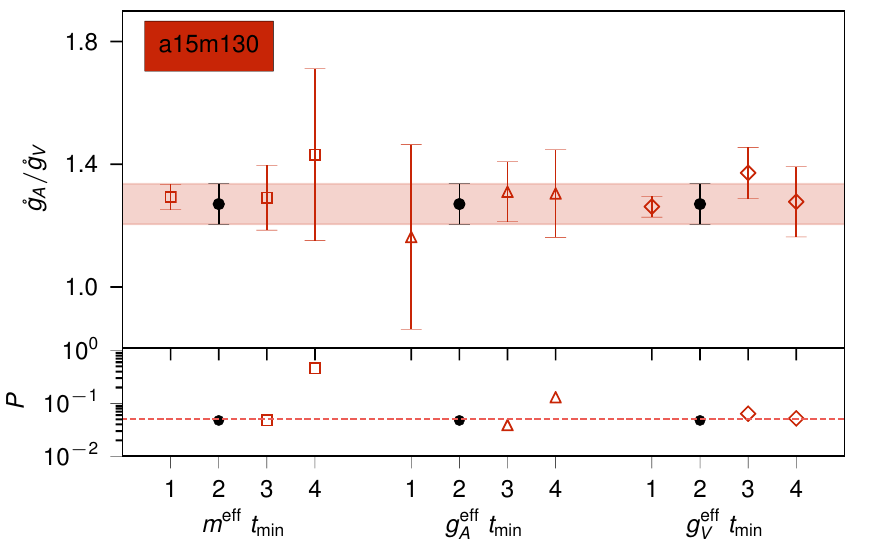}
\caption{\label{fig:stability_I}
{\textbf{Correlator fit $t_{\textrm{min}}$ stability study I.}} Analogous to Extended Data Fig.~\ref{fig:correlator_fitcurves}e. Solid circles accompanied by shaded bands are the preferred simultaneous fits. Varying fit regions for the two-point correlator ($\medsquare$), and axial ($\medtriangleup$), and vector ($\meddiamond$) effective derivatives are presented. Corresponding $P$-values are presented, with the dashed red line at $p=0.05$ discriminating statistical significance of the fit results. Uncertainties are one s.e.m.}
}\end{figure*}

\begin{figure*}[h]{\docfont
\includegraphics[width=0.49\textwidth]{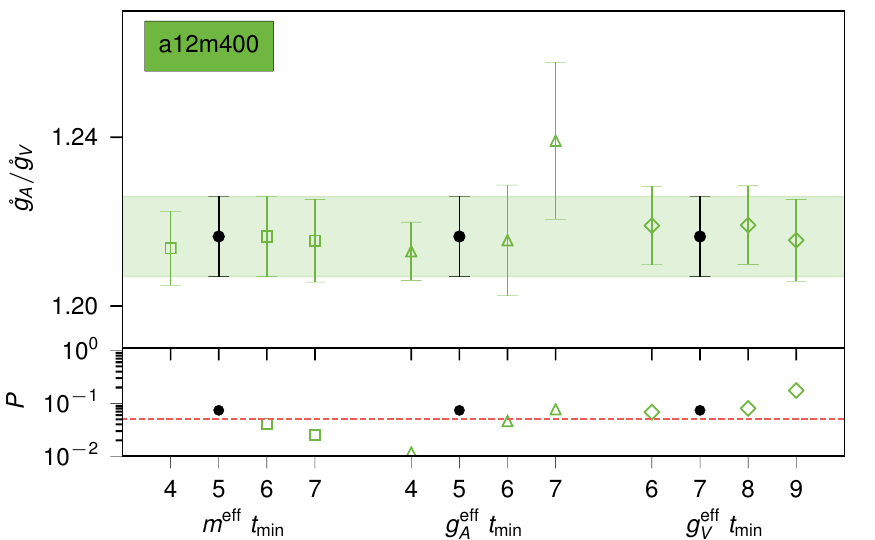}
\includegraphics[width=0.49\textwidth]{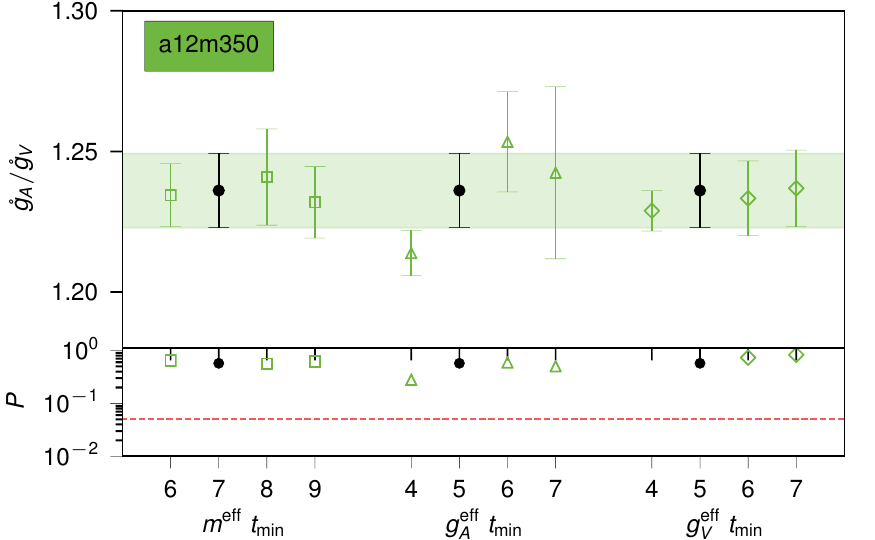}
\includegraphics[width=0.49\textwidth]{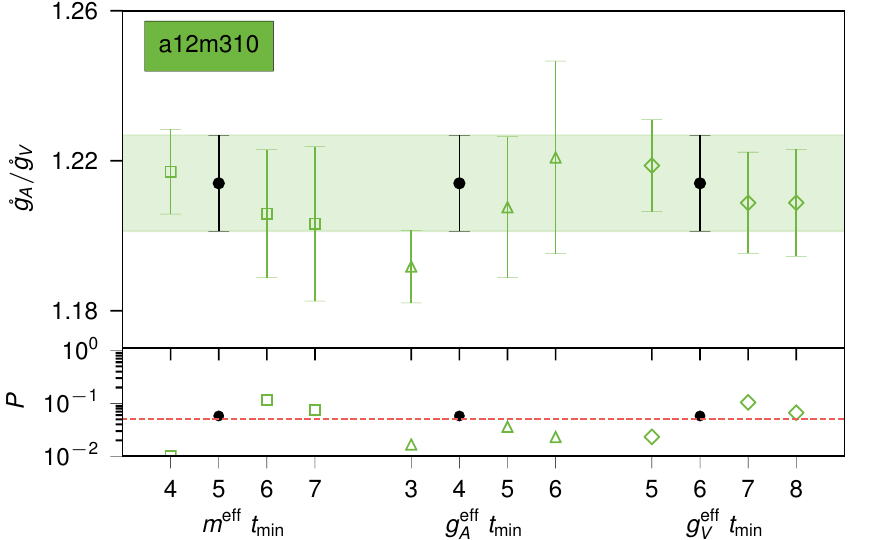}
\includegraphics[width=0.49\textwidth]{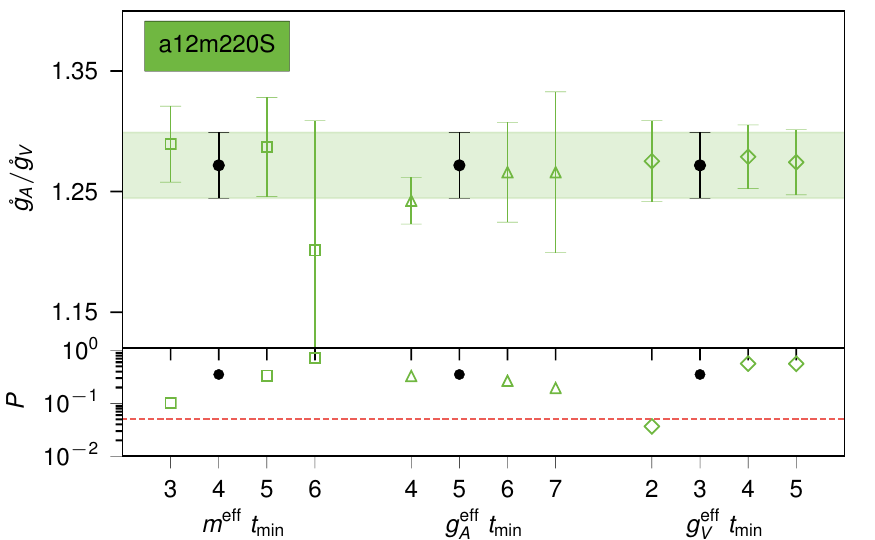}
\includegraphics[width=0.49\textwidth]{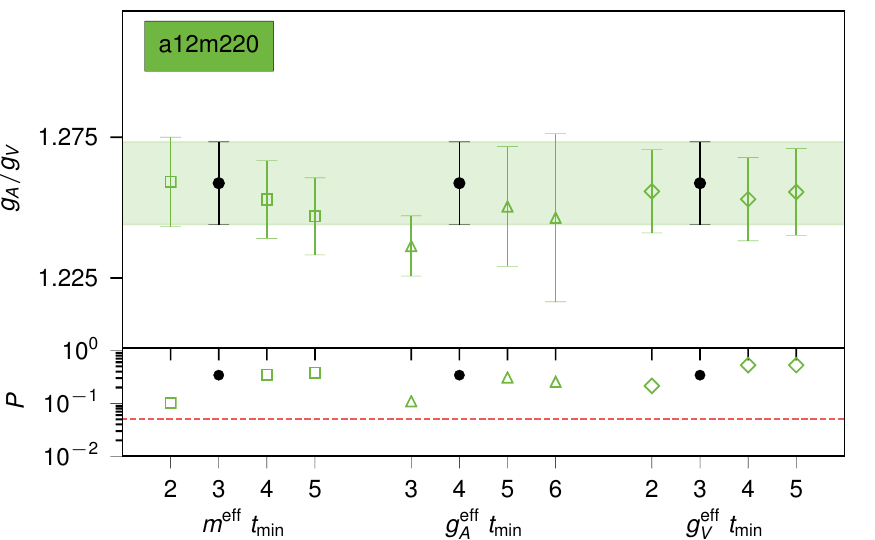}
\includegraphics[width=0.49\textwidth]{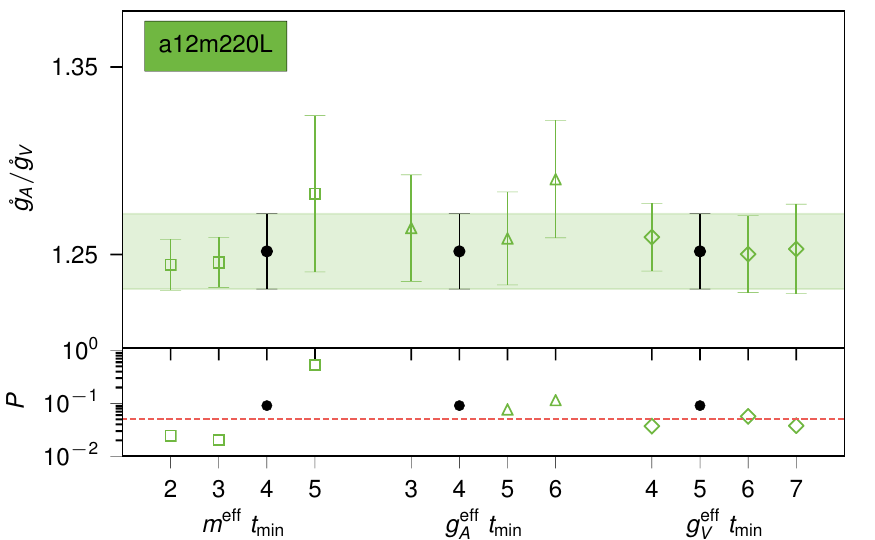}
\includegraphics[width=0.49\textwidth]{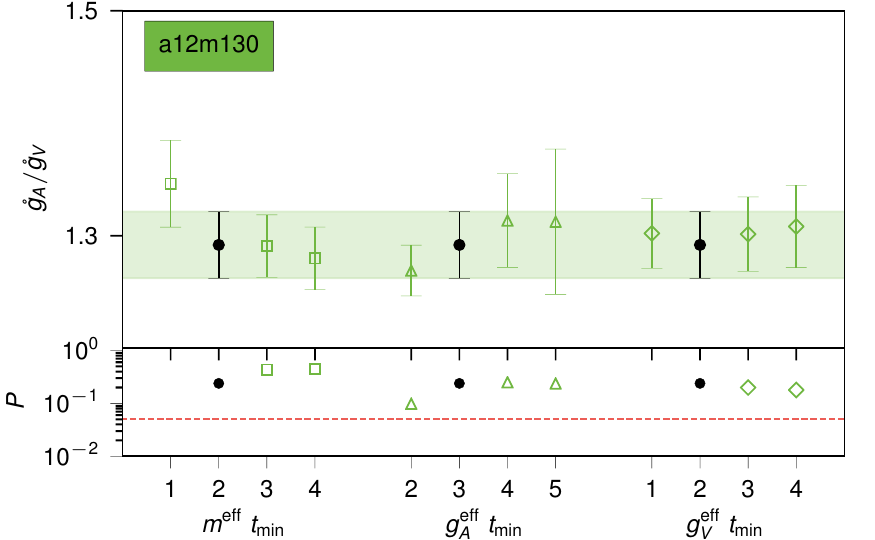}
\caption{\label{fig:stability_II}
{\textbf{Correlator fit $t_{\textrm{min}}$ stability study II.}} Analogous to Extended Data Fig.~\ref{fig:correlator_fitcurves}e and Supplemental Fig.~\ref{fig:stability_I} for the remaining ensembles.
Uncertainties are one s.e.m.}
}\end{figure*}

\begin{figure*}[h]{\docfont
\includegraphics[width=0.49\textwidth]{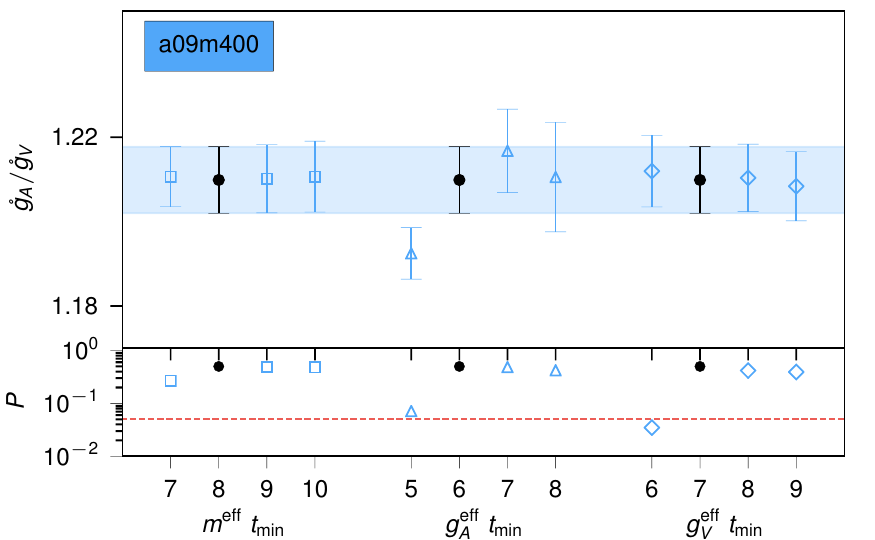}
\includegraphics[width=0.49\textwidth]{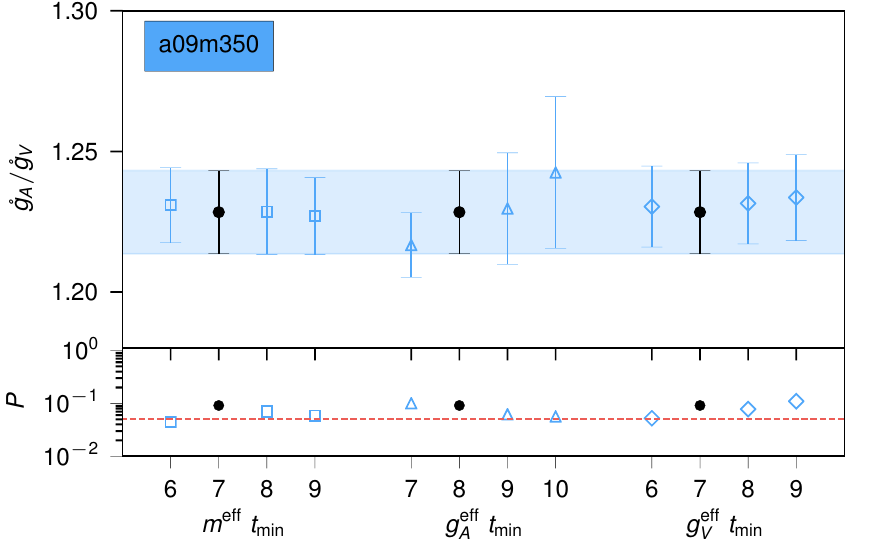}
\includegraphics[width=0.49\textwidth]{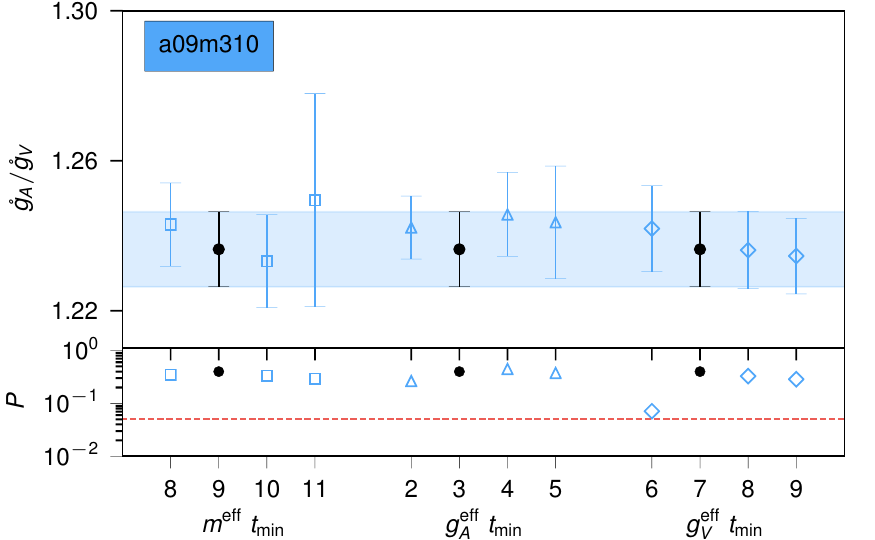}
\includegraphics[width=0.49\textwidth]{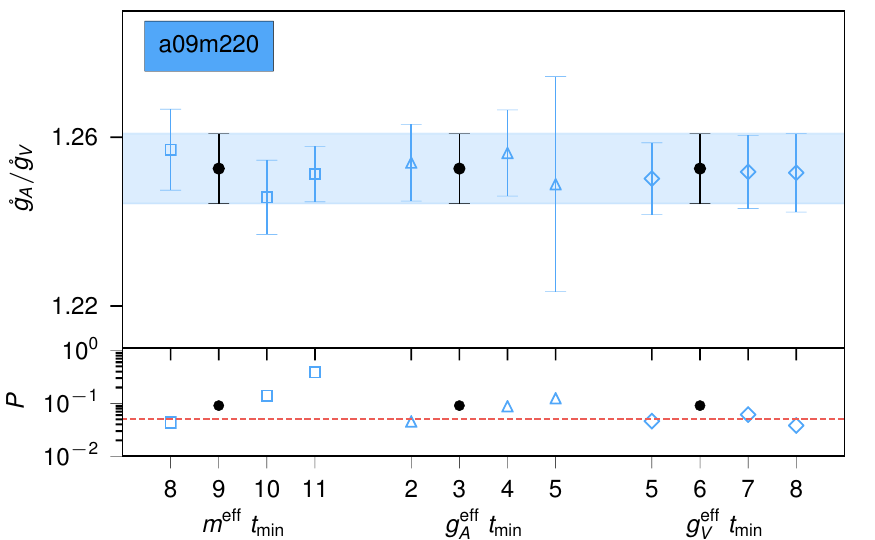}
\caption{\label{fig:stability_III}
{\textbf{Correlator fit $t_{\textrm{min}}$ stability study III.}} Analogous to Extended Data Fig.~\ref{fig:correlator_fitcurves}e and Supplemental Fig.~\ref{fig:stability_I} for the remaining ensembles.
Uncertainties are one s.e.m.}
}\end{figure*}

\begin{figure*}[h]{\docfont
	\includegraphics[width=0.49\textwidth]{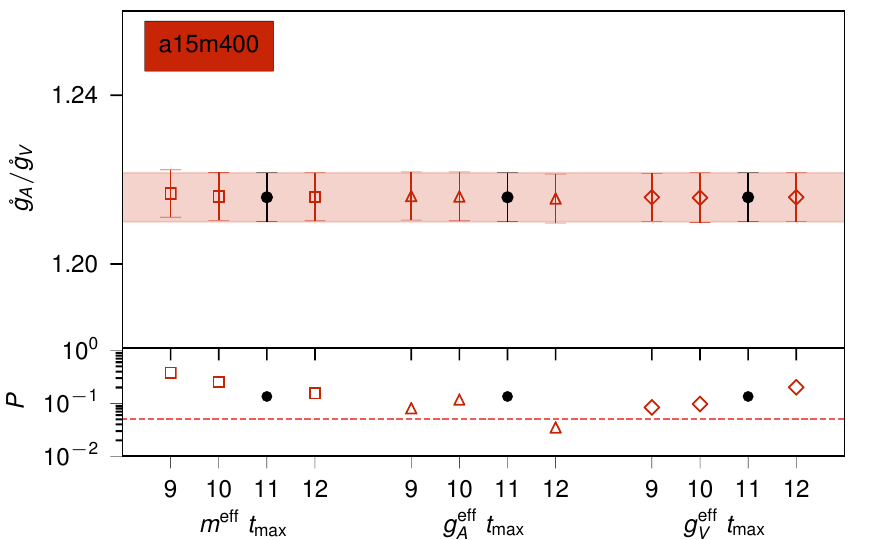}
	\includegraphics[width=0.49\textwidth]{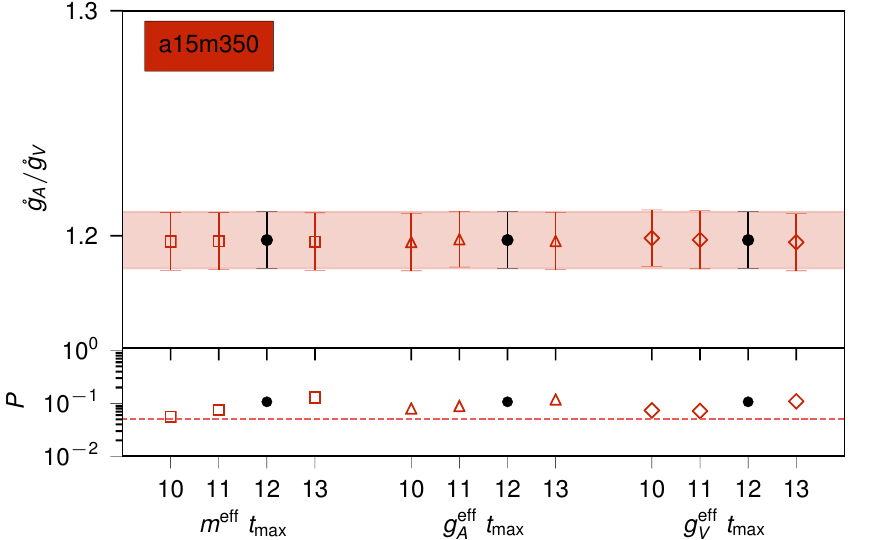}
	\includegraphics[width=0.49\textwidth]{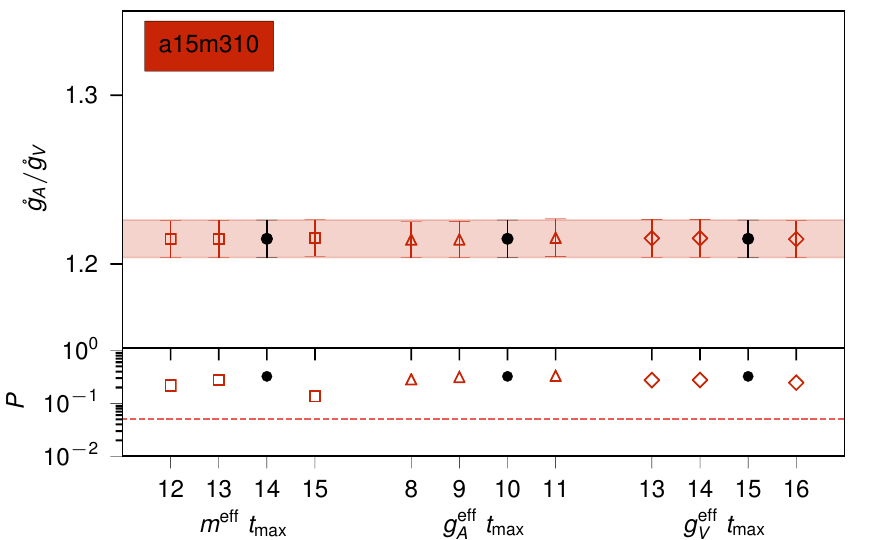}
	\includegraphics[width=0.49\textwidth]{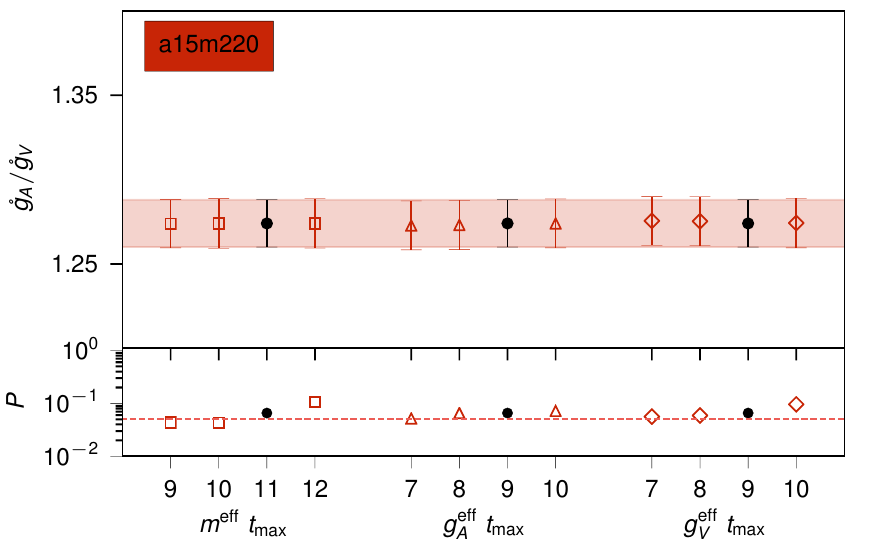}
	\includegraphics[width=0.49\textwidth]{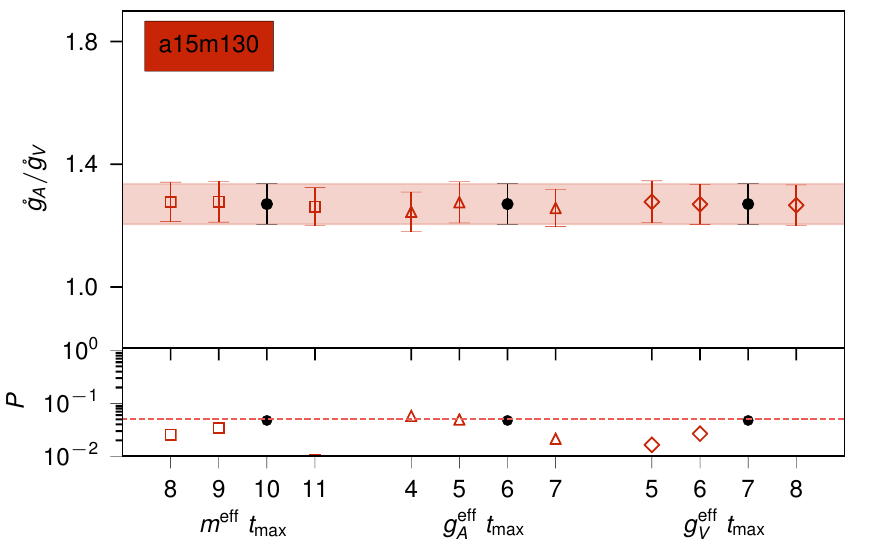}
	\caption{\label{fig:tmax_stability_I}
		{\textbf{Correlator fit $t_{\textrm{max}}$ stability study I.}} Analogous to Extended Data Fig.~\ref{fig:correlator_fitcurves}f. Solid symbols accompanied by shaded bands are the preferred simultaneous fits. Varying fit regions for the two-point correlator ($\medsquare$), and axial ($\medtriangleup$), and vector ($\meddiamond$) effective derivatives are presented. Corresponding $P$-values are presented, with the dashed red line at $p=0.05$ discriminating statistical significance of the fit results. Uncertainties are one s.e.m.}
}\end{figure*}

\begin{figure*}[h]{\docfont
	\includegraphics[width=0.49\textwidth]{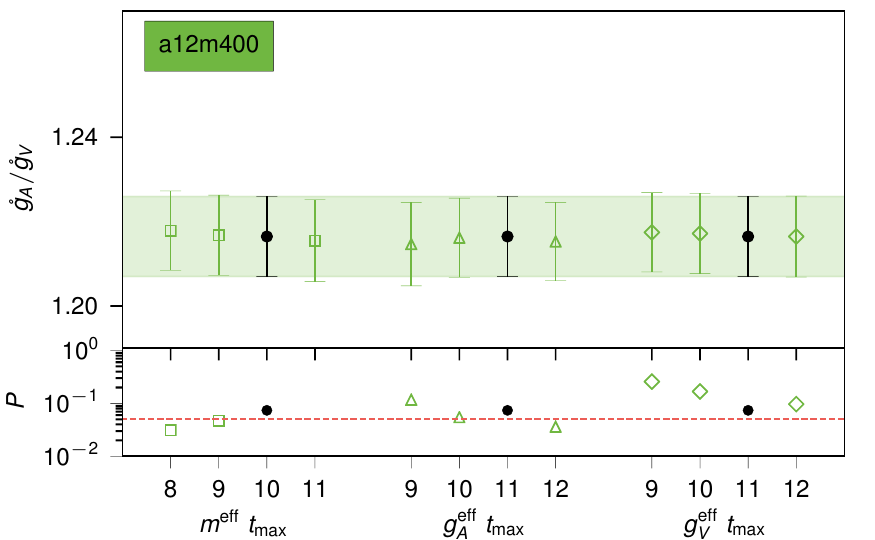}
	\includegraphics[width=0.49\textwidth]{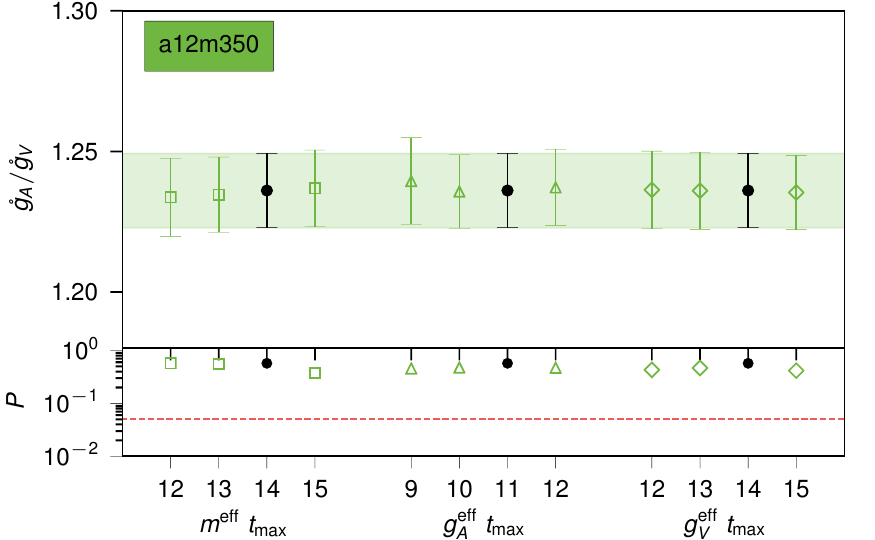}
	\includegraphics[width=0.49\textwidth]{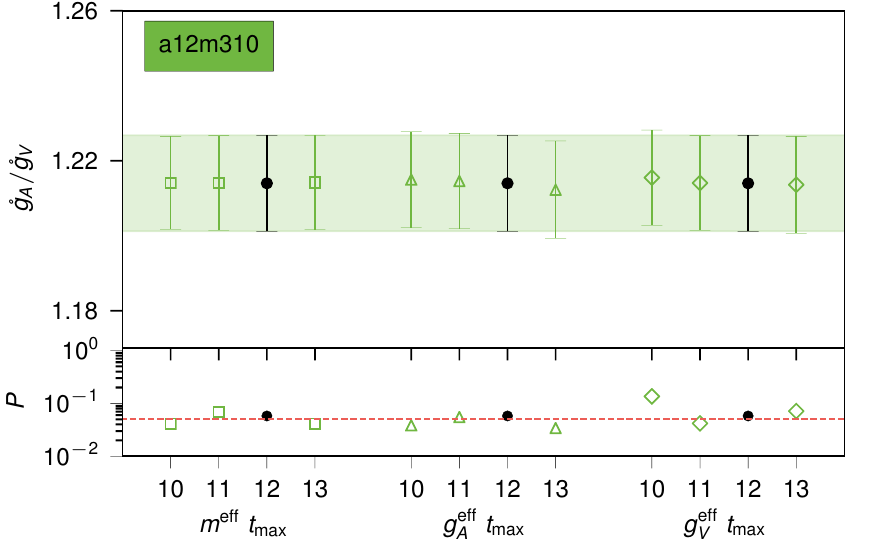}
	\includegraphics[width=0.49\textwidth]{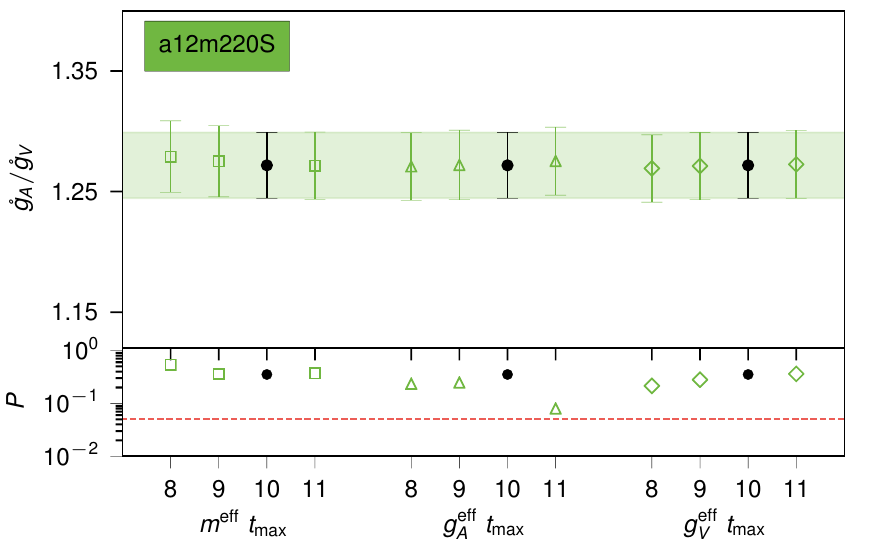}
	\includegraphics[width=0.49\textwidth]{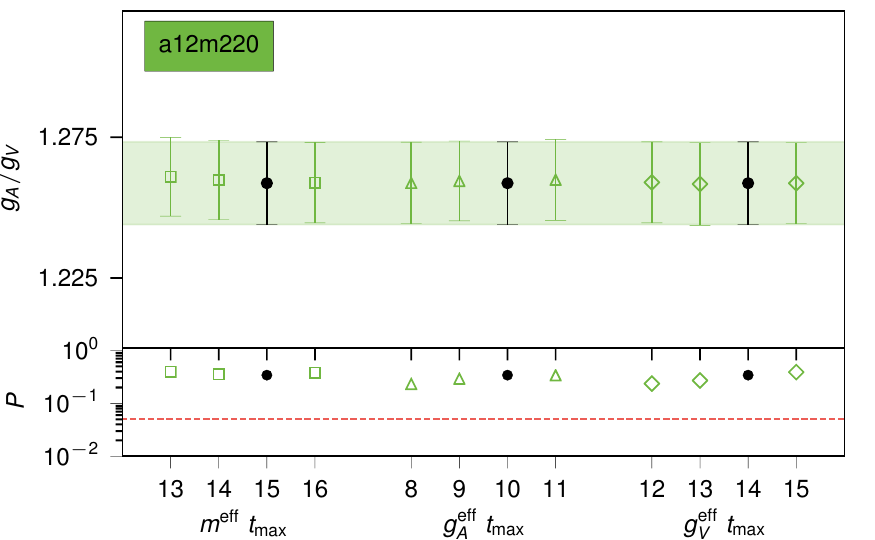}
	\includegraphics[width=0.49\textwidth]{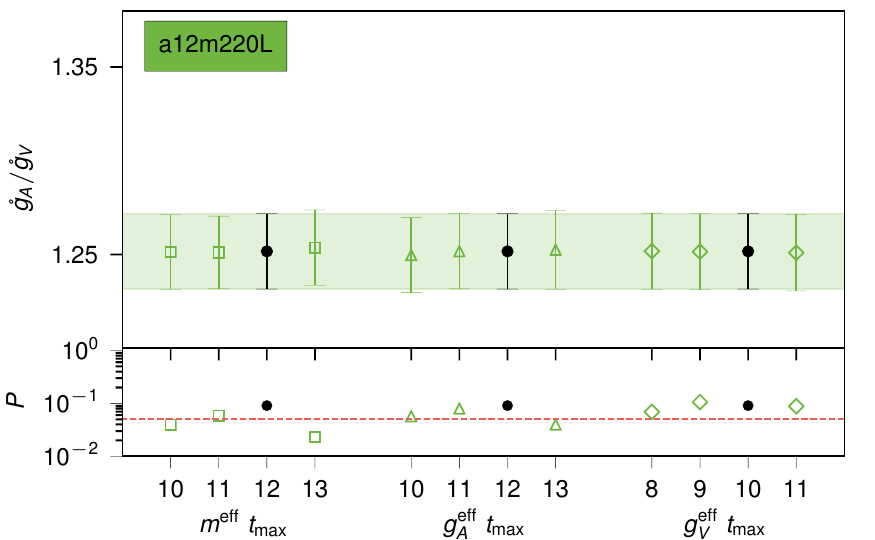}
	\includegraphics[width=0.49\textwidth]{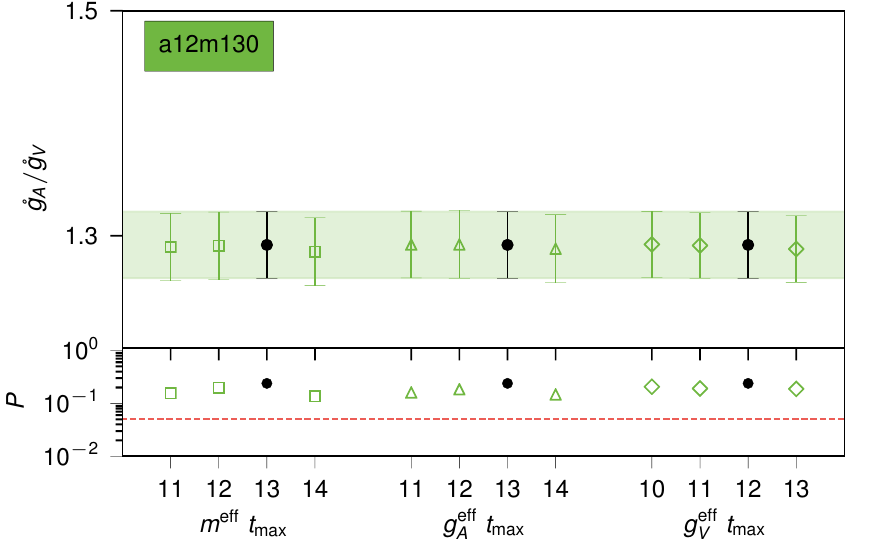}
	\caption{\label{fig:tmax_stability_II}
		{\textbf{Correlator fit $t_{\textrm{max}}$ stability study II.}} Analogous to Extended Data Fig.~\ref{fig:correlator_fitcurves}f and Supplemental Fig.~\ref{fig:tmax_stability_I} for the remaining ensembles.
Uncertainties are one s.e.m.}
}\end{figure*}

\begin{figure*}[h]{\docfont
	\includegraphics[width=0.49\textwidth]{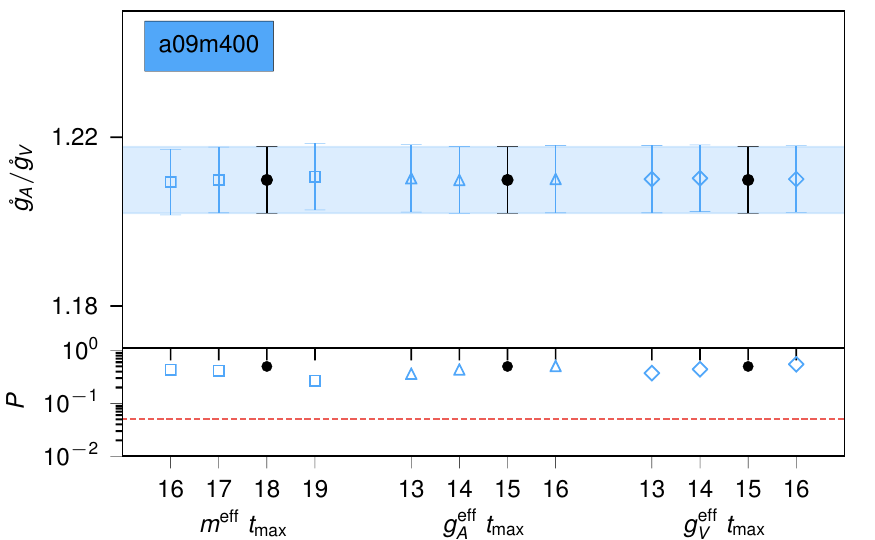}
	\includegraphics[width=0.49\textwidth]{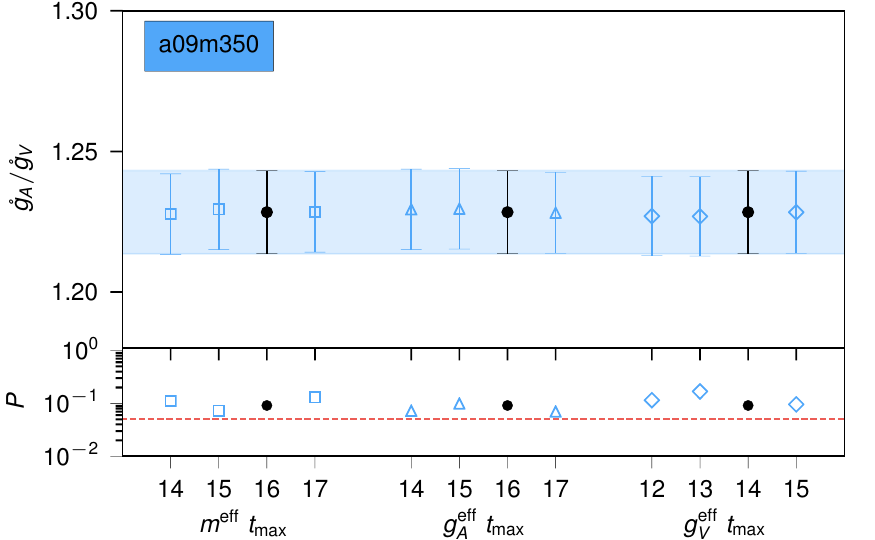}
	\includegraphics[width=0.49\textwidth]{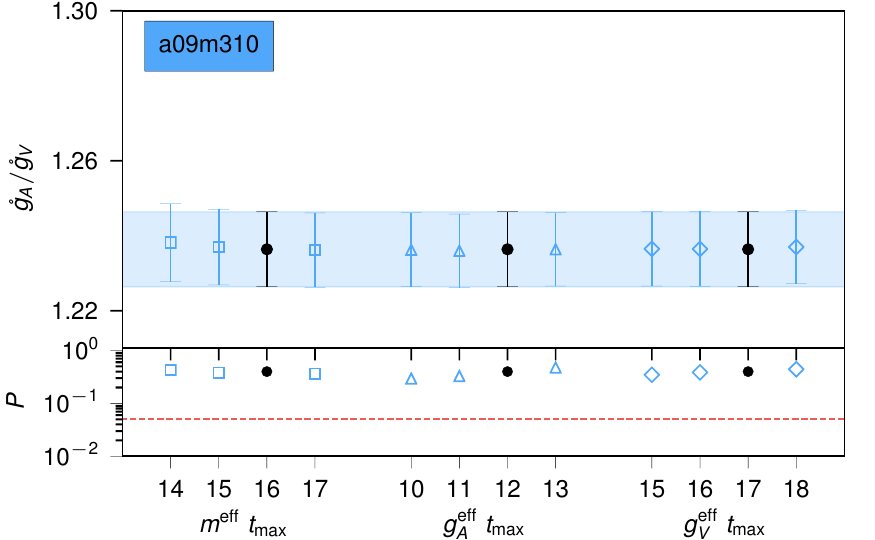}
	\includegraphics[width=0.49\textwidth]{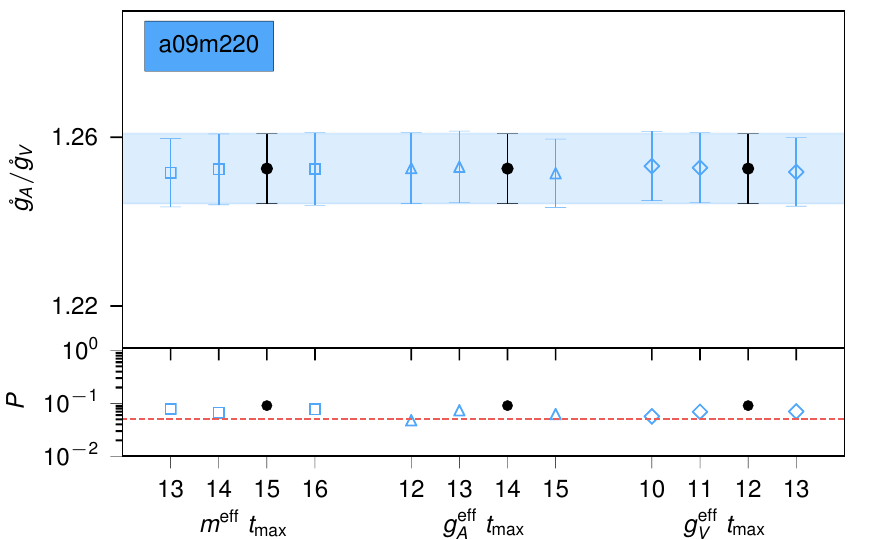}
	\caption{\label{fig:tmax_stability_III}
		{\textbf{Correlator fit stability study III.}} Analogous to Extended Data Fig.~\ref{fig:correlator_fitcurves}f and Supplemental Fig.~\ref{fig:tmax_stability_I} for the remaining ensembles.
Uncertainties are one s.e.m.}
}\end{figure*}

\begin{figure*}[h]{\docfont
	\includegraphics[width=0.49\textwidth]{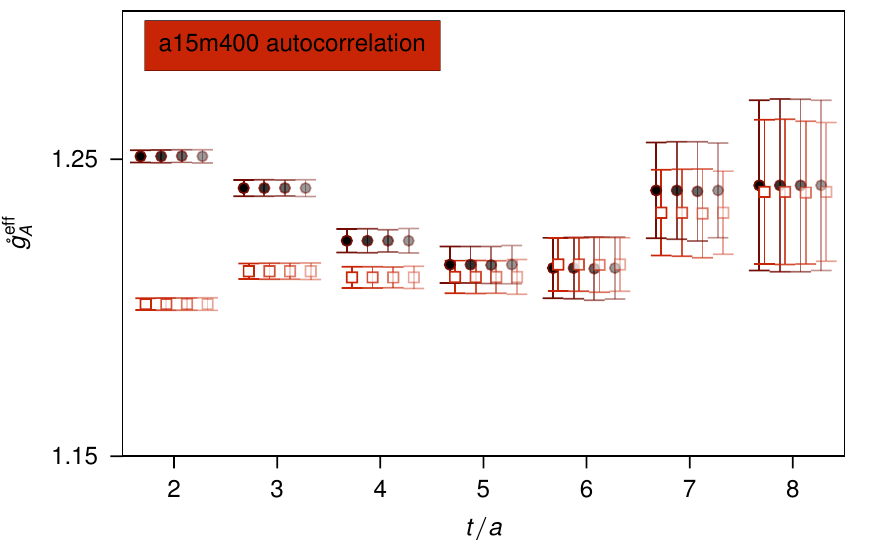}
	\includegraphics[width=0.49\textwidth]{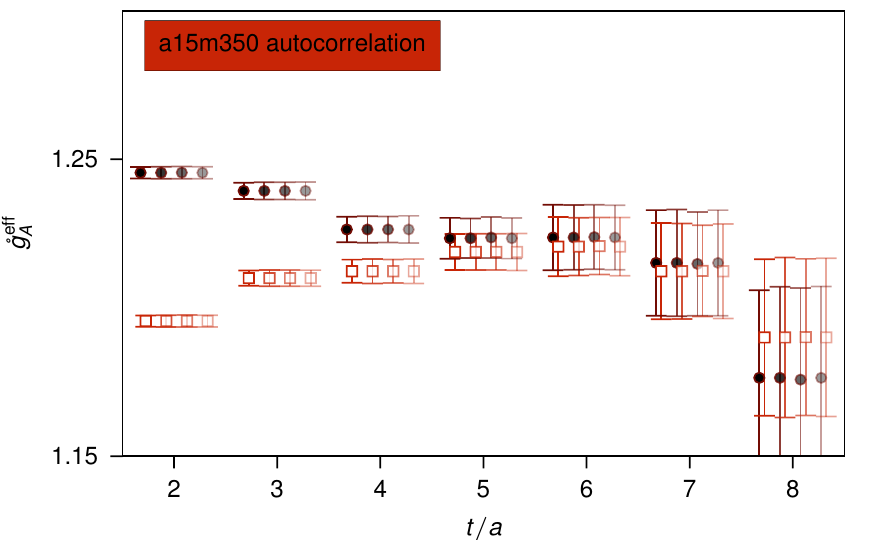}
	\includegraphics[width=0.49\textwidth]{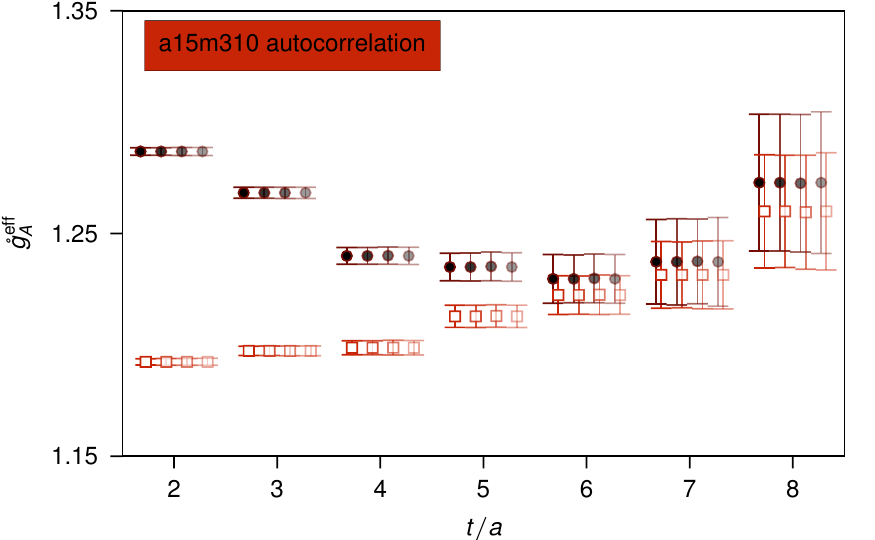}
	\includegraphics[width=0.49\textwidth]{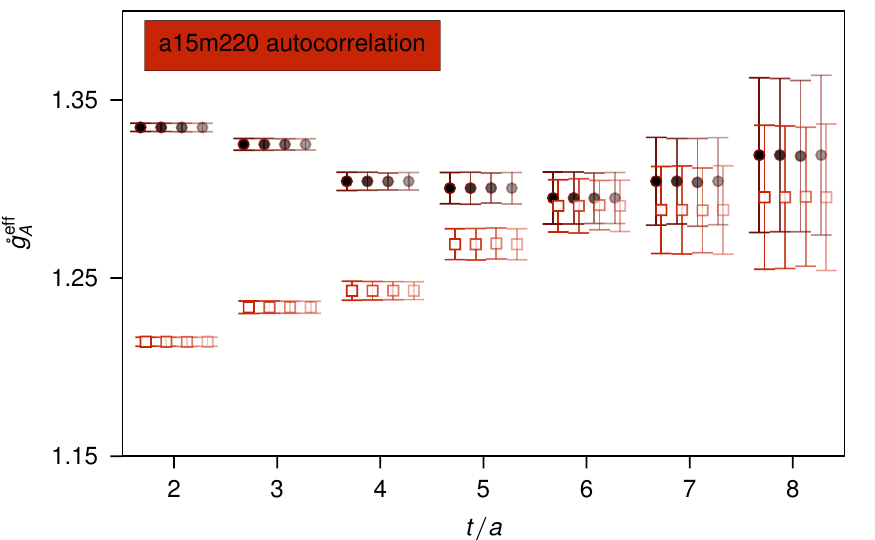}
	\includegraphics[width=0.49\textwidth]{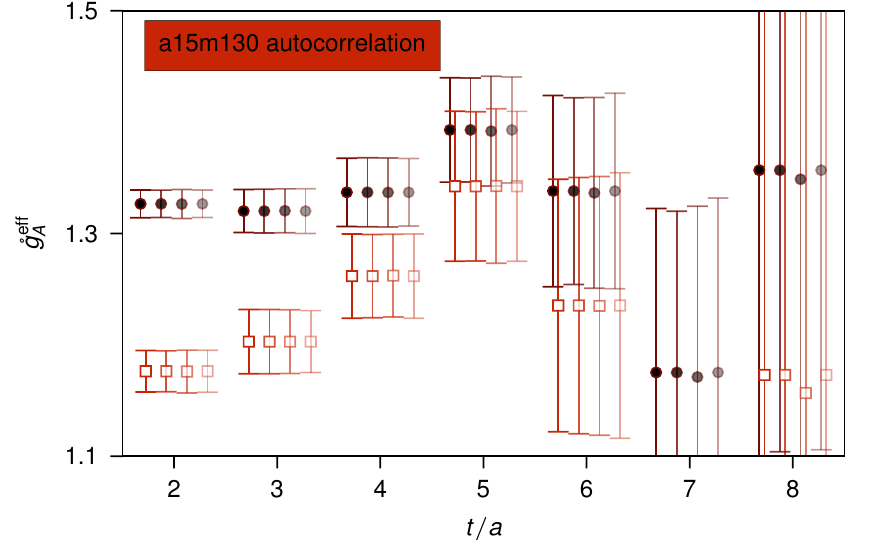}
	\caption{\label{fig:autocorrelation_I}
		{\bf{Autocorrelation study I.}} Analogous to Extended Data Fig.~\ref{fig:flow_auto_study}c and d. Uncertainties are one s.e.m.}
}\end{figure*}

\begin{figure*}[h]{\docfont
	\includegraphics[width=0.49\textwidth]{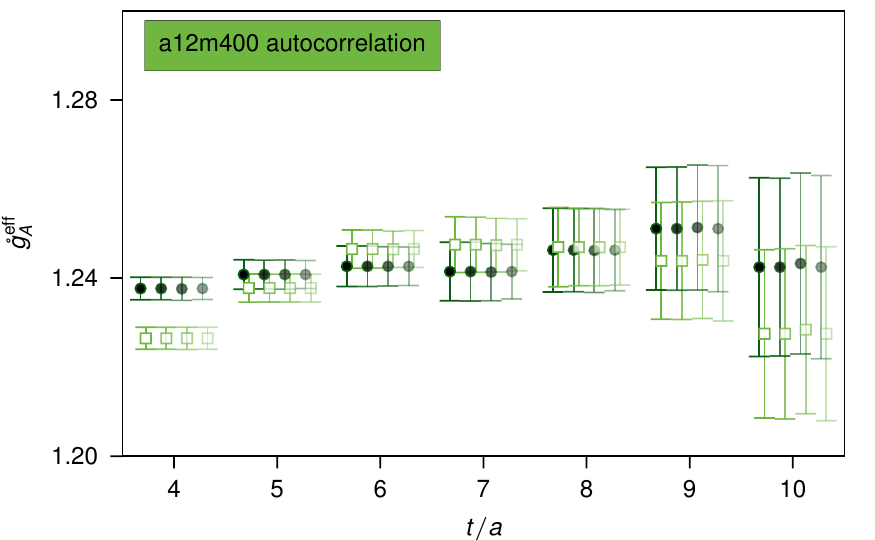}
	\includegraphics[width=0.49\textwidth]{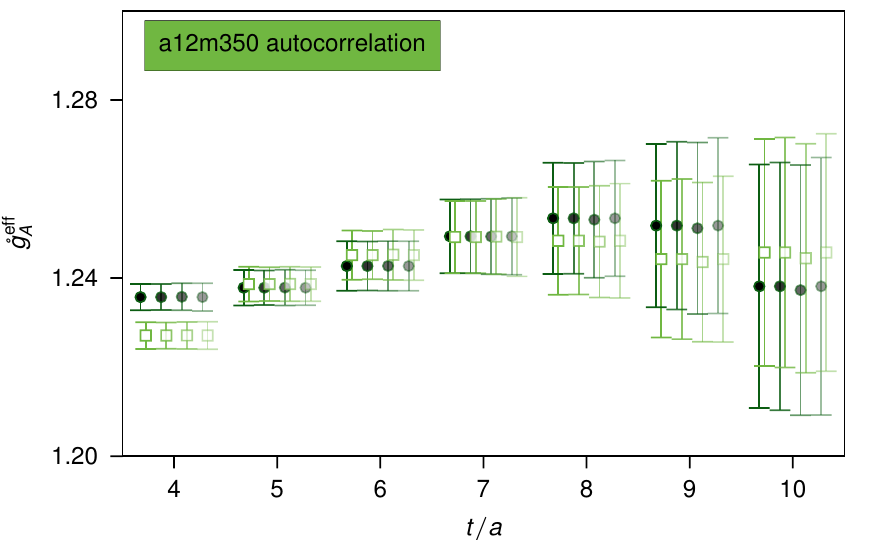}
	\includegraphics[width=0.49\textwidth]{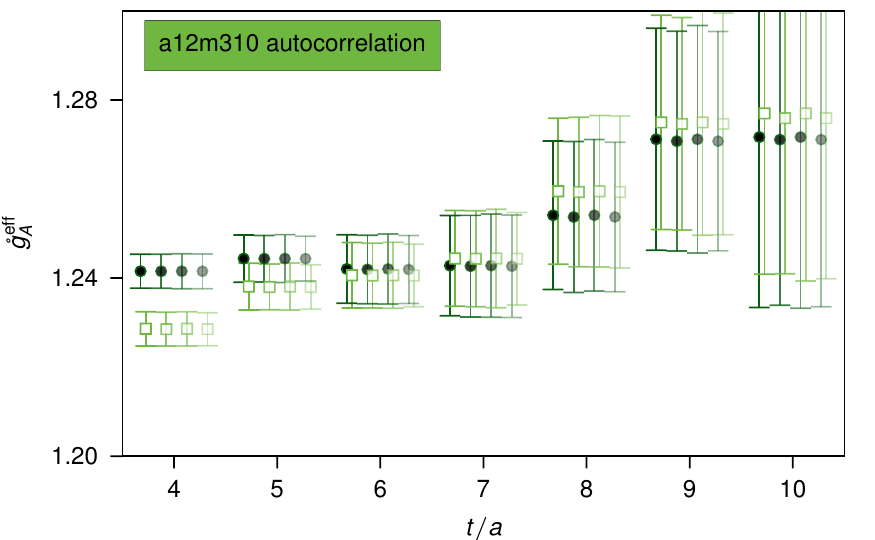}
	\includegraphics[width=0.49\textwidth]{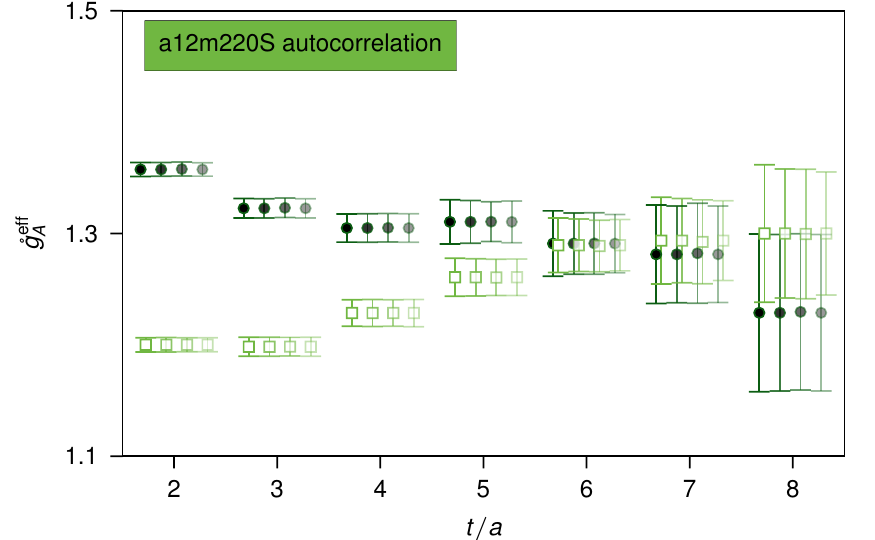}
	\includegraphics[width=0.49\textwidth]{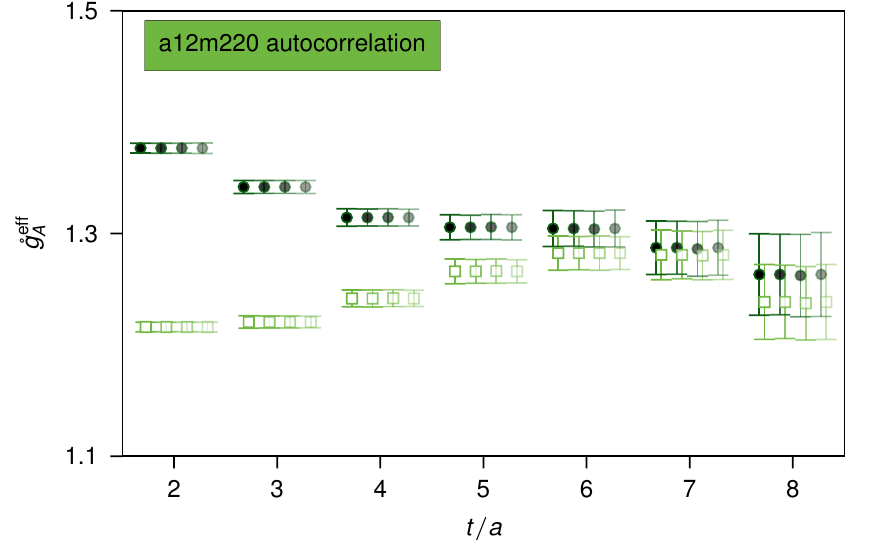}
	\includegraphics[width=0.49\textwidth]{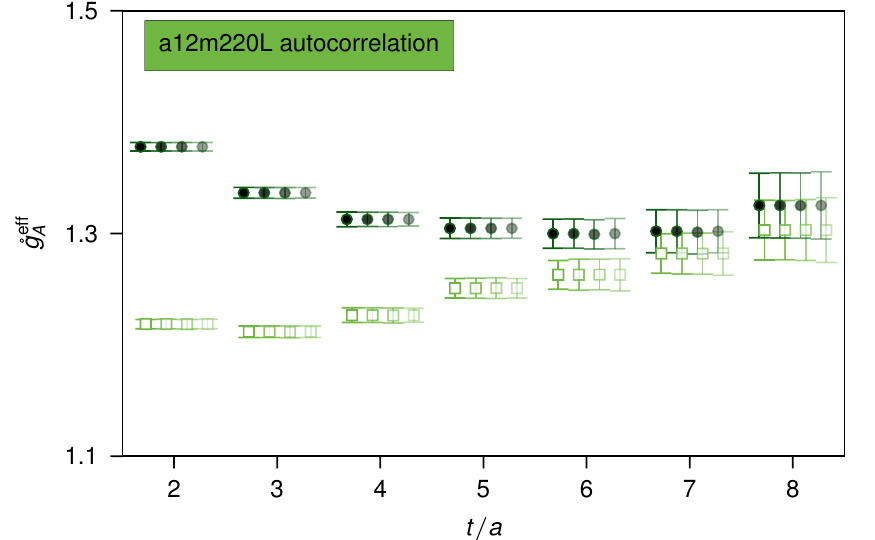}
	\includegraphics[width=0.49\textwidth]{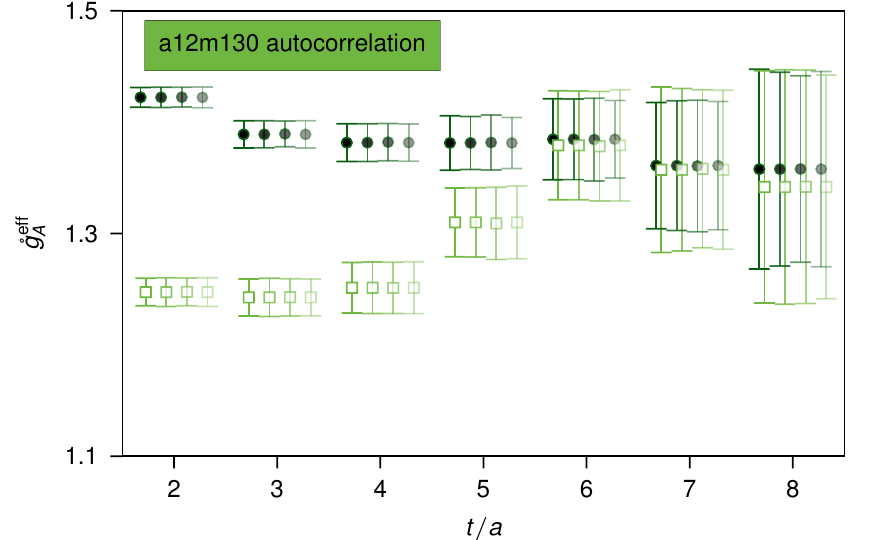}
	\caption{\label{fig:autocorrelation_II}
		{\bf{Autocorrelation study II.}} Analogous to Extended Data Fig.~\ref{fig:flow_auto_study}c and d. Uncertainties are one s.e.m.}
}\end{figure*}

\begin{figure*}[h]{\docfont
	\includegraphics[width=0.49\textwidth]{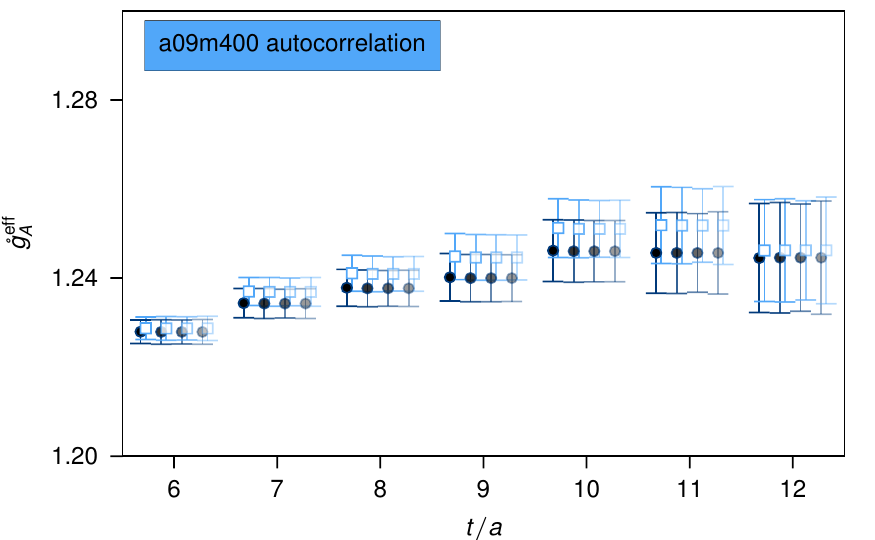}
	\includegraphics[width=0.49\textwidth]{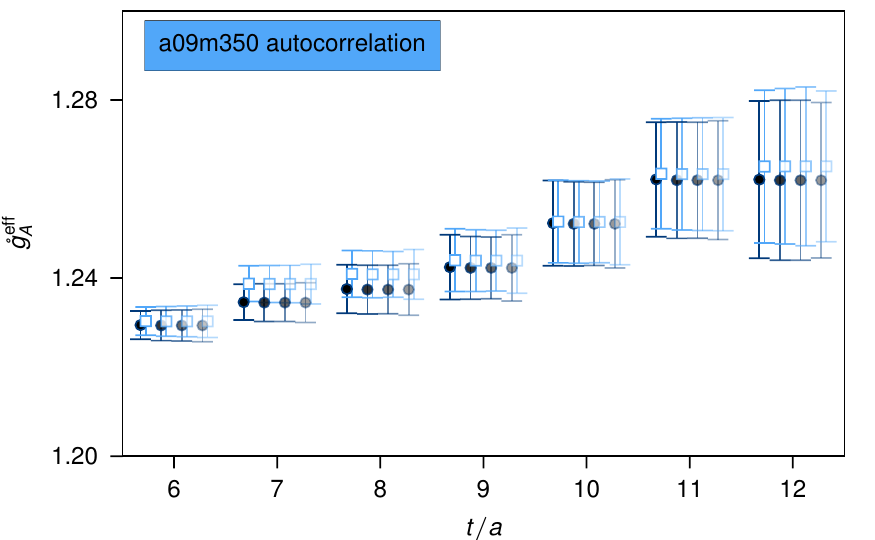}
	\includegraphics[width=0.49\textwidth]{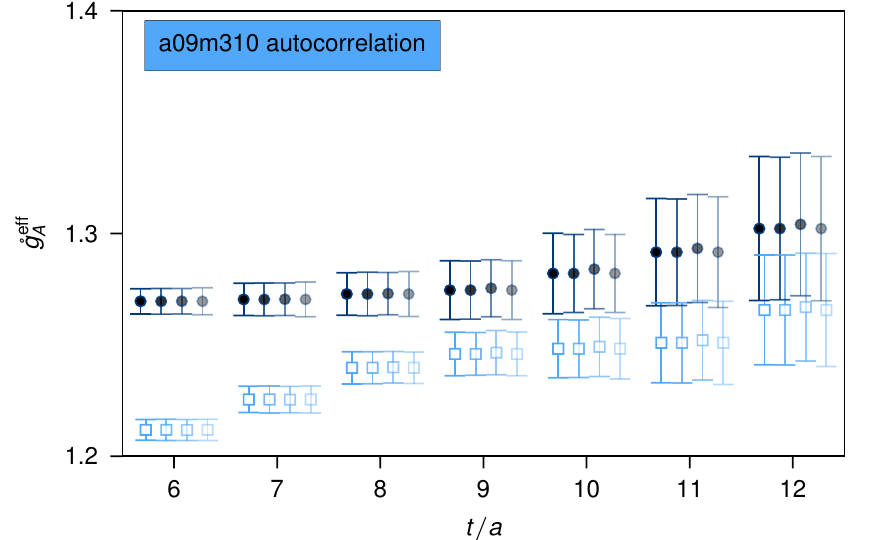}
	\includegraphics[width=0.49\textwidth]{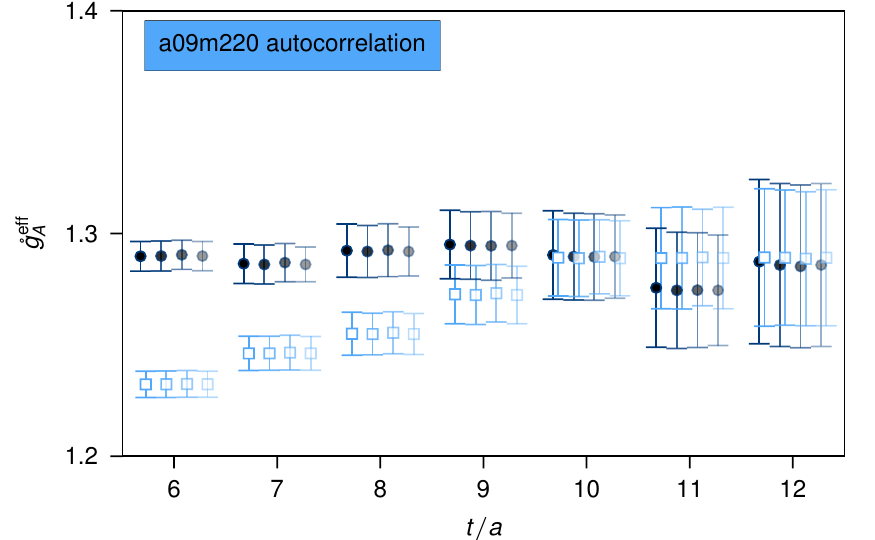}
	\caption{\label{fig:autocorrelation_III}
		{\bf{Autocorrelation study III.}} Analogous to Extended Data Fig.~\ref{fig:flow_auto_study}c and d. Uncertainties are one s.e.m.}
}\end{figure*}

\begin{figure*}[h]{\docfont
\includegraphics[width=0.49\textwidth]{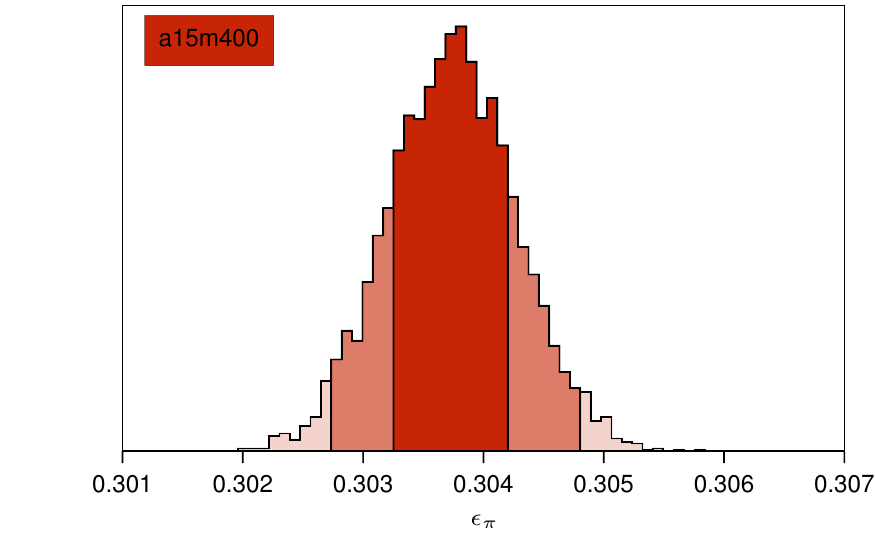}
\includegraphics[width=0.49\textwidth]{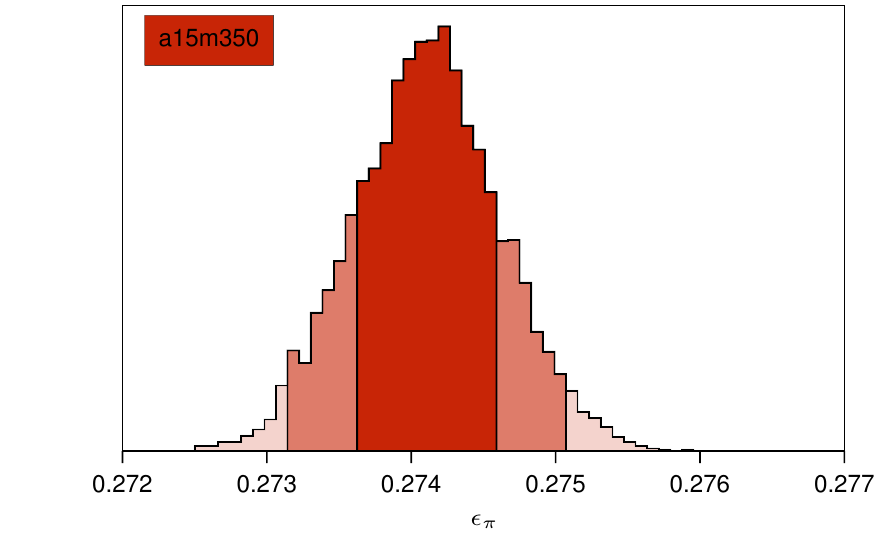}
\includegraphics[width=0.49\textwidth]{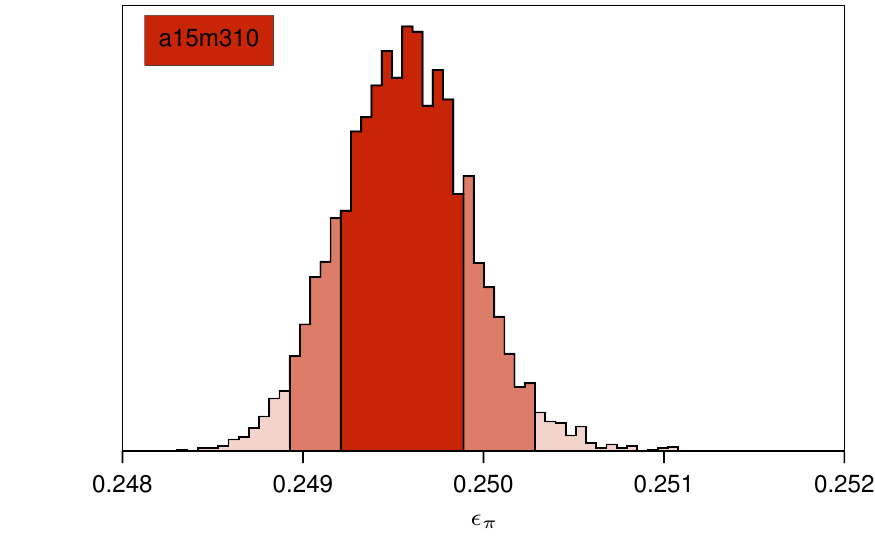}
\includegraphics[width=0.49\textwidth]{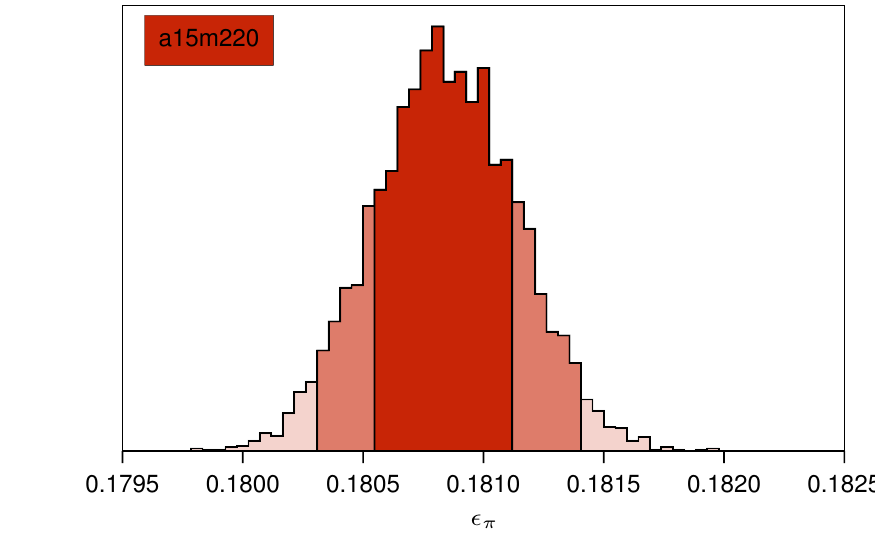}
\includegraphics[width=0.49\textwidth]{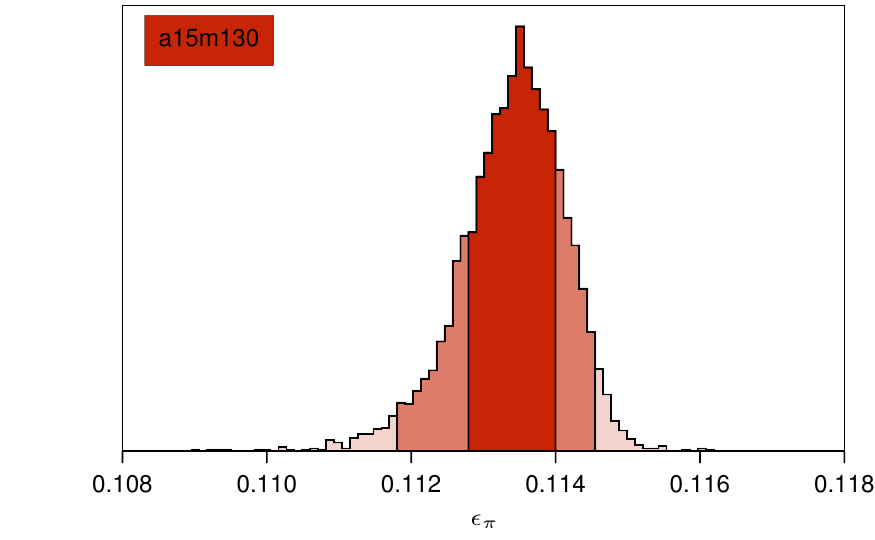}
\caption{\label{fig:epi_I}
{\textbf{Histograms I for $\epsilon_\pi$.}} Analogous to Extended Data Fig.~\ref{fig:correlator_fitcurves}d. The inner shaded regions correspond to the 68\% and 95\% confidence intervals.
}
}\end{figure*}

\begin{figure*}[h]{\docfont
\includegraphics[width=0.49\textwidth]{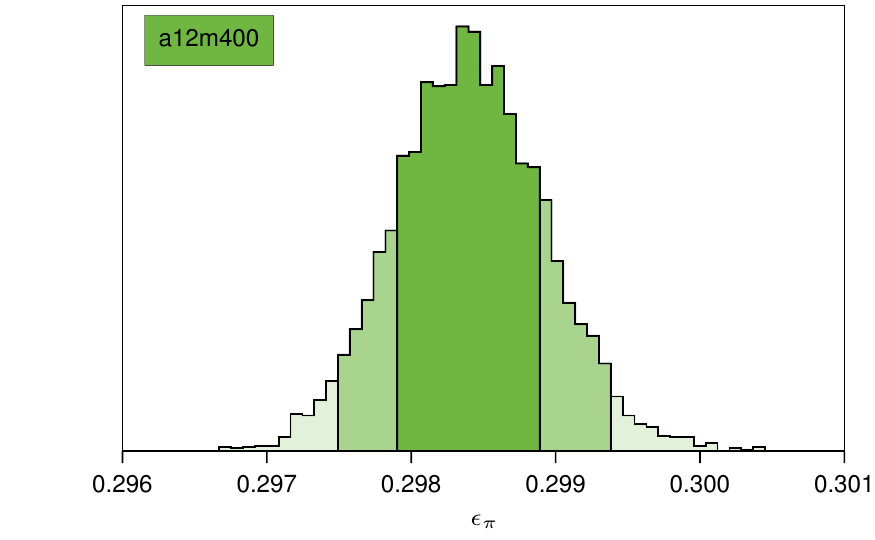}
\includegraphics[width=0.49\textwidth]{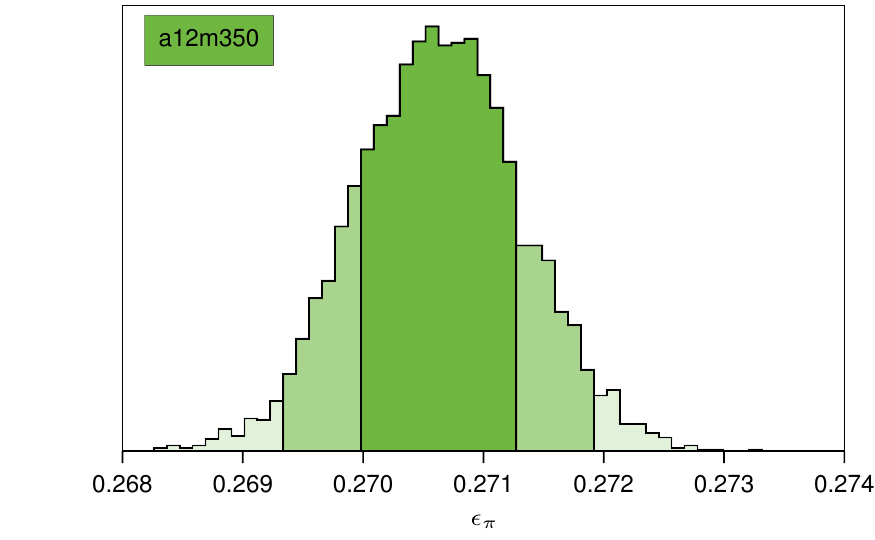}
\includegraphics[width=0.49\textwidth]{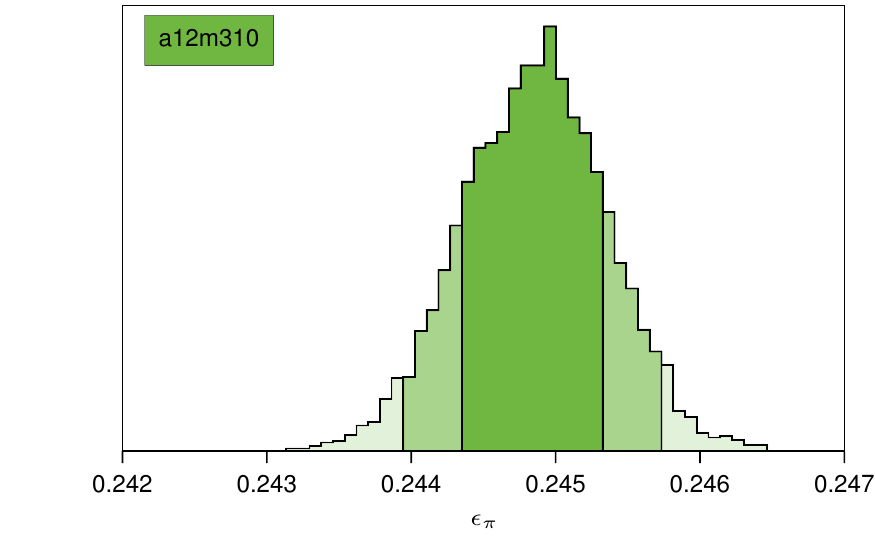}
\includegraphics[width=0.49\textwidth]{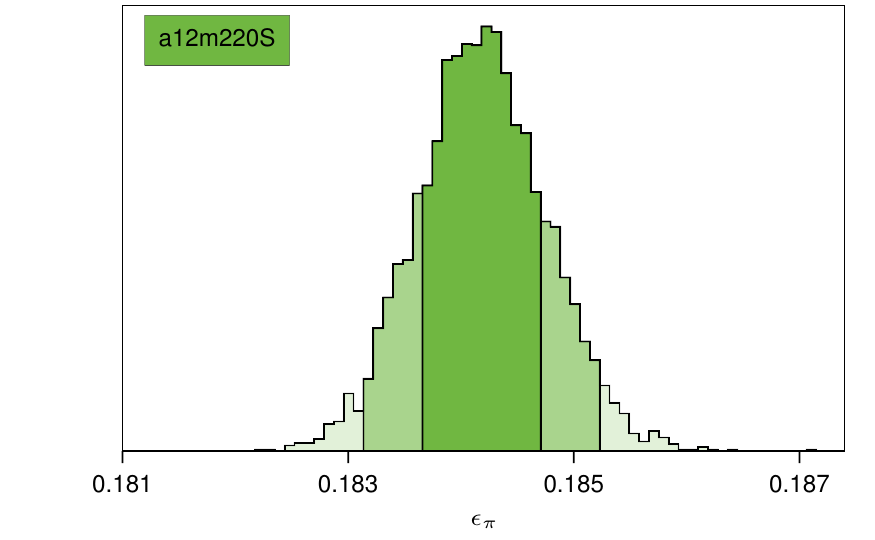}
\includegraphics[width=0.49\textwidth]{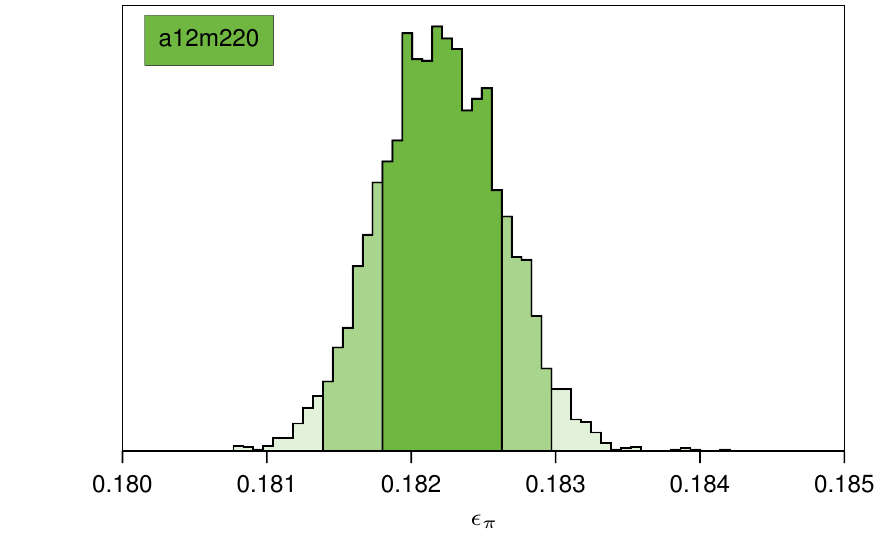}
\includegraphics[width=0.49\textwidth]{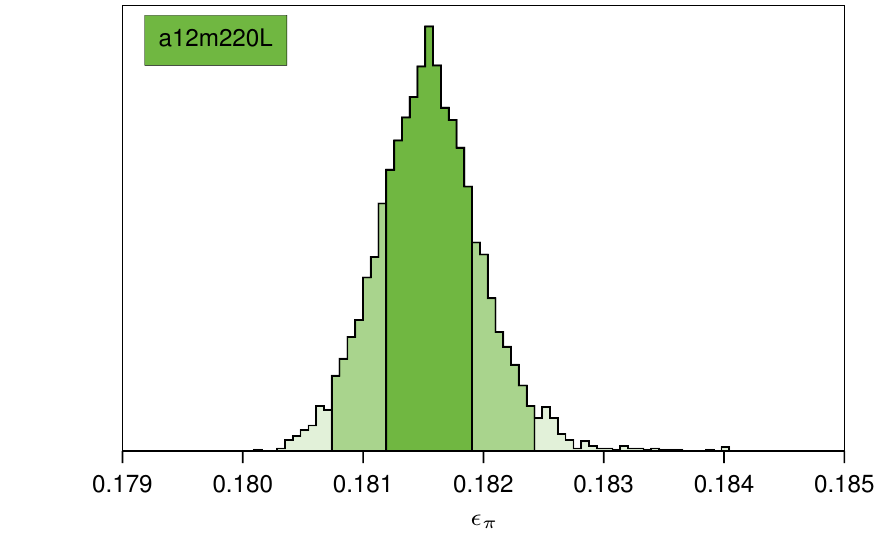}
\includegraphics[width=0.49\textwidth]{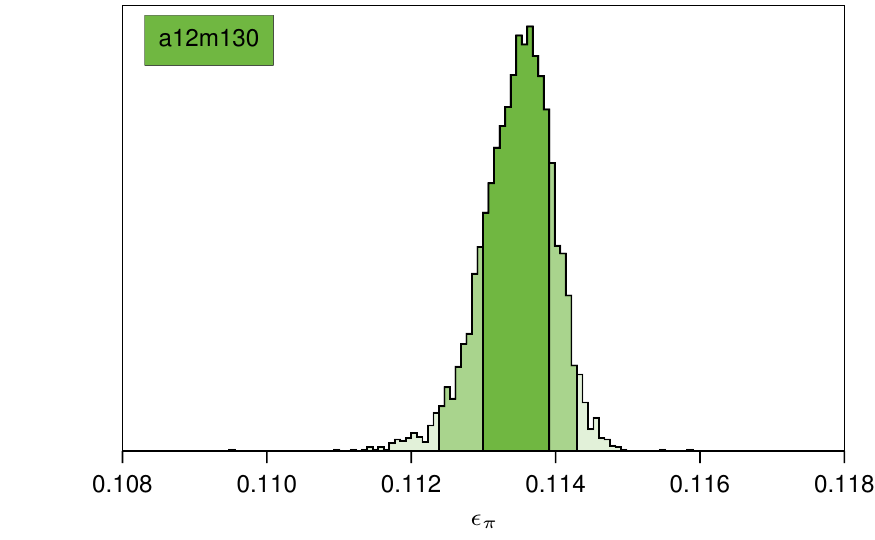}
\caption{\label{fig:epi_II}
{\textbf{Histograms II for $\epsilon_\pi$.}} Analogous to Extended Data Fig.~\ref{fig:correlator_fitcurves}d and Supplemental Fig.~\ref{fig:epi_I} for the remaining ensembles.}
}\end{figure*}

\begin{figure*}[h]{\docfont
\includegraphics[width=0.49\textwidth]{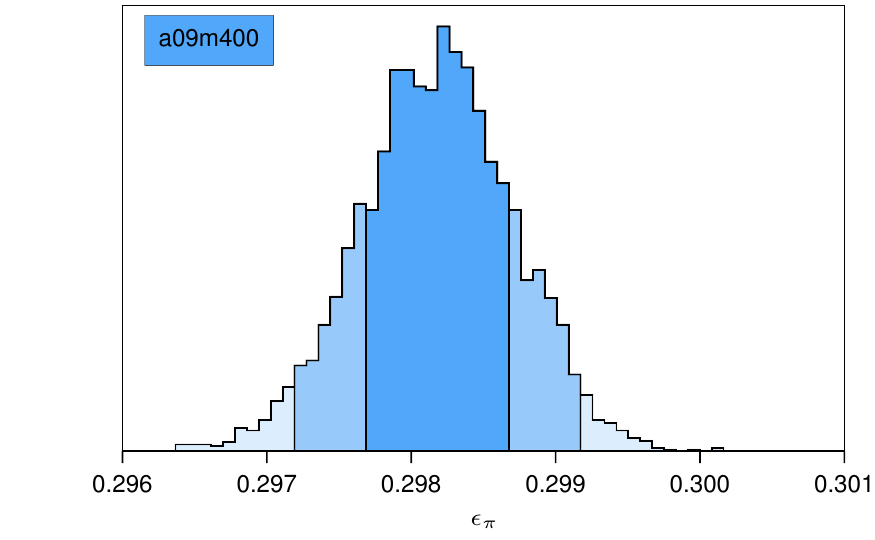}
\includegraphics[width=0.49\textwidth]{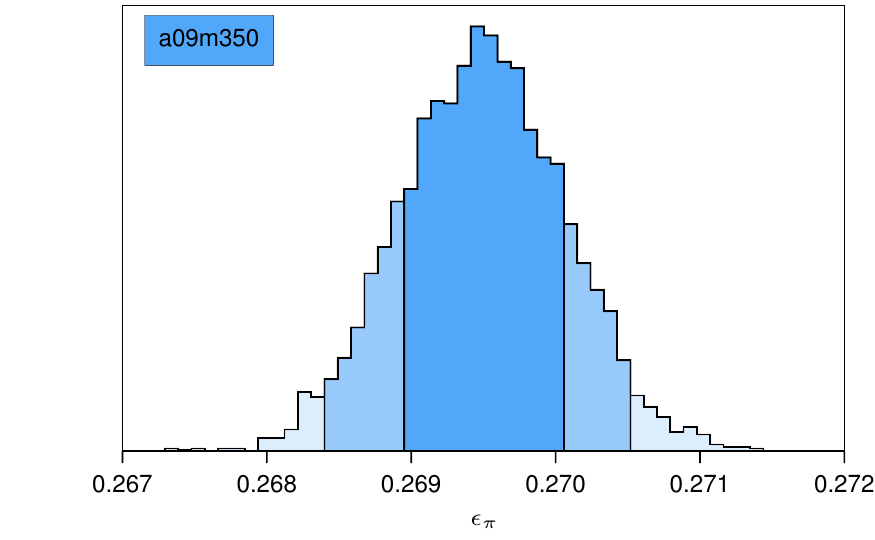}
\includegraphics[width=0.49\textwidth]{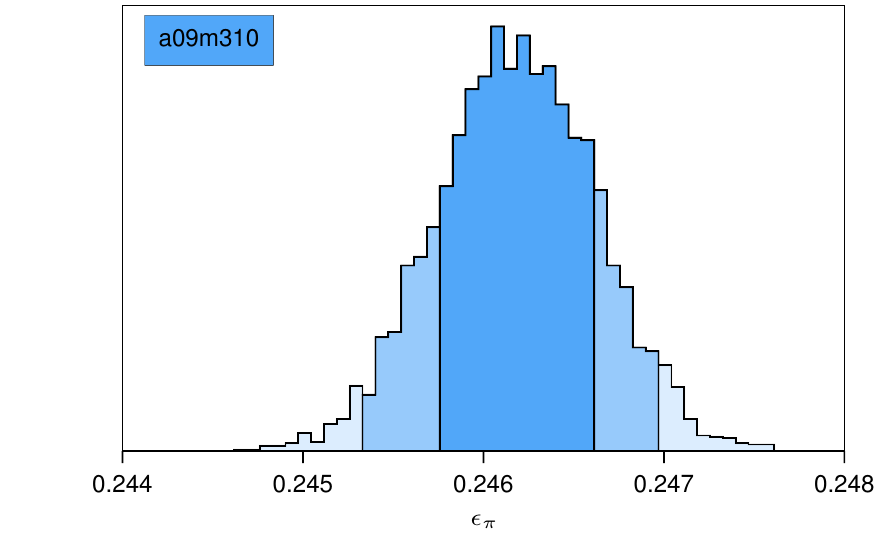}
\includegraphics[width=0.49\textwidth]{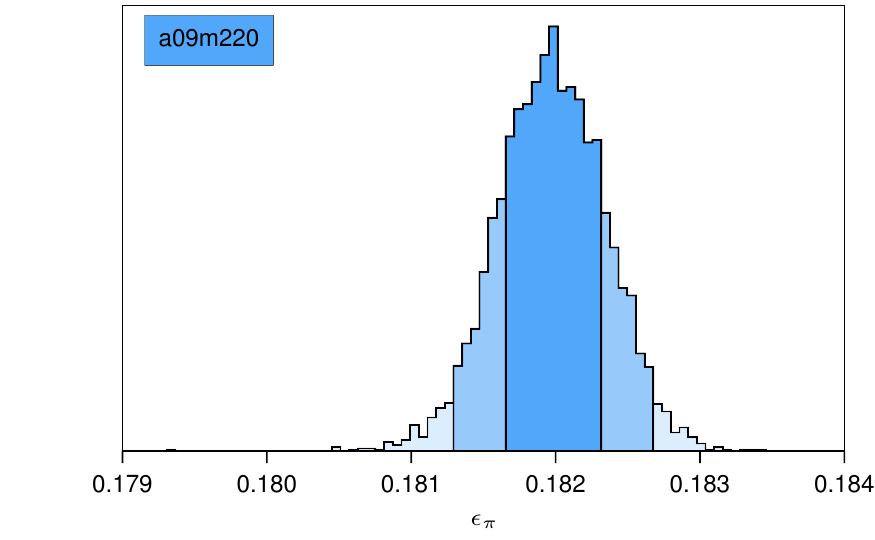}
\caption{\label{fig:epi_II}
{\textbf{Histograms III for $\epsilon_\pi$.}} Analogous to Extended Data Fig.~\ref{fig:correlator_fitcurves}d and Supplemental Fig.~\ref{fig:epi_I} for the remaining ensembles.}
}\end{figure*}

\clearpage
\bibliographystyleM{revtex_nature}
\bibliographyM{c51_bib}

\end{document}